%% file: paper.tex
\documentclass[sigconf, nonacm]{acmart}
\AtBeginDocument{%
  }
\usepackage{pifont}
\usepackage{pgfplotstable}
\usepackage{paralist}
\usepackage{amsmath}
\usepackage{multirow}
\usepackage{shortcuts}
\usepackage{epsfig,endnotes}
\usepackage{tikz}
\usepackage{pgfplots}
\usepackage[ruled,vlined,linesnumbered]{algorithm2e}

\usepackage[noend]{algpseudocode}

\usepackage{amsfonts}
\usepackage{url}
\usepackage{tikz}
\usepackage{array}
\usepackage{booktabs,adjustbox}
\usepackage{pstricks}
\usetikzlibrary{calc,positioning,shapes.geometric,shapes.symbols,shadows,arrows}
\usepackage{wasysym}
\usepackage{shortcuts}
\usepackage{paralist}

\usepackage[justification=justified]{caption}
\usetikzlibrary{arrows}
\usetikzlibrary{patterns}
\usepackage{times}
\usepackage{listings}
\usepackage{multicol}
\usepackage{lipsum}
\usepackage{adjustbox}
\usepackage[font=bf,skip=\baselineskip]{caption}
\usepackage{tabularx}
\usepackage{multirow}
\usepackage{shortcuts}

\usepackage[english]{babel}
\usepackage{listings}
\lstdefinestyle{cstyle}{
  language=C,
  basicstyle=\footnotesize\ttfamily,
  keywordstyle=\color{blue!55!black}\bfseries,
  commentstyle=\color{green!40!black}\itshape,
  stringstyle=\color{orange!65!black},
  identifierstyle=\color{black},
  numberstyle=\tiny\color{gray},
  columns=fullflexible,
  keepspaces=true,
  xleftmargin=5pt,
  numbersep=3pt,
  escapechar=|,
  showstringspaces=false
}
\definecolor{codebg}{RGB}{248,248,248}       
\definecolor{codekw}{RGB}{0,128,128}          
\definecolor{codestr}{RGB}{160,50,30}         

\lstdefinestyle{mintedlike}{
  language=C,
  basicstyle=\footnotesize\ttfamily,
  keywordstyle=\color{codekw},
  stringstyle=\color{codestr},
  backgroundcolor=\color{codebg},
  showstringspaces=false,
  xleftmargin=5pt,
  aboveskip=0pt,
  belowskip=0pt,
  breaklines=true,
  columns=fullflexible,
  keepspaces=true,
}
\usepackage{subcaption}
\usepackage{minted}
\usepackage{enumitem}
\setitemize{noitemsep,topsep=0pt,parsep=0pt,partopsep=0pt}

\usepackage{adjustbox}
\usepackage{xurl}

\newcommand{\ma}[1]{\textcolor{red}{#1}}

\AtBeginDocument{%
  }

\usepackage[all]{nowidow}

\usepackage{caption}
\usepackage{hyperref}

\hypersetup{
  colorlinks=true,
  linkcolor=blue,
  citecolor=green,
  filecolor=magenta,
  urlcolor=cyan
}

\usepackage{wrapfig}
\usepackage{tikz}
\usepackage{enumitem}

 \newcommand{\parag}[1]{\vspace{0.05in}\noindent{\bf{#1}.}}
 
\begin{document}

\newcounter{rcounter}
\newcommand{\result}[1]{\refstepcounter{rcounter}
\vspace{0.5em}\noindent\fbox{%
    \parbox{0.95\linewidth}{%
  \vspace{0.3em}{\bf
  {Finding~\arabic{rcounter}:~}} {#1}
\vspace{0.3em}
  }
}
\vspace{0.5em}

}
\newcommand{\sys}{{\sc REBench}\xspace}
\title{\sys: A Comprehensive Benchmark for LLM-based Type Inference and Name Recovery in Stripped Binaries}

\title{\sys: A Procedural, Fair-by-Construction Benchmark for LLMs on Stripped-Binary Types and Names}
\subtitle{{\LARGE \bf (Extended Version)}}








\author{Jun Yeon Won}
\affiliation{%
  \institution{Ohio State University}
  \city{Columbus}
  \state{OH}
  \country{United States}
}
\email{won.126@osu.edu}

\author{Xin Jin}
\affiliation{%
  \institution{Ohio State University}
  \city{Columbus}
  \state{OH}
  \country{United States}
}
\email{jin.967@osu.edu}

\author{Shiqing Ma}
\affiliation{%
  \institution{University of Massachusetts Amherst}
  \city{Amherst}
  \state{MA}
  \country{United States}
}
\email{shiqingma@umass.edu}

\author{Zhiqiang Lin}
\affiliation{%
  \institution{Ohio State University}
  \city{Columbus}
  \state{OH}
  \country{United States}
}
\email{zlin@cse.ohio-state.edu}

\begin{abstract}
    \input{paper/00_abs.tex}
\end{abstract}

\maketitle

\renewcommand{\thefootnote}{\fnsymbol{footnote}}
\footnotetext[1]{This paper is an extended version of the work published in AIWare 2026, titled `REBench: A Procedural, Fair-by-Construction Benchmark for LLMs on Stripped-Binary Types and Names.'}

\section{Introduction}\label{sec:intro}
\input{paper/01_introduction.tex}

\section{Background \& Motivation}\label{sec:back}
\input{paper/02_background.tex}

\input{paper/03_overview.tex}

\section{\sys Design}\label{sec:dataset}
\input{paper/04_1_data-curation}

\input{paper/04_2_metrics_definition}





\section{Discussion \& Future Work}\label{sec:discussion}
\input{paper/08_future_work.tex}

\section{Conclusion}\label{sec:conclusion}

\input{paper/10_conclusion.tex}

\section{Data Availability}
The data normalization process can be reproduced using the provided scripts, which include implementations for both {\sf Ghidra} and {\sf IDA Pro}. 
The corresponding scripts and datasets are available at: \url{https://github.com/OSUSecLab/REBench} and \url{https://zenodo.org/records/19899116}

\bibliographystyle{ACM-Reference-Format}
\bibliography{paper}







\end{document}

%% file: paper/00_abs.tex
%
Large Language Models (LLMs) have achieved remarkable progress in recent years, driving their adoption across a wide range of domains, including computer security. 
In reverse engineering, LLMs are increasingly applied to critical tasks such as function and variable name recovery and type inference. 
However, despite the rapid growth of research in this area, progress has been hindered by the absence of a standardized dataset. 
Existing studies rely on disparate datasets, preprocessing pipelines, and evaluation metrics, making fair comparisons between approaches difficult and obscuring a clear understanding of LLM capabilities in binary analysis.
To address these challenges, we present \sys, a comprehensive benchmark dataset for evaluating LLMs on binary reverse engineering tasks. 
\sys consolidates a superset of existing datasets, comprising hundreds of millions of lines of source code and a diverse collection of binaries spanning multiple architectures and optimization levels. 
\sys adopts a knowledge-base-driven methodology that stores byte-level stack information to generate ground truth, ensuring that task difficulty is preserved while maintaining universal applicability. 
This design enables fair evaluation across tasks while avoiding simplifications that could bias results.
As a use case, we apply \sys to measure the reverse engineering performance of LLMs and the result demonstrates difficulties in complex tasks. 

%% file: paper/01_introduction.tex
Large Language Models (LLMs) have recently gained significant attention due to rapid advances in model architectures, training techniques, and the increasing availability of large-scale datasets. 
Built on transformer architectures~\cite{vaswani2017attention}, LLMs have demonstrated remarkable success in natural language processing (NLP) tasks such as translation and question answering, with widely known examples including OpenAI's ChatGPT~\cite{chatGPT}. 
The adoption of transfer learning~\cite{torrey2010transfer} has further expanded their applications across diverse domains, including healthcare~\cite{lee2020biobert,huang2019clinicalbert} and finance~\cite{yang2023finchain}, where they have improved both efficiency and automation.

In the field of computer security, researchers are increasingly exploring LLMs for binary reverse engineering tasks, including function summarization~\cite{jin2023binary}, symbol name recovery in stripped binaries~\cite{xu2023lmpa,jiang2025beyond,xie2024resym}, and type inference, as LLMs can interpret human-readable representations, such as source code.
These developments highlight the potential of LLMs to advance program analysis and support security-critical activities such as malware analysis, vulnerability discovery, and software maintenance. 
Although early results are promising, evaluating the effectiveness of LLMs in this domain remains a fundamental challenge.

The conventional approach to applying LLMs in reverse engineering involves fine-tuning pre-trained models on task-specific datasets. 
Although this strategy can substantially improve performance, it also introduces variability depending on the goals and experimental setups of individual studies. 
For example, \textsc{SymGen}~\cite{jiang2025beyond} reports an F1 score of 0.351 on an x64 dataset when reproducing 
\textsc{AsmDepictor}~\cite{kim2023transformer}, which itself reports only 0.05 under the same comparison. 
Meanwhile, \textsc{AsmDepictor} claims a considerably higher F1 score of 0.715 on its own evaluation dataset, revealing significant inconsistencies across studies.

A key barrier to resolving such discrepancies is the absence of a standardized dataset. 
Existing works evaluate models on heterogeneous datasets covering different processor architectures (e.g., x86, ARM, MIPS), compiler optimization levels (e.g., \texttt{O0}–\texttt{O3}), and program families (e.g., coreutils, binutils). 
Even when datasets overlap, inconsistencies arise from differences in preprocessing (e.g., assembly vs. decompiled code), naming conventions (e.g., memory vs. mem), and evaluation metrics. 
For instance, some studies consider only exact string matches when evaluating predicted variable names, while others award partial credit for semantically similar predictions. 
This lack of consistency prevents the community from fairly and systematically assessing progress.

To address these challenges, we present \sys, the first comprehensive dataset for evaluating LLMs on binary reverse engineering tasks. 
\sys is designed to be broad, standardized, and extensible. 
It consolidates datasets previously used in reverse engineering research~\cite{xu2023improving,jin2022symlm,pei2020trex}, covering four major processor architectures (x86, x64, ARM 32-bit, MIPS 32-bit) and multiple compiler optimization levels (\texttt{O0}–\texttt{O3}). 
To ensure compatibility across analysis pipelines, we convert all binaries into a universal format using a general-purpose decompiler. 
Crucially, \sys introduces a knowledge-base-driven data normalization process that preserves task difficulty while ensuring consistent ground-truth labeling across datasets. 
After eliminating both duplicated and semantically trivial functions, \sys contains 29.94 million lines of decompiled code on the x64 architecture alone.

\parag{Contributions} We make the following contributions. 
\begin{packeditemize}
    \item \textbf{Development of \sys:} We introduce \sys, a large-scale benchmark dataset normalized into a universal input format, enabling fair evaluation of LLMs in binary reverse engineering.

    \item \textbf{Reflection of Real-world Experiment Setup:} \sys preserves the inherent difficulty of reverse engineering by closely following real-world experimental settings, avoiding oversimplifications that could cause bias in evaluation.

    \item \textbf{Completeness of Ground Truth:} We construct a knowledge base by leveraging byte-level stack layout information, ensuring complete and reliable ground truth that overcomes limitations of existing decompilers.
    

\end{packeditemize}


%% file: paper/02_background.tex
\subsection{Problem Definition}\label{sec:problem}
The field of binary analysis is rapidly evolving, driven in part by the growing application of LLMs. 
Establishing a standardized benchmark to evaluate LLMs in this domain is not only beneficial but essential for understanding their true capabilities. 
In this work, we focus on two fundamental reverse engineering tasks, name recovery and type inference, both of which are especially challenging in the context of stripped binaries, where symbol information has been removed. 
This scenario closely mirrors real-world reverse engineering tasks encountered in security-critical domains such as malware analysis and vulnerability research.

\begin{packeditemize}
    \item {\bf Name Recovery} seeks to predict human-readable, semantically meaningful names for functions, arguments, and variables, thereby improving code comprehension.
    \item {\bf Type Inference} aims to identify the data types of variables, return values, and function arguments, which is critical for reasoning about program semantics.
\end{packeditemize}
%
%
%

%% file: paper/03_overview.tex
\label{overview}

\subsection{Related Works}\label{sec:rw}
Binary reverse engineering involves analyzing and reconstructing machine code to understand program behavior, structure, and functionality. 
In recent years, research in this area has increasingly incorporated machine learning techniques, and more recently, LLMs. 
We provide an overview of related work, focusing on the two primary tasks addressed in this paper, name recovery and type inference, along with the datasets commonly used to evaluate them. 
We also briefly introduce additional reverse engineering tasks to illustrate the breadth of research in this field.

\parag{(I) Type Inference} 
Several works have explored automatic type inference in binaries:
\begin{packeditemize}
\item \textsc{StateFormer}~\cite{pei2021stateformer} employs generative state modeling for type inference across 33 datasets, evaluating performance via exact name matching.
\item \textsc{DIRTY}~\cite{chen2022augmenting} uses a transformer-based model for type inference, restricting its dataset to binaries containing fewer than one million decompiled functions.
\item \textsc{Osprey}~\cite{zhang2021osprey} applies probabilistic techniques to two benchmarks comprising 106 programs, with outcomes also measured by exact name matching.
\item \textsc{Debin}~\cite{he2018debin} employs structured prediction algorithms for type inference, evaluating on $9,000$ binaries, although the specific list of binaries is not detailed.
\end{packeditemize}

\parag{(II) Name Recovery} 
Recovering meaningful function and variable names has been advanced significantly by machine learning:
\begin{packeditemize}
    \item \textsc{SymLM}~\cite{jin2022symlm} leverages transformer models for function name recovery across 27 datasets and employs CodeWordNet to compute semantic distances between names.
    \item \textsc{NERO}~\cite{david2020neural}, uses sequential models for name recovery across 99 datasets and assesses performance through exact name matching.
    \item \textsc{VarBERT}~\cite{pal2024len} performs exact name matching on the \textsc{DIRE}~\cite{lacomis2019dire}, which contains 3.1 million decompiled functions collected from GitHub.

    \item \textsc{ReSym}~\cite{xie2024resym} and \textsc{SymGen}~\cite{jiang2025beyond} fine-tune models to recover variable and function names via decompiled code, respectively. 
\end{packeditemize}

\parag{(III) Others} 
Two recent works generate function summaries from decompiled code using LLM~\cite{jin2023binary,shang2024far}.
Both operate directly on unmodified decompiled code and focus on a single task, making them comparatively simpler than multi-task approaches.


%
%

\subsection{Observations}\label{sec:insight}
Our analysis of existing research underscores the critical need for a standardized dataset for evaluating LLMs on reverse engineering tasks. 
Without such a dataset, ensuring clear, consistent, and fair comparisons of model performance is difficult. 
We identify three major issues in current practices: variability in datasets, inconsistency in evaluation metrics, and incompleteness of decompiler output.

\parag{Observation 1: Variability in Datasets} 
A comprehensive review of recent studies reveals that the datasets used for evaluating LLMs in reverse engineering vary significantly in both scope and composition. 
Some works rely on small, narrowly defined datasets, while others evaluate across many datasets, reflecting considerable diversity in data sources. 
For example, \textsc{DeepBinDiff}~\cite{duan2020deepbindiff} evaluates its framework on three well-known binary sets: coreutils, diffutils, and findutils. 
In contrast, \textsc{SymLM} and \textsc{SymGen} incorporate 27 datasets, forming a superset of {\sc DeepBinDiff}'s collection but excluding ImageMagick, a dataset frequently used in other studies~\cite{pei2020trex,yang2021codee}.

This inconsistency creates a major barrier to progress. 
The lack of a standardized dataset makes it difficult to compare results across studies meaningfully, limiting collaborative research and slowing real-world applicability. 
Moreover, architectural coverage also varies across studies: {\sc Debin} targets three architectures (x86, x64, ARM), while {\sc SymLM} extends this to include MIPS. 
Such discrepancies highlight the fragmented nature of current evaluation practices. 
A universal benchmark dataset would resolve these issues by providing a consistent foundation for evaluation, enabling objective model comparison, and supporting the training of future security researchers in binary analysis.

\parag{Observation 2:  Inconsistency in Evaluation Metrics} 
Most existing studies evaluate LLM performance using traditional metrics such as precision, recall, and F1 score, the majority of which rely on exact string matches between predicted outputs and ground-truth labels. 
However, this purely syntactic approach fails to capture semantic similarity, leading to systematic underestimation of LLM capabilities. 
For instance, Feitelson et al.~\cite{feitelson2020developers} found that the probability of two developers independently selecting the same name for an identical function is only 6.9\%. 
Moreover, developers frequently abbreviate or modify names for conciseness~\cite{hindle2016naturalness}. 
In such cases, string matching unfairly penalizes predictions that are semantically correct but syntactically different. 
For example, {\tt compute} and {\tt calculate} are semantically equivalent, yet exact-match criteria would label the prediction as incorrect.

These shortcomings are evident in {\sc Debin}, which enforces strict string-based evaluation, overlooking meaningful semantic matches. 
By contrast, {\sc SymLM} and {\sc NERO} adopt semantic-aware evaluation strategies, enabling a more accurate assessment of model performance. 
This disparity highlights the critical need for evaluation frameworks that incorporate semantic similarity, whether through embeddings, lexical resources, or domain-specific keywords, rather than relying solely on exact string comparison. 
Addressing this gap is essential for building reliable and realistic benchmarks for LLMs in reverse engineering.

\parag{Observation 3: Incompleteness of Decompiler Output} 
State-of-the-art decompilers still exhibit errors when generating decompiled code, particularly in reconstructing data structures, largely due to the absence of debug symbols. 
As a result, decompiled code generated with debug symbols often differs substantially from that produced without them. 
Prior work~\cite{banerjee2021variable} has therefore relied on binaries containing debug symbols to generate decompiled code. 
While this approach improves the accuracy of the ground truth, it also simplifies the reverse engineering task, since the resulting code preserves richer semantic information than code derived from stripped binaries. 
For instance, developers can readily identify variables within a data structure and infer its components from debug-enriched output. 
However, this setup does not reflect real-world scenarios, where debug symbols are typically unavailable. 
Consequently, it is essential to select binaries that both mirror real-world conditions and enable the construction of complete, reliable ground truth.

\ignore{
\section{\sys Overview}\label{sec:overview}

To address the lack of standardized datasets and evaluation methodologies for binary analysis tasks, we present \sys, a comprehensive benchmark for assessing LLMs. 
\sys is specifically designed to evaluate the effectiveness of LLMs on reverse engineering tasks, with the broader goal of establishing consistent and reproducible evaluation criteria. 
By constructing a high-quality dataset, standardizing preprocessing pipelines, and adopting semantic-aware evaluation strategies, \sys enables accurate and fair comparisons.

\parag{Step 1: Data Collection} 
We begin by identifying the essential properties of a high-quality dataset for evaluating LLMs in reverse engineering. 
This involves reviewing prior works and surveying publicly available datasets to curate a collection that reflects both research diversity and real-world relevance.

\parag{Step 2: Data Compilation}
Once collected, the datasets are compiled into executable binaries across multiple processor architectures and multiple optimization levels. 
This diversity is critical for mimicking real-world reverse engineering scenarios, where binaries originate from heterogeneous platforms and tool-chains.

\parag{Step 3: Data Normalization}
To ensure compatibility across studies, we normalize the compiled datasets into a universal, self-designed input format. 
This abstraction preserves the difficulty of the underlying tasks while making the data adaptable for multiple LLM-based or non-LLM-based analysis pipelines. 
Normalization prevents task simplification and guarantees that symbol recovery and type inference remain as challenging as they are in real-world binaries.

\parag{Step 4: Semantic-aware Evaluation Metrics}
Finally, \sys introduces standardized evaluation criteria. 
For name recovery, we employ semantic distance metrics to capture meaningful similarities between predicted and ground-truth names, avoiding the difficulty of strict string matching. 
For type inference, we apply exact matching to reflect the strict semantics of type systems in C and related languages. 
}
%
%

\ignore{
\smallskip
\noindent\textbf{Results:} We have collected a list of open-source LLMs and evaluated them using our benchmark.
With the prepared datasets, we apply various experiment setups to comprehensively understand the capabilities. 
This detailed comparative analysis of their performances not only highlights the capabilities and limitations of each LLM but also fosters a better understanding of their practical applications in reverse engineering tasks. 
}

%% file: paper/04_1_data-curation.tex

Selecting appropriate datasets and compiling source code into executable binaries is a crucial step in preparing a usable benchmark. 
Since LLM-based methods require structured input such as assembly code or decompiled code, \sys abstracts symbols from these representations to construct a universal input format. 
This abstraction ensures flexibility, allowing the dataset to be customized for diverse downstream tasks while preserving the difficulty of real-world reverse engineering challenges.

\subsection{Dataset Collection}\label{sec:selection}
To ensure the quality, representativeness, and usability of \sys, we adopt a principled dataset selection process guided by the following criteria:

\begin{itemize}[noitemsep,topsep=0pt,parsep=0pt,partopsep=0pt,leftmargin=*]
    \item \textbf{Coverage of Existing Work.} \sys collects datasets from recent reverse engineering studies to maximize relevance and inclusivity. 
    This ensures that the benchmark reflects the diversity of prior research and enables direct comparison with existing approaches.

    \item \textbf{Architectural and Compiler Optimization Diversity.} We include binaries compiled for different architectures and optimization levels, mimicking real-world scenarios where binary analysis must handle code from heterogeneous platforms and toolchains.

    \item \textbf{Accessibility.} We prioritize open, publicly available datasets, making \sys easily adoptable by both academic researchers and practitioners. Accessibility encourages collaboration and ensures reproducibility. \looseness=-1


    \item \textbf{Popularity.} We select projects that are widely used and recognized in the community. Popular datasets frequently serve as baselines in reverse engineering studies, increasing the utility and credibility of \sys.

    \item \textbf{Ethical Sourcing and Licensing.} We prioritize ethical considerations in dataset acquisition to ensure responsible and transparent data usage, respecting privacy and intellectual property rights. 
    We select datasets with permissive licenses that allow redistribution, promoting legal and ethical use and facilitating broader adoption of the benchmark.

\end{itemize}
Following these guidelines, we conducted a comprehensive review of existing work and surveyed open-source projects. 
Our focus is on programs implemented in C or C++, the most widely used languages in system-level programming and reverse engineering. 
To maintain openness and reproducibility, we exclude commercial or proprietary datasets, such as SPEC2017, that restrict redistribution. 
We also filter out binaries lacking debug symbols, such as firmware without source code, since ground-truth symbol information is necessary for constructing reliable evaluation datasets.

\subsection{Dataset Compilation}
Compiling the selected datasets requires manual effort due to diverse build requirements and dependencies. 
To ensure consistency, we adopt the standard {\tt configure} and {\tt make} workflow, which is widely used across open-source projects. 
The {\tt configure} process generates files such as {\tt Makefile} based on the given configurations, and the {\tt make} process compiles the binaries. 
This approach provides the flexibility to compile each dataset for multiple target architectures (x64, x86, ARM 32-bit, and MIPS 32-bit) and optimization levels ({\tt O0}–{\tt O3}). 
For each dataset, we generate two sets of binaries that serve distinct purposes:

\begin{packeditemize}
    \item {\bf Debug-symbol Binaries.} These binaries preserve ground-truth information such as variable names, function names, and types. They are used to construct the knowledge base and to map predictions back to the ground truth.
    \item {\bf Stripped Binaries.} These binaries have all symbol information removed, reflecting real-world scenarios where reverse engineers must infer missing information. They provide the decompiled or assembly code inputs supplied to LLMs in downstream tasks.
\end{packeditemize}

\subsection{Dataset Normalization} 
The dataset normalization process is a non-trivial but essential step in building \sys. 
Our goal is to design a universal input format applicable to diverse reverse engineering tasks while preserving the inherent difficulty of stripped binary analysis. 
Since stripped binaries lack symbol information, we must systematically collect and reconstruct ground truth for each symbol before normalizing the dataset. 
All examples described in this section are generated using the Ghidra~\cite{ghidra} decompiler.

\begin{figure*}[th!]
    \begin{subfigure}{0.32\textwidth}
    \begin{lstlisting}[style=cstyle]
typedef struct {
  int a;
  short b;
  unsigned int c;
} data;

int sum() {
  data val;
  scanf("%d %d %d", 
    &val.a, &val.b, 
    &val.c);
  return val.a + 
    val.b + val.c;
}
\end{lstlisting}
        \caption{Original C Code}\label{fig:ex_orig}
    \end{subfigure}
    \begin{subfigure}{0.32\textwidth}
\begin{lstlisting}[style=cstyle]
int sum(void){
  data val;
    
  scanf("%d %d %d", &val, 
    &val.b, &val.c);  
    
  return val.c + 
    val.b + val.a;
}

|\phantom{word}|
|\phantom{word}|
|\phantom{word}|
|\phantom{word}|
\end{lstlisting}
\caption{Decompiled Code w Debug Symbols}\label{fig:eg_with}
\end{subfigure}
\begin{subfigure}{0.32\textwidth}
        \begin{lstlisting}[style=cstyle]
int FUNC(void){
  int local_1c;
  short local_18 [2];
  int local_14;
    
  scanf("%d %d %d", 
    &local_1c, local_18, 
    &local_14);
    
  return local_14 + 
    local_18[0] + 
    local_1c;
}
|\phantom{word}|
\end{lstlisting}
\caption{Decompiled Code wo Debug Symbols}\label{fig:ex_without}
\end{subfigure}
  \caption{Differences of Decompiled Code Between With and Without Debug Symbols.}\label{fig:ex}
\end{figure*}

\parag{Input Format Selection} 
Using raw binary code (i.e., sequences of \texttt{0}s and \texttt{1}s) is impractical for LLMs, as it carries no explicit semantic information. 
Instead, following prior work, we adopt decompiled code and assembly code as input representations, since LLMs can leverage the richer vocabulary of keywords present in these formats, a key distinction from traditional machine learning techniques:

\begin{packeditemize}
    \item {\bf Assembly Code} provides low-level semantics (e.g., instruction mnemonics, registers) and is particularly useful for tasks such as primitive type inference and function name recovery~\cite{jin2022symlm}. 
    However, identifying variables within assembly code requires additional analysis techniques. 
    Its limited token vocabulary also makes it amenable to token-based machine learning methods.

    \item {\bf Decompiled Code} offers high-level, human-readable pseudocode that exposes constructs such as functions, variables, and control flow~\cite{jiang2025beyond,xie2024resym}. 
    While imperfect, it provides richer semantic context, making it well suited for both name recovery and type inference. \looseness=-1
\end{packeditemize}
Together, these representations enable \sys to support a wide range of reverse engineering tasks while maintaining fidelity to real-world workflows.

\parag{Decompiled Code Selection and Challenges} 
Decompiled code selection is a pivotal phase of our methodology, as it directly dictates the realism of the benchmark and the integrity of the ground truth. 
A primary challenge lies in ensuring that the normalization process does not inadvertently trivialize the reverse engineering task.

As illustrated in~\autoref{fig:ex}, the original source code (\autoref{fig:ex_orig}) defines a user-defined structure comprising three variables.
When decompiled from a binary containing debug symbols, the output preserves the original structure and variable names, offering a direct mapping to the ground truth. 
Conversely, stripped binaries lack these symbols, leading to significantly different decompiled outputs. 
As shown in~\autoref{fig:eg_with} and~\autoref{fig:ex_without}, the decompiler ``flattens'' the structure into three independent variables, discarding the high-level abstraction of the composite type. 
This creates a structural mismatch between the source-level ground truth and the decompiled representation. \looseness=-1

Consequently, recovering the ground truth requires rigorous reconstruction. 
Our framework, \sys, must reconcile the original data structures and variable names by inspecting stack offsets and memory ranges rather than relying solely on decompiler-generated identifiers. 
Neglecting this step would underestimate the complexity of the task and risk introducing artificial ``shortcuts'' that inflate model performance. \looseness=-1

\begin{algorithm}[t!]
\footnotesize
\caption{Building Knowledge Base of Every Symbol from Binary}\label{build_kb}
\DontPrintSemicolon
    \KwIn{Unstripped Binary ($B$)}
    \KwOut{Knowledge-Base ($KB$)}

    $f \gets first\_func(B)$ \; \label{line1}
    $KB \gets \{\}$ \;

    \While{$True$}{
        \If{$f == NULL$}
        {
            $break$ \;
        }
        
        $sym \gets first\_sym(f)$ \; \label{line2}
        $KB[f.entry\_addr] \gets \{\}$ \;

        \While{$True$}
        {
        \If{$sym == NULL$}
        {
            $break$ \;
        }
        
        \For {$i = stack\_offset(sym)$ \bf{to} $stack\_offset(sym) + size(sym)$}
        {
            $KB[f.entry\_addr][kb] = \{i : [sym.name, sym.type] \}$ \;
        }

        $sym \gets next\_sym(f)$
        }

    $f \gets next\_func(B)$ \;
    }
    
\end{algorithm}

\parag{Abstracting Symbols via a Knowledge Base} 
To mitigate these challenges, we employ a knowledge-base (KB) approach for symbol abstraction. \looseness=-1

\begin{packeditemize}
    \item {\bf KB Construction.} The process begins with the construction of a comprehensive KB containing every symbol from an unstripped version of the binary, as detailed in~\autoref{build_kb}.
    We initiate the process by decompiling the executable with debug symbols and systematically traversing each function from its entry point (\autoref{line1}). 
    For every function, we identify the initial symbols (\autoref{line2}) and extract their associated stack offsets and memory footprints. 
    These symbols are then mapped at the byte level within a dictionary.
     For example, consider a variable of type {\tt int} stored at stack indices 0 to 3; since the int type occupies 4 bytes, the KB will record the variable's name together with its type ({\tt int}) across stack indices 0 to 3.
     This granular mapping is essential because decompilers often fail to recognize complex data structures, instead flattening them into raw size allocations. 
     By mapping symbols at the byte level, we preserve their semantic context for downstream analysis. 
     This procedure is repeated for all symbols within a function and indexed by the function's entry address in the KB.

    \item {\bf KB Utilization.} Once the KB is populated, we analyze the stripped binary to generate the corresponding decompiled and assembly code (\autoref{use_kb}). 
    At this stage, we identify all detectable symbols and replace them with uniform placeholders (e.g., {\tt VAR1}). 
    Simultaneously, we query the KB to retrieve the original names and types. 
    A similar replacement logic is applied to function names, which are substituted with generic identifiers (e.g., {\tt FUNC1}). 
    While~\autoref{build_kb} and~\autoref{use_kb} omit the granular details of function renaming for brevity, the process involves recording callee entry addresses to ensure that inter-procedural relationships and ground truth names remain recoverable during the analysis of stripped binaries.
\end{packeditemize}

\begin{algorithm}[t!]
\footnotesize
\caption{Utilizing Knowledge Base to Build Ground Truth}\label{use_kb}
\DontPrintSemicolon
    \KwIn{Knowledge-Base ($KB$), Stripped binary ($B$)}
    \KwOut{Decompiled Code}

    $f \gets first\_func(B)$ \;

    \While{$True$}{
        \If{$f == NULL$}
        {
            $break$ \;
        }
        
        $sym \gets first\_sym(f)$ \;
        $idx \gets 1$ \;

        \While{$True$}
        {
        \If{$sym == NULL$}
        {
            $break$ \;
        }

        $sym.name, sym.type = VAR_{idx}, TYPE_{idx}$ \; \label{use_line1}
        $sym.name\_gt, sym.type\_gt = KB[f.entry\_addr][kb][stack\_addr(sym)]$
        
        $idx++$\;
        $sym \gets next\_sym(f)$
        }
    
    $generate\_decompile\_code(f)$\;
    $f \gets next\_func(B)$ \;
    }
    
\end{algorithm}

%

%

\begin{figure*}[tb!]
    \begin{subfigure}{0.32\textwidth}
    \begin{lstlisting}[style=cstyle]
int main(int argc) {
  int sum;
  int number1;
  int number2;

  printf("Enter two 
    integers: ");
  scanf("%d %d", 
    &number1, 
    &number2);
  sum = number1 + 
        number2;
  printf("%d + %d = %d", 
    number1, 
    number2, 
    sum);
  return 0;
}
\end{lstlisting}
        \caption{Original C Code}\label{fig:addc}
    \end{subfigure}
    \begin{subfigure}{0.32\textwidth}
\begin{lstlisting}[style=cstyle]
int main(int argc){
  int sum;
  int number1;
  int number2;
    
  printf("Enter two 
    integers: ");
  __isoc99_scanf("%d %d", 
    &number1, &number2);  
  sum = num1 + num2;
  printf("%d + %d = %d", 
    number1, number2, 
    sum);
    
    return 0;
}
|\phantom{word}|
|\phantom{word}|
\end{lstlisting}
\caption{Decompiled Code}\label{fig:adddecompiled}
\end{subfigure}
\begin{subfigure}{0.32\textwidth}
\begin{lstlisting}[style=cstyle]
TYPE1 FUNC1(TYPE2 VAR1){
  TYPE3 VAR2;
  TYPE4 VAR3;
  TYPE5 VAR4;
    
  FUNC2("Enter two 
    integers: ");
  FUNC3("%d %d", 
    &VAR3, &VAR4);
  VAR2 = VAR3 + VAR4;
  FUNC2("%d + %d = %d", 
    VAR3, VAR4, VAR2);
    
  return 0;
}
|\phantom{word}|
|\phantom{word}|
|\phantom{word}|
\end{lstlisting}

\caption{Our Final Input}\label{fig:addabstract}
\end{subfigure}
  \caption{The Process of Normalization. The Example with Debug Symbols is Used For Better Understanding.}\label{fig:add}
\end{figure*}

\autoref{fig:add} demonstrates this workflow using a basic addition function.
While~\autoref{fig:addc} shows the original C code and~\autoref{fig:adddecompiled} shows the standard decompiled output,~\autoref{fig:addabstract} depicts the final version with abstracted symbols. 
In a real-world scenario, the intermediate step (\autoref{fig:addabstract}) would not retain original identifiers. 
Instead, we use the raw decompiled output to build the KB, cross-referencing abstracted placeholders ({\tt VAR, TYPE}) with their original forms in the database to provide a reliable, automated mapping for model evaluation.\looseness=-1

\parag{Limitations of DWARF-only Ground Truth}
A possible counter-argument is that DWARF information from unstripped binaries could serve as a direct ground truth. 
However, our methodology relies on the decompiler's interpretation of the code. 
Since the decompiler itself utilizes DWARF information to reconstruct the stack layout, using DWARF directly as the sole ground truth would be redundant and potentially overlook the idiosyncratic ways a decompiler reinterprets symbol relationships.

Moreover, relying strictly on DWARF may oversimplify the evaluation of variable name recovery. 
Decompilers often generate a different number of variables than those present in the original source code. 
Forcing an LLM to match the exact count and layout dictated by DWARF, rather than what is actually visible in the decompiled code, would fail to accurately measure the model's performance in a realistic reverse engineering context.

\parag{Filtering Decompiled Code}
To ensure the integrity of our dataset, we implement a rigorous filtering pipeline. 
This process addresses two primary concerns: the removal of duplicate functions and the exclusion of external library wrappers.

\begin{packeditemize}
    \item {\bf Redundancy and Data Leakage:} Because our dataset aggregates multiple projects that may share common utility functions or dependencies, eliminating duplicates is essential. 
    Failure to do so would result in data leakage, where functions encountered during fine-tuning reappear during evaluation, leading to inflated and biased performance metrics. 
    We avoid relying solely on function names for de-duplication, as identical names often correspond to different implementations across different versions or projects. 
    Instead, we compute code similarity across both assembly and decompiled representations to identify and discard redundant entries.

    \item {\bf External and Trivial Functions:} A significant portion of a binary's decompiled code often consists of external library calls (e.g., PLT stubs). 
    These functions are typically incomplete, containing only a few instructions to invoke a shared library, which provides insufficient context for meaningful evaluation. 
    To address this, we exclude functions that fall below a specific line-count threshold (e.g., fewer than 10 lines of decompiled code). This heuristic offers two key advantages, 1) It effectively filters out external wrappers that lack internal logic, and 2) It eliminates trivial functions (such as simple getters or counter increments) that offer limited semantic depth. 
    By removing these, we ensure the benchmark focuses on complex code that meaningfully challenges the model's ability to infer original identifiers and types. \looseness=-1
\end{packeditemize}

\parag{Adapting to Downstream Tasks} 
The \sys abstracted representation is designed for high versatility. 
By using a uniform, indexed placeholder system, our format can be easily customized for various binary analysis tasks. 
We illustrate this adaptability through four representative applications:

\begin{packeditemize}
    \item {\bf Single Function Name Recovery.} This task involves predicting a target function's name based on its decompiled or assembly context. 
    To facilitate this, we restore all internal symbols while leaving the target function name abstracted. The placeholder acts as a functional equivalent to the {\tt [MASK]} token in masked language models (MLMs). 
    Task difficulty can be modulated by varying the ratio of abstracted-to-original symbols, effectively simulating different levels of stripping (from partially to fully stripped binaries).

    \item {\bf Data Structure Reconstruction.} This task requires recovering user-defined types from ``flattened'' variables. 
    While decompilers often break composite structures into individual stack variables, the \sys ground truth maintains the original associations and memory offsets. 
    By analyzing variable usage patterns and inter-procedural interactions, models can be trained to re-aggregate these flattened elements into their original composite types. \looseness=-1

    \item {\bf Function Similarity.} \sys supports function-level similarity analysis directly. 
    While our dataset minimizes overlaps, the presence of multiple versions of the same project allows for testing model robustness across compiler variations. 
    This capability can be naturally extended to binary-level similarity through lightweight aggregation of function-level embeddings. \looseness=-1

    \item {\bf Control Flow Graph Recovery.} Since \sys preserves function boundaries and call relationships, it provides a foundational layer for CFG reconstruction. 
    While static analysis alone may struggle with indirect jumps or calls, our enriched representation, which assigns unique, consistent identifiers to every function and variable, ensures that cross-references and dependencies are preserved for more advanced post-processing techniques.

\end{packeditemize}
Ultimately, \sys provides a robust and flexible framework that not only enhances the accuracy of standard recovery tasks but also broadens the scope of automated program understanding in reverse engineering. \looseness=-1


%% file: paper/04_2_metrics_definition.tex
\subsection{Evaluation Metrics}
To rigorously assess the performance of LLMs or other frameworks in reverse engineering, we employ distinct metrics for type inference and name recovery that account for both the strictness of C semantics and the flexibility of natural language.

\parag{Type Inference} 
The C programming language consists of a finite set of primitive types (e.g., {\tt char, int, float}) and composite types (e.g., {\tt struct, union}) constructed from those primitives. 
We employ exact type matching for our evaluation. This strictness is necessary because even closely related types, such as unsigned int and signed int, exhibit fundamentally different behaviors under C semantics. 
For instance, {\tt unsigned int} overflow is well-defined and wraps around to zero, whereas {\tt signed int} overflow is undefined and may lead to unpredictable compiler optimizations or vulnerabilities. 
Consequently, semantic equivalence cannot be assumed; exact matching ensures the most reliable measurement of type reconstruction accuracy.

\parag{Name Recovery} 
As discussed in \S\ref{sec:insight}, names inferred by LLMs may be semantically correct even if they do not provide a character-for-character match with the ground truth. 
For example, {\tt string} and {\tt str} or {\tt memory} and {\tt mem} are functionally equivalent in a reverse engineering context. 
To address this, we move beyond exact string matching and adopt a semantic similarity metric based on the {\sc SymLM} method, which leverages CodeWordNet to compute the semantic distance between sub-tokens. 
The process involves:

\begin{enumerate}[noitemsep,topsep=0pt,parsep=0pt,partopsep=0pt,leftmargin=*]
    \item {\bf Decomposition: } Breaking each identifier into constituent sub-names (e.g., {\tt extract\_trim\_name} into {\tt extract, trim, name}). \looseness=-1 

    \item {\bf Semantic Mapping: } Measuring the similarity between corresponding sub-names using CodeWordNet.

    \item {\bf Aggregate Scoring: } Calculating a similarity score that credits the model for synonyms while penalizing missing or incorrect semantic components.
\end{enumerate}

\parag{Example}
Suppose the ground truth is {\tt extract\_trim\_name} and the inferred name is {\tt get\_file\_name}.
\begin{packeditemize}
    \item {\tt name} is an exact match.
    \item {\tt extract} and {\tt get} are identified as semantically related synonyms. \looseness=-1
    \item {\tt trim}(ground truth) and {\tt file}(inferred) are recognized as distinct, resulting in a mismatch.
\end{packeditemize}
In this case, the evaluation identifies 2 True Positives (TP) and 1 False Negative (FN), yielding an F1 score of 0.8 (1.0 Precision and 0.66 Recall).

\begin{figure}[t!]
\centering
\begin{subfigure}{0.45\textwidth}
\centering
\begin{minipage}{0.75\linewidth}
            \begin{lstlisting}[style=cstyle]
// int main(int argc)
|\colorbox{cyan}{int} \colorbox{cyan}{main}(\colorbox{pink}{char} \colorbox{pink}{count})|
{
    |\colorbox{cyan}{int} \colorbox{pink}{index}|;  // int sum;
    |\colorbox{cyan}{int} \colorbox{cyan}{num1}|;   // int number1;
    |\colorbox{cyan}{int} \colorbox{cyan}{num2}|;   // int number2;

    //printf("Enter two integers: ");
    |\colorbox{cyan}{printf}|("Enter two integers: ");
    
    //scanf("%d %d",
    //      &number1, &number2);
    |\colorbox{cyan}{scanf}|("%d %d", &num1, &num2);
    
    // sum = number1 + number2;
    |\colorbox{pink}{index} = \colorbox{cyan}{num1} + \colorbox{cyan}{num2}|;
    
    // printf("%d + %d = %d", 
    // number1, number2, sum);
    |\colorbox{cyan}{printf}|("%d + %d = %d", 
    |\colorbox{cyan}{num1}, \colorbox{cyan}{num2}, \colorbox{pink}{index}|);
    
    return 0;
}
\end{lstlisting}
\end{minipage}
\end{subfigure}
    \caption{Example of LLM Inference Results. Comments are the Original Code. \colorbox{cyan}{Blue} Boxes are Correctly Inferred and \colorbox{pink}{Red} Boxes are not.}\label{fig:example}
\end{figure}

\input{table/stats}

\parag{Evaluation Metrics}
We report Precision, Recall, and F1 score as our primary metrics to provide a comprehensive view of LLM performance. 
While our framework also supports Area Under the Curve (AUC) and Mean Average Precision (mAP), we omit them from the primary discussion as they reflect similar performance trends. 
The \sys architecture remains modular, allowing for the integration of additional metrics as needed.

\parag{Evaluation Example}
\autoref{fig:example} illustrates the performance of a CodeLlama model when tasked with recovering the addition function previously described in~\autoref{fig:add}.

\begin{packeditemize}
    \item {\bf Type Inference: } The model correctly identifies four out of five type tokens, incorrectly inferring {\tt char} instead of {\tt int} for {\tt TYPE2}. This results in a type F1 score of 0.8.

    \item {\bf Function Name Recovery: } The model achieves a perfect F1 score of 1.0, correctly identifying the high-level intent. 

    \item {\bf Variable Name Recovery: } The model infers {\tt count, index, num1}, and {\tt num2}. 
    While none are exact string matches for the original {\tt number1} and {\tt number2}, \sys utilizes CodeWordNet to validate {\tt num1} and {\tt num2} as semantically correct. 
    With two semantic matches out of four predicted variables, the model achieves a Variable Name F1 score of 0.5.
\end{packeditemize}

\section{Experiments}
\subsection{Models}\label{sec:models}
\input{table/models}
We evaluate seven mainstream code LLMs as listed in~\autoref{llms}, which specifies each model's name, parameter count, and publication year.
We select LLMs that are widely deployed for programming tasks~\cite{popularllm} and well-documented in the literature~\cite{li2023starcoder,guo2025deepseek}.
To ensure a fair comparison among existing models, we select models of similar sizes, ranging from 13B to 15B parameters.
We exclude ChatGPT~\cite{chatGPT} because it is not freely available via API, and its parameter counts (175 billion for GPT-3 and 1.76 trillion for GPT-4) are not adequate for comparison with mid-sized open-source models.
Similarly, any larger and commercially restricted models are excluded from our measurement.
%

The selected models span diverse architectures and training methodologies: CodeLlama~\cite{roziere2023code} and Llama2~\cite{touvron2023llama} represent the Llama family; WizardCoder~\cite{luo2023wizardcoder} uses Evol-Instruct for code-specific tuning; DeepSeek-R1~\cite{guo2025deepseek} applies reinforcement learning on top of the Qwen2.5 architecture; Qwen2.5~\cite{qwen2.5} and Qwen3~\cite{qwen3technicalreport} allow evaluating generational improvements within the same family; and Phi4 Unsloth~\cite{unsloth} is optimized for efficient inference.

\subsection{Prompts and Input Templates}\label{sec:prompts}
We use a single prompt for all LLM measurements, asking models to perform inference on the provided code.
We prepare the prompt based on the format used in previous works~\cite{xie2024resym,jiang2025beyond}.
The prompt provides task instructions alongside the input code (decompiled or assembly), explicitly indicating which placeholders require inference.
For example, all placeholders in the decompiled code are included in the prompt, ensuring that the model clearly understands which tokens to predict.

\begin{figure}[tb!]
\centering
\includegraphics[width=0.7\linewidth]{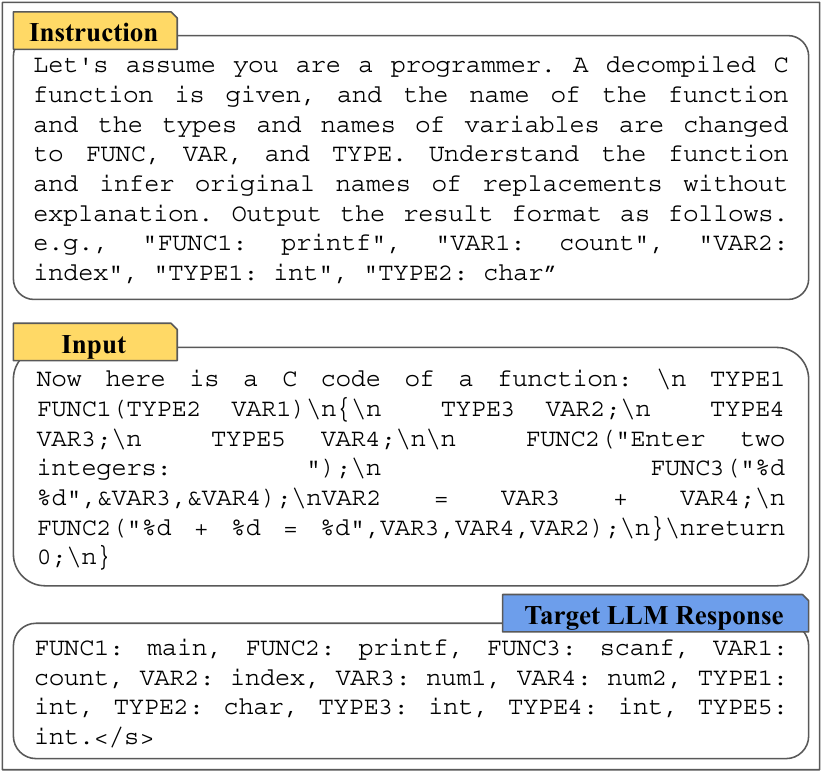}
\caption{The Prompt Used to Infer Types and Names}
\label{fig:inference}
\end{figure}

It is plausible that each model requires different prompts to produce an optimal response.
However, to ensure a fair comparison, we use the identical prompt content across all models, varying only the model-specific template format.
Since each LLM requires a model-specific input template, we define templates for each evaluated model.
For example, CodeLlama follows the format \texttt{<s>[INST]$\backslash$n<<SYS>>\{instruction\}\\$\backslash$n<</SYS>>$\backslash$n$\backslash$n\{input\}[/INST]}.
As a result, we locate the identical instruction and input following the input template described by the document of each LLM. 

\subsection{Inference Pipeline}\label{sec:inference_pipeline}
\parag{Response Generation}
We initialize each pre-trained LLM using the Hugging Face Transformers API~\cite{transformerAPI}.
For each function, the model receives the normalized decompiled or assembly code along with the prompt from \autoref{fig:inference}, specifying which placeholders require inference.
We set the maximum input context length to $1{,}024$ tokens and the maximum output length to $1{,}024$ tokens.
All models are loaded in 4-bit quantization (QLoRA~\cite{dettmers2023qlora}) on a single NVIDIA A100 (80\,GB) GPU to fasten the process.
Due to differences in instruction-following ability, responses often deviate from the requested format, include extraneous explanations, or vary between short tokens and full sentences.
Moreover, we disable any randomization on responses from LLM, such as temperature, top\_k, and top\_p, to minimize the randomness and compare the performance consistently.

\subsection{Fine-tuning Strategy}\label{sec:finetuning}
To assess LLMs beyond zero-shot inference, we fine-tune them on our reverse engineering dataset.
Full parameter fine-tuning is computationally expensive for large models, so we adopt Low-Rank Adaptation (LoRA)~\cite{hu2022lora,dettmers2023qlora}.
LoRA freezes the base model weights and introduces trainable low-rank matrices, which are orders of magnitude smaller than the full parameter set.
Only these matrices are updated during fine-tuning.
This approach drastically reduces resource requirements, allowing us to fine-tune multiple models efficiently while preserving the representational capacity of the base LLMs.

For each model, we prepare a training set by randomly selecting $10{,}000$ functions per architecture and optimization pair, independent of the $2{,}500$-sample test set, which prevents data leakage during the fine-tuning process.
This yields a total of $160{,}000$ training functions across all architecture--optimization combinations, with a training-to-testing split of 80\% and 20\%.
Moreover, instead of fine-tuning the model on each architecture along with each optimization level, we fine-tuned the whole dataset at once to measure the generality of each LLM. 
We train each model using the LoRA configuration with a rank of $8$, an alpha value of $16$, and a batch size of $128$.
The learning rate is set to $3 \times 10^{-4}$ with a cosine scheduler.
All fine-tuning experiments are conducted on a NVIDIA A100 GPU.
We adopt 3 epochs for fine-tuning process, which demonstrated that LoRA fine-tuning converges rapidly for code-related tasks within the first few hundred steps.
%
%
This standardized fine-tuning setup ensures a fair comparison across models, as all models are fine-tuned with identical hyperparameters and training data.

\subsection{Experiment Setup}\label{sec:exps}

\parag{Name Recovery and Type Inference via Decompiled Code}
We first evaluate the ability of LLMs to recover both function and variable names, as well as to infer variable types, using the full set of $40{,}000$ samples.
Decompiled code serves as input because it retains richer semantic information than assembly code, enabling a more comprehensive assessment of each LLM across all architectures and optimization levels.
This approach follows prior work such as \textsc{VarBERT}~\cite{pal2024len} and \textsc{ReSym}~\cite{xie2024resym}, which also rely on decompiled code for variable name recovery.
By using the same setup for both name recovery and type inference, we obtain a deeper understanding of LLM performance in semantic program analysis.

\parag{Function Name Recovery via Assembly Code}
Assembly code provides a faithful representation of the execution process of a function and is therefore suitable for function name recovery.
However, recovering variable names from assembly is significantly more difficult due to the lack of semantic context, as variables are represented only as register values and memory offsets.
Accordingly, this experiment focuses exclusively on predicting function names from assembly, again using the $40{,}000$ sample dataset.
Unlike prior work~\cite{jin2022symlm}, we do not incorporate program state information.
In summary, the assembly-based setup evaluates only function name recovery.

\parag{Name Recovery With and Without CodeWordNet}
To measure the contribution of CodeWordNet, which plays a central role in our name recovery evaluation, we run comparative experiments with and without it.
These experiments are restricted to the x64 architecture to illustrate general trends without requiring a full-scale analysis across all architectures.
Only the F1 score is reported, as it effectively reflects the trends in both precision and recall observed in earlier experiments.

\parag{Impact of Model Size on Task Performance}
To investigate whether larger models yield better performance, we evaluate three variants of CodeLlama with different parameter sizes: 7B, 13B, and 34B.
The 70B model is excluded due to GPU memory limitations.
Similar to the CodeWordNet experiments, this analysis is conducted on the x64 architecture, with F1 score as the primary metric.

\parag{Impact of Model Architecture and Training Quality}
To disentangle the effects of model architecture from training data quality, we compare models that share the same underlying architecture but differ in their training procedures.
Specifically, we compare DeepSeek-R1, which is a fine-tuned variant of Qwen2.5 with reinforcement learning, against the original Qwen2.5 model.
We also compare models across two consecutive generations of the same family (Qwen2.5 vs.\ Qwen3) to assess whether newer model versions consistently improve performance on reverse engineering tasks.

\parag{Evaluation on Real-World Firmware Using Two Decompilers}
Finally, to test LLMs in a realistic reverse engineering setting, we evaluate them on real-world firmware binaries.
We use ARM-based IoT firmware images from the P2IM dataset~\cite{feng2020p2im}, comprising microcontroller-unit (MCU) firmware from multiple vendors.
We extract $500$ functions from these firmware images for evaluation.
All symbols are stripped, and the binaries are decompiled into pseudocode using two decompilers: {\sf Ghidra} and {\sf IDA Pro}.
This allows us to compare performance across decompiler outputs and examine the impact of decompilation quality on model predictions.
Each LLM is evaluated on all tasks (name recovery and type inference) using both versions of the decompiled code.
This experiment is particularly important because it simulates the typical workflow of a reverse engineer, who must analyze unknown firmware without access to ground truth.

\subsection{Results}
\label{eval_result}

\begin{packeditemize}
    \item {\bf \autoref{tbl:nameresult}, ~\ref{tbl:varnameresult}}: Name recovery performance across architectures and optimization.
    \item {\bf \autoref{tbl:typeresult}}: Type inference performance.
    \item {\bf \autoref{tbl:asmresult}}: Function name recovery from assembly code.
    \item {\bf \autoref{fig:ablation}}: Impact of CodeWordNet on function name recovery performance.
    \item {\bf \autoref{fig:diffsize}}: Comparison of CodeLlama model sizes on name recovery and type inference.
    \item {\bf \autoref{tbl:realworld}}: Performance on real-world firmware with decompiled code from two different decompilers.
\end{packeditemize}

\input{table/func_result}
\input{table/var_result}

\parag{LLM Performance on Name Recovery}
For function name recovery, both Llama2 and CodeLlama perform worse than other models across all architectures.
Qwen3 and Qwen2.5 consistently achieve the highest F1 scores on both x86 and ARM architectures.
Moreover, every model performs worse on MIPS architecture compared to others.
Furthermore, every model obtains better F1 score on x86 than others. 

For variable name recovery, similar to function name recovery, both CodeLlama and Llama2 perform worst among all evaluated models.
However, the trend of WizardCoder becomes different from the previous task.
It achieves higher F1 scores on x64 and ARM architectures.
Similar to the previous observation, every model struggles with the MIPS architecture.

\result{Qwen2.5 and Qwen3 perform well compared to others on name recovery, significantly better than CodeLlama and Llama2.}

\result{All LLMs struggle with the MIPS architecture.}

Moreover, variable name recovery reveals an interesting trend.
Every LLM except for CodeLlama and Llama2 achieves a higher F1 score (over 0.1) on 64 architecture with {\tt O0} optimization.
However, at {\tt O1-O3}, all models obtain substantially lower F1 scores, around 0.05.
A similar trend occurs on ARM, although the differences are more marginal than on x64.
This performance gap between {\tt O0} and higher optimization levels is expected, as compiler optimizations such as inlining, constant propagation, and register allocation fundamentally alter the structure of decompiled code, removing semantic cues that models rely on for name prediction.

\result{All models achieve lower F1 scores at {\tt O1-O3} optimization levels compared to {\tt O0}.}

\input{table/type_result}

\parag{LLM Performance on Type Inference}
Type inference proves even more challenging, with all models performing worse than on name recovery.
Due to uniformly low performance across all models, it is difficult to determine which model performs best.
Even with the marginal differences, CodeLlama consistently achieves the lowest F1 scores among all models, including Llama2.
The difficulty can be attributed to strict evaluation criteria requiring exact type matches, as well as the frequent use of user-defined types in real-world code, which complicates inference further.
Additionally, type inference requires understanding the data flow and memory layout of the program, which demands deeper program analysis capabilities than name recovery.

\result{All models perform worse on type inference compared to name recovery.}

\parag{LLM Performance on Function Name Recovery via Assembly Code}
As shown in~\autoref{tbl:asmresult}, Phi4 Unsloth achieves a higher F1 score on x64 and x86 compared to ARM and MIPS.
Every model achieves a lower F1 score on ARM and MIPS.
This is likely attributable to the prevalence of x86 code in LLM training data, as x64 and x86 architectures are more widely represented than ARM and MIPS in publicly available repositories and documentation.

\input{table/asm_result}

\result{Assembly-based function name recovery yields F1 scores comparable to decompiled code-based recovery, with top models achieving F1 scores of 0.05-0.10 on x86 architectures.}

More interestingly, every model achieves a higher F1 score as optimization level increases.
This trend is noteworthy, as it contrasts with the opposite pattern observed in decompiled code-based tasks.
A plausible explanation is that compiler optimization simplifies and regularizes assembly code patterns, making them easier for LLMs to interpret.
Specifically, higher optimization levels tend to produce shorter, more canonical instruction sequences by eliminating redundant operations, which may align better with patterns seen during pre-training.

\result{In contrast to decompiled code-based function name recovery, LLMs generally achieve higher F1 scores at {\tt O1-O3} than at {\tt O0}}.

\input{table/real_world}
\parag{LLM Performance on Real-World Firmware with Two Decompilers}
On real-world firmware, all models again achieve near-zero F1 scores on function name recovery and variable type inference (\autoref{tbl:realworld}).
However, the F1 score on variable name recovery is similar to the previous experiments, which is higher than the function name.
A plausible explanation is that variable names in real-world code tend to follow common naming conventions that models can partially learn from pre-training data.
In contrast, function names are more domain-specific and unique to the firmware's purpose.
For example, {\tt HAL\_delay}, a widely used function in IoT devices related to the hardware abstraction layer, does not appear in our training data, making it difficult for LLMs to predict such domain-specific function names.

\result{All models achieve comparable F1 scores only on the variable name recovery task.}

Notably, no model achieves a higher F1 score on variable name recovery from decompiled code generated by {\sf IDA Pro}.
This is likely because the models are fine-tuned on decompiled code produced by {\sf Ghidra}.
This suggests that LLMs do not generalize well across different decompiler output formats, as each decompiler produces syntactically distinct pseudocode with different naming conventions, control flow representations, and type annotations.

\result{LLMs may struggle to generalize across decompiled code from different decompilers.}

\parag{Impact of CodeWordNet}
As shown in \autoref{fig:ablation}, removing CodeWordNet causes a sharp drop in performance due to strict string matching penalties.
Without CodeWordNet, semantically correct predictions such as \texttt{calc\_sum} for the ground truth \texttt{compute\_total} would be scored as completely incorrect, despite conveying the same meaning.
This demonstrates that while LLMs often generate semantically similar names, CodeWordNet is essential for fair evaluation.

\input{table/ablation}

\parag{Impact of LLM Model Size vs.\ Training Quality}
Our observations show that training quality is more important than parameter count.
For example, Qwen3 (14B) outperforms WizardCoder (15B) on function name recovery, despite having fewer parameters.

\result{Larger model size does not guarantee better performance; training data quality is a more decisive factor.}

\parag{Impact of LLM Model Architecture vs.\ Training Quality}
Comparisons of models with identical architectures further highlight the importance of data quality.
DeepSeek-R1, a fine-tuned variant of Qwen2.5, consistently underperforms Qwen2.5 in name recovery, despite identical architecture and parameter count.
This suggests that the reinforcement learning-based training procedure used by DeepSeek-R1 may not be well-suited for the pattern recognition required in reverse engineering tasks.

\result{Training data quality significantly impacts performance, even among models with the same architecture.}

\input{table/diff_size}
\parag{Performance of CodeLlama Across Different Model Sizes}
Comparing the 7B, 13B, and 34B variants of CodeLlama (\autoref{fig:diffsize}) shows little difference in performance.
The larger model tends to obtain a higher F1 score on function name recovery (34B $>$ 13B $>$ 7B in most settings).
However, this is not always guaranteed, as shown in the F1 score on variable name recovery, where even the smallest model outperforms the two larger models.
This result suggests diminishing returns from scaling within the same model family, particularly when the base training data does not adequately cover reverse engineering tasks.

\result{Larger variants of the same model family usually perform better, but do not always yield better results.}

\parag{Newer Version of LLM}
We evaluate two versions of Qwen in our experiments.
Qwen2.5 and Qwen3 were introduced in 2024 and 2025, respectively.
Although Qwen3 is newer than Qwen2.5, it does not always outperform the older version.
This indicates that model updates do not uniformly improve performance across all tasks; improvements in general-purpose capabilities may not transfer to specialized domains like reverse engineering.

\result{A newer version of an LLM does not always guarantee better performance.}

\parag{LLM Performance by Release Year}
CodeLlama and Llama2, both released in 2023, consistently underperform compared to newer models.
Similarly, WizardCoder, one of the oldest models in our evaluation, tends to underperform.
Models released after 2024 show significantly improved performance, though the most recent model does not always outperform all others.
We hypothesize that newer models may have been exposed to decompiled code during pre-training, given the growing popularity of LLM-based software security tools and the increasing availability of reverse engineering datasets in public repositories.

\parag{Overall LLM Performance}
Across all tasks, most LLMs demonstrate limited capability, with the majority achieving F1 scores below 0.1.
Recent models (post-2024) perform reasonably well on name recovery, but all models struggle to achieve high F1 scores on type inference.
Furthermore, all models perform poorly on MIPS and generally on higher optimization levels in decompiled code-based tasks.
These results suggest that while LLMs have potential for reverse engineering assistance, significant improvements in training methodology and data are needed before they can reliably support these tasks.

\subsection{Further Analysis}
In this section, we provide a deeper analysis of LLM behavior and practical challenges encountered during our evaluation.

\ignore{
\paragraph{Failure Mode and Type Category Analysis}
To understand \emph{why} models fail, we analyze error patterns across function complexity and type categories.
We categorize functions by variable count: simple ($\leq$3), medium (4-8), and complex ($>$8).
For name recovery, all models achieve higher F1 scores on simple functions (0.12-0.15) than complex ones (0.03-0.06), confirming that function complexity is a key factor (see \autoref{tbl:failure} in the Appendix for the full breakdown).
For type inference, performance remains uniformly low ($<$0.05) regardless of complexity, suggesting the challenge lies in the task itself.
We also observe that models tend to predict common, high-frequency names (e.g., {\tt main}, {\tt init}, {\tt buf}) regardless of context, indicating a training data frequency bias.
Breaking down type inference by category (\autoref{tbl:typecategory} in the Appendix), models achieve the highest accuracy on primitive types ({\tt int}, {\tt char}), moderate accuracy on pointer types, and near-zero accuracy on user-defined types (e.g., {\tt struct sockaddr\_in}), which are project-specific and rarely appear in pre-training data.
}

\parag{Base LLM Evaluation}
The capability of base LLMs without fine-tuning warrants further investigation.
Prompt engineering is particularly challenging because each LLM generates responses in different formats, often ignoring the requested structure.
As a result, extensive post-processing is required to extract the predicted results.

While an auxiliary LLM could be used to parse and extract data from responses,
this approach may bias the results in unpredictable ways.
Moreover, the additional LLM can affect the performance of target LLM, which could cause inconsistent comparison. 
Consequently, the severe inconsistency in response formats necessitates multiple post-processing stages.

\input{table/finetune}
\parag{Time Spent on Fine-tuning}
As shown in~\autoref{tbl:finetime}, most models require approximately $6{,}000$ seconds (1.66 hours) to complete 200 fine-tuning steps.
Notably, WizardCoder requires less time to finish the 200 steps.

Initially, we planned to include the DeepSeek-V2 model to compare its performance with DeepSeek-R1.
However, DeepSeek-V2 takes $56{,}640$ seconds, over nine times longer than the average.
Consequently, we excluded DeepSeek-V2 due to its prohibitive overhead given our extensive experimental workload.
Further investigation revealed that this extended duration is largely due to prompt template design.
When we modified CodeLlama's prompt template, we observed a similar increase in training time, confirming the strong influence of template structure on training efficiency.
Unfortunately, not all models provide well-documented templates, making it difficult to select effective ones or verify their legitimacy.
Moreover, even though the model comes with the document, it does not always provide the expected consequences. 

\result{Fine-tuning time is highly sensitive to prompt templates; poor template design can increase training duration by up to nine times.}

\parag{Responses from LLMs}
Despite being instructed to produce outputs in a strict format without explanations, no LLM consistently adhered to the required structure.
For example, DeepSeek-R1 frequently generated a chain-of-thought reasoning before producing an answer.
Moreover, Llama2 always provides a response starting with ``Sure!'' word, even though data used in fine-tuning does not contain the word.
In some cases, the reasoning was so long that the model failed to return the final output within the token limit.
Hence, we modified the prompt to make the model skip the chain-of-thinking process.
However, it is unclear whether suppressing this reasoning process may affect performance. \looseness=-1

Moreover, many models exhibit a common pattern of predicting the same variable name for multiple distinct variables.
DeepSeek-R1 sometimes predicts the variable name as {\tt local\_\#}, which is one of the symbols automatically generated by the {\sf Ghidra} decompiler.
This pattern is also frequently observed in other models, such as Qwen2.5, Qwen3, and Phi4.
Furthermore, models do not consistently follow the output format learned during fine-tuning, suggesting that the fine-tuning signal for format compliance may be insufficient at 200 steps.

\result{LLMs usually fail to follow consistent output formatting unless assisted by fine-tuning or post-processing.}

\parag{Additional Performance Reasoning}
In many cases, LLMs fail to generate the complete name or type requested by the prompt, which inherently lowers F1 scores.
For example, although a prompt asks for ten variable names, LLMs often miss some of them, which can lead to a lower F1 score due to reduced recall.
Moreover, due to their tendency to produce extended reasoning, models sometimes exceed output size limits, causing truncation before the answer is generated.
A few models, such as CodeLlama, further complicate evaluation by producing multiple candidate names for a single function, making it unclear which prediction should be considered the final answer.
These observations highlight the importance of output format compliance and suggest that more structured decoding strategies, such as constrained generation or grammar-guided decoding, could improve both the reliability and accuracy of LLM outputs in reverse engineering tasks.

%% file: table/stats.tex
\begin{table*}
\caption{\bf Statistics of \sys Before and After Filtering Process. Numbers are in Millions.}
\label{tbl:stat}
\centering
\footnotesize
\setlength{\tabcolsep}{4pt} 
\scalebox{0.87}{
\begin{tabular}{lrrrr|rrrr|rrrr|rrrr} 
\hline
\multirow{2}{*}{\textbf{Count (m)}} & \multicolumn{4}{c|}{\textbf{x64 }}                                                         & \multicolumn{4}{c|}{\textbf{x86 }}                                                         & \multicolumn{4}{c|}{\textbf{ARM}}                                                          & \multicolumn{4}{c}{\textbf{MIPS}}                                                          \\ 
\cline{2-17}
                                    & \textbf{O0}          & \textbf{O1}          & \textbf{O2}          & \textbf{O3}           & \textbf{O0}          & \textbf{O1}          & \textbf{O2}          & \textbf{O3}           & \textbf{O0}          & \textbf{O1}          & \textbf{O2}          & \textbf{O3}           & \textbf{O0}          & \textbf{O1}          & \textbf{O2}          & \textbf{O3}           \\ 
\hline
\multicolumn{17}{c}{\textbf{\textbf{Before Filtering Process}}}                                                                                                                                                                                                                                                                                                                                                                 \\ 
\hline
\textbf{LoC}                        & 48.82                & 43.47                & 43.67                & 50.59                 & 40.41                & 36.01                & 36.29                & 41.50                 & 11.11                & 9.11                 & 8.81                 & 10.25                 & 14.86                & 14.03                & 13.43                & 14.51                 \\
\textbf{Func}                       & 4.64                 & 3.99                 & 3.97                 & 4.18                  & 5.20                 & 4.30                 & 4.27                 & 4.46                  & 0.95                 & 0.72                 & 0.66                 & 0.66                  & 0.93                 & 0.63                 & 0.59                 & 0.56                  \\
\textbf{Var}                        & 5.92                 & 2.28                 & 2.22                 & 2.51                  & 6.14                 & 3.75                 & 3.84                 & 4.26                  & 1.83                 & 0.87                 & 0.83                 & 0.91                  & 2.14                 & 1.09                 & 1.08                 & 1.09                  \\ 
\hline
\multicolumn{17}{c}{\textbf{\textbf{After Filtering Process}}}                                                                                                                                                                                                                                                                                                                                                                  \\ 
\hline
\textbf{LoC}                        & 29.94                & 28.29                & 29.11                & 36.88                 & 26.19                & 24.37                & 25.31                & 30.88                 & 7.99                & 6.87                 & 6.85                 & 8.34                 & 9.77                & 10.10                & 9.75                & 11.02                 \\
\textbf{Func}                       & 2.87                 & 2.45                 & 2.51                 & 2.88                  & 3.23                 & 2.79                 & 2.83                 & 3.14                  & 0.67                 & 0.52                 & 0.49                 & 0.51                  & 0.73                 & 0.50                 & 0.48                 & 0.48                  \\
\textbf{Var}                        & 3.65                 & 1.65                 & 1.61                 & 1.94                  & 4.06                 & 2.49                 & 2.66                 & 3.16                  & 1.40                 & 0.70                 & 0.67               & 0.75                  & 1.61                 & 0.88                 & 0.90                 & 0.93                  \\ 
\hline
\end{tabular}}
\end{table*}

\ignore{
\begin{table}[t!]
\centering
\footnotesize
\setlength{\tabcolsep}{4pt} 
\scalebox{0.9}{
\begin{tabular}{lrrrr|rrrr|rrrr|rrrr} 
\hline
                 & \multicolumn{4}{c|}{\textbf{x64}}                    & \multicolumn{4}{c|}{\textbf{x86}}                    & \multicolumn{4}{c|}{\textbf{ARM}}                     & \multicolumn{4}{c}{\textbf{MIPS}}                      \\ 
\cline{2-17}
                 & \textbf{O0} & \textbf{O1} & \textbf{O2} & \textbf{O3} & \textbf{O0} & \textbf{O1} & \textbf{O2} & \textbf{O3} & \textbf{O0} & \textbf{O1} & \textbf{O2} & \textbf{O3} & \textbf{O0} & \textbf{O1} & \textbf{O2} & \textbf{O3}  \\ 
\cline{2-17}
\textbf{LoC}     & 48.82       & 43.47       & 43.67       & 50.59       & 40.41       & 36.01       & 36.29       & 41.50       & 11.11       & 9.11        & 8.81        & 10.25       & 14.86       & 14.03       & 13.43       & 14.51        \\
\textbf{\# Func} & 4.64        & 3.99        & 3.97        & 4.18        & 5.20        & 4.30        & 4.27        & 4.46        & 0.95        & 0.72        & 0.66        & 0.66        & 0.93        & 0.63        & 0.59        & 0.56         \\
\textbf{\# Var}  & 5.92        & 2.28        & 2.22        & 2.51        & 6.14        & 3.75        & 3.84        & 4.26        & 1.83        & 0.87        & 0.83        & 0.91        & 2.14        & 1.09        & 1.08        & 1.09         \\
\hline
\end{tabular}}
\caption{\bf Statistics of \sys Before Filtering Process. The unit size is a million.}
\label{tbl:stat}
\end{table}
}

%% file: table/models.tex
\begin{table}[tb!]
\caption{List of Evaluated LLMs.}\label{llms}
\centering
\footnotesize
\scalebox{0.87}{
\begin{tabular}{lrrlrr}
\hline
\textbf{LLMs}                                               & \textbf{Size} & \textbf{Years} & \textbf{LLMs}                                                 & \textbf{Size} & \textbf{Years} \\ \hline
CodeLlama~\cite{roziere2023code}      & 13B           & 2023           & Llama2~\cite{touvron2023llama}          & 13B           & 2023           \\
WizardCoder~\cite{luo2023wizardcoder} & 15B           & 2023           & Phi4 Unsloth~\cite{unsloth}                                                  & 14B           & 2024           \\
Qwen 2.5~\cite{qwen2.5}               & 14B           & 2024           & Deepseek-R1-Qwen~\cite{guo2025deepseek} & 14B           & 2025           \\
Qwen 3~\cite{qwen3technicalreport}                                                      & 14B           & 2025           &  &            &             \\ \hline
\end{tabular}}
\end{table}

\ignore{
\begin{table}[tb!]
\centering
\caption{List of evaluated LLMs. \ma{you can make it 6 column table to save some space}}\label{llms}
\begin{tabular}{llll}
\toprule
\textbf{LLMs}        & \textbf{Size}                    & \textbf{Years} \\ \midrule
CodeGen~\cite{nijkamp2022codegen}     & 16B       & 2022     \\
CodeGen2~\cite{nijkamp2023codegen2}    & 16B                  & 2022               \\
Llama2~\cite{}      & 13B            & 2023                \\
CodeLlama~\cite{}   & 13B            & 2023             \\
CodtT5Plus~\cite{wang2023codet5plus}  & 16B & 2023              \\
StarCoder~\cite{li2023starcoder}   & 15.5B              & 2023              \\
WizardCoder~\cite{luo2023wizardcoder} & 15B             & 2023                \\ \bottomrule
\end{tabular}
\end{table}
}

%% file: table/func_result.tex
\ignore{
\begin{table*}[tb!]
\centering
\footnotesize
\scalebox{0.51}{
\begin{tabular}{lllllll|llllll|llllll|llllll} 
\hline
\multicolumn{1}{c}{\multirow{4}{*}{\textbf{Model}}} & \multicolumn{24}{c}{\textbf{Function Name Recovery Performance}}                                                                                                                                                                                                                                                                                                                                                                                                                                                                                                                                                                                                                                                                                                                                                                                                                                  \\ 
\cline{2-25}
\multicolumn{1}{c}{}                                & \multicolumn{6}{c|}{\textbf{O0}}                                                                                                                                                                                       & \multicolumn{6}{c|}{\textbf{O1}}                                                                                                                                                                                       & \multicolumn{6}{c|}{\textbf{O2}}                                                                                                                                                                                       & \multicolumn{6}{c}{\textbf{O3}}                                                                                                                                                                                        \\ 
\cline{2-25}
\multicolumn{1}{c}{}                                & \multicolumn{2}{c}{\textbf{Prec }}                                    & \multicolumn{2}{c}{\textbf{Recall }}                                  & \multicolumn{2}{c|}{\textbf{F1 }}                                      & \multicolumn{2}{c}{\textbf{Prec }}                                    & \multicolumn{2}{c}{\textbf{Recall }}                                  & \multicolumn{2}{c|}{\textbf{F1}}              & \multicolumn{2}{c}{\textbf{Prec }}                                    & \multicolumn{2}{c}{\textbf{Recall}}                                   & \multicolumn{2}{c|}{\textbf{F1}}                                       & \multicolumn{2}{c}{\textbf{Prec}}                                     & \multicolumn{2}{c}{\textbf{Recall}}                                   & \multicolumn{2}{c}{\textbf{F1}}                                        \\ 
\cline{2-25}
\multicolumn{1}{c}{}                                & \multicolumn{1}{c}{\textbf{Orig}} & \multicolumn{1}{c}{\textbf{Fine}} & \multicolumn{1}{c}{\textbf{Orig}} & \multicolumn{1}{c}{\textbf{Fine}} & \multicolumn{1}{c}{\textbf{Orig}} & \multicolumn{1}{c|}{\textbf{Fine}} & \multicolumn{1}{c}{\textbf{Orig}} & \multicolumn{1}{c}{\textbf{Fine}} & \multicolumn{1}{c}{\textbf{Orig}} & \multicolumn{1}{c}{\textbf{Fine}} & \multicolumn{1}{c}{\textbf{Orig}} & \multicolumn{1}{c|}{\textbf{Fine}} & \multicolumn{1}{c}{\textbf{Orig}} & \multicolumn{1}{c}{\textbf{Fine}} & \multicolumn{1}{c}{\textbf{Orig}} & \multicolumn{1}{c}{\textbf{Fine}} & \multicolumn{1}{c}{\textbf{Orig}} & \multicolumn{1}{c|}{\textbf{Fine}} & \multicolumn{1}{c}{\textbf{Orig}} & \multicolumn{1}{c}{\textbf{Fine}} & \multicolumn{1}{c}{\textbf{Orig}} & \multicolumn{1}{c}{\textbf{Fine}} & \multicolumn{1}{c}{\textbf{Orig}} & \multicolumn{1}{c}{\textbf{Fine}}  \\ 
\hline
\multicolumn{25}{c}{\bf x64}                                                                                                                                                                                                                                                                                                                                                                                                                                                                                                                                                                                                                                                                                                                                                                                                                                                                                                                                \\ 
\hline
CodeLlama  & 0.05 & 0.10 & 0.11 & 0.09 & 0.07 & 0.10 & 0.05 & 0.13 & 0.11 & 0.11 & 0.07 & 0.12 & 0.06 & 0.14 & 0.12 & 0.12 & 0.08 & 0.13 & 0.05 & 0.13 & 0.12 & 0.12 & 0.07 & 0.12 \\
Llama2    & 0.02 & 0.05 & 0.04 & 0.06 & 0.03 & 0.05 & 0.02 & 0.07 & 0.05 & 0.08 & 0.03 & 0.08 & 0.02 & 0.07 & 0.06 & 0.08 & 0.03 & 0.07 & 0.02 & 0.07 & 0.05 & 0.08 & 0.03 & 0.07   \\
Deepseek-V2  & 0.02 & 0.10 & 0.05 & 0.11 & 0.03 & 0.11 & 0.03 & 0.12 & 0.06 & 0.12 & 0.04 & 0.12 & 0.03 & 0.12 & 0.06 & 0.12 & 0.04 & 0.12 & 0.03 & 0.12 & 0.06 & 0.13 & 0.04 & 0.12   \\
Deepseek-R1   & 0.05 & 0.15 & 0.09 & 0.13 & 0.06 & 0.14 & 0.04 & 0.17 & 0.09 & 0.16 & 0.06 & 0.16 & 0.04 & 0.18 & 0.10 & 0.16 & 0.06 & 0.17 & 0.04 & 0.18 & 0.08 & 0.16 & 0.05 & 0.17  \\
Qwen  & 0.06 & 0.15 & 0.08 & 0.14 & 0.07 & 0.15 & 0.06 & 0.18 & 0.07 & 0.16 & 0.06 & 0.17 & 0.06 & 0.18 & 0.08 & 0.16 & 0.07 & 0.17 & 0.06 & 0.18 & 0.08 & 0.17 & 0.07 & 0.18    \\
WizardCoder & 0.02 & 0.14 & 0.06 & 0.12 & 0.03 & 0.13 & 0.02 & 0.15 & 0.04 & 0.13 & 0.03 & 0.14 & 0.02 & 0.15 & 0.05 & 0.14 & 0.03 & 0.14 & 0.02 & 0.15 & 0.05 & 0.14 & 0.03 & 0.15     \\ 
\hline
\multicolumn{25}{c}{\bf x86}                                                                                                                                                                                                                                                                                                                                                                                                                                                                                                                                                                                                                                                                                                                                                                                                                                                                                                                                \\ 
\hline
CodeLlama   & 0.08 & 0.14 & 0.16 & 0.13 & 0.11 & 0.14 & 0.10 & 0.15 & 0.17 & 0.14 & 0.13 & 0.15 & 0.09 & 0.16 & 0.17 & 0.14 & 0.12 & 0.15 & 0.10 & 0.16 & 0.18 & 0.15 & 0.13 & 0.16    \\
Llama2   & 0.03 & 0.09 & 0.07 & 0.11 & 0.04 & 0.10 & 0.03 & 0.10 & 0.07 & 0.12 & 0.04 & 0.11 & 0.03 & 0.09 & 0.08 & 0.11 & 0.04 & 0.10 & 0.03 & 0.10 & 0.07 & 0.13 & 0.05 & 0.11    \\
Deepseek-V2  & 0.05 & 0.16 & 0.08 & 0.18 & 0.06 & 0.17 & 0.05 & 0.17 & 0.08 & 0.19 & 0.06 & 0.18 & 0.05 & 0.18 & 0.08 & 0.19 & 0.06 & 0.18 & 0.05 & 0.18 & 0.09 & 0.20 & 0.06 & 0.19    \\
Deepseek-R1  & 0.05 & 0.37 & 0.09 & 0.35 & 0.06 & 0.36 & 0.05 & 0.38 & 0.09 & 0.36 & 0.07 & 0.37 & 0.05 & 0.37 & 0.09 & 0.35 & 0.07 & 0.36 & 0.06 & 0.37 & 0.10 & 0.36 & 0.08 & 0.36  \\
Qwen  & 0.09 & 0.36 & 0.12 & 0.32 & 0.10 & 0.34 & 0.09 & 0.36 & 0.12 & 0.33 & 0.11 & 0.34 & 0.09 & 0.35 & 0.13 & 0.32 & 0.11 & 0.33 & 0.10 & 0.36 & 0.14 & 0.33 & 0.12 & 0.34    \\
WizardCoder  & 0.05 & 0.35 & 0.11 & 0.31 & 0.07 & 0.33 & 0.05 & 0.33 & 0.09 & 0.30 & 0.06 & 0.32 & 0.05 & 0.32 & 0.09 & 0.30 & 0.06 & 0.31 & 0.05 & 0.33 & 0.11 & 0.31 & 0.07 & 0.32   \\ 
\hline
\multicolumn{25}{c}{\bf ARM}                                                                                                                                                                                                                                                                                                                                                                                                                                                                                                                                                                                                                                                                                                                                                                                                                                                                                                                                \\ 
\hline
CodeLlama  & 0.05 & 0.06 & 0.09 & 0.04 & 0.06 & 0.05 & 0.05 & 0.07 & 0.09 & 0.05 & 0.06 & 0.06 & 0.05 & 0.07 & 0.09 & 0.05 & 0.06 & 0.06 & 0.06 & 0.07 & 0.11 & 0.06 & 0.07 & 0.06  \\
Llama2  & 0.02 & 0.03 & 0.05 & 0.03 & 0.03 & 0.03 & 0.02 & 0.04 & 0.05 & 0.04 & 0.03 & 0.04 & 0.02 & 0.04 & 0.04 & 0.04 & 0.03 & 0.04 & 0.03 & 0.04 & 0.05 & 0.05 & 0.04 & 0.04  \\
Deepseek-V2  & 0.03 & 0.07 & 0.04 & 0.07 & 0.03 & 0.07 & 0.03 & 0.07 & 0.04 & 0.07 & 0.03 & 0.07 & 0.03 & 0.07 & 0.04 & 0.07 & 0.03 & 0.07 & 0.03 & 0.08 & 0.05 & 0.08 & 0.04 & 0.08   \\
Deepseek-R1 & 0.04 & 0.08 & 0.07 & 0.07 & 0.05 & 0.08 & 0.05 & 0.09 & 0.09 & 0.08 & 0.06 & 0.09 & 0.05 & 0.10 & 0.08 & 0.09 & 0.06 & 0.09 & 0.05 & 0.10 & 0.09 & 0.09 & 0.07 & 0.10  \\
Qwen   & 0.05 & 0.09 & 0.06 & 0.09 & 0.05 & 0.09 & 0.06 & 0.09 & 0.06 & 0.09 & 0.06 & 0.09 & 0.06 & 0.10 & 0.06 & 0.10 & 0.06 & 0.10 & 0.06 & 0.12 & 0.07 & 0.11 & 0.07 & 0.12 \\
WizardCoder  & 0.03 & 0.08 & 0.05 & 0.07 & 0.04 & 0.07 & 0.03 & 0.08 & 0.05 & 0.07 & 0.03 & 0.08 & 0.03 & 0.09 & 0.05 & 0.08 & 0.04 & 0.08 & 0.03 & 0.10 & 0.06 & 0.08 & 0.04 & 0.09  \\ 
\hline
\multicolumn{25}{c}{\bf MIPS}                                                                                                                                                                                                                                                                                                                                                                                                                                                                                                                                                                                                                                                                                                                                                                                                                                                                                                                               \\ 
\hline
CodeLlama   & 0.03 & 0.04 & 0.05 & 0.03 & 0.04 & 0.04 & 0.02 & 0.04 & 0.05 & 0.03 & 0.03 & 0.04 & 0.03 & 0.04 & 0.05 & 0.04 & 0.03 & 0.04 & 0.03 & 0.04 & 0.05 & 0.03 & 0.03 & 0.03  \\
Llama2   & 0.02 & 0.03 & 0.03 & 0.03 & 0.02 & 0.03 & 0.02 & 0.03 & 0.03 & 0.03 & 0.02 & 0.03 & 0.02 & 0.03 & 0.03 & 0.03 & 0.02 & 0.03 & 0.01 & 0.03 & 0.03 & 0.03 & 0.02 & 0.03    \\
Deepseek-V2   & 0.01 & 0.04 & 0.01 & 0.04 & 0.01 & 0.04 & 0.01 & 0.04 & 0.01 & 0.04 & 0.01 & 0.04 & 0.01 & 0.04 & 0.01 & 0.04 & 0.01 & 0.04 & 0.01 & 0.04 & 0.01 & 0.04 & 0.01 & 0.04    \\
Deepseek-R1  & 0.02 & 0.06 & 0.04 & 0.05 & 0.03 & 0.05 & 0.02 & 0.06 & 0.04 & 0.05 & 0.03 & 0.05 & 0.02 & 0.06 & 0.03 & 0.05 & 0.02 & 0.05 & 0.02 & 0.06 & 0.04 & 0.05 & 0.03 & 0.06  \\
Qwen  & 0.03 & 0.06 & 0.04 & 0.05 & 0.03 & 0.05 & 0.02 & 0.06 & 0.03 & 0.05 & 0.03 & 0.05 & 0.02 & 0.06 & 0.03 & 0.05 & 0.03 & 0.06 & 0.02 & 0.06 & 0.03 & 0.05 & 0.03 & 0.05  \\
WizardCoder  & 0.01 & 0.06 & 0.03 & 0.05 & 0.02 & 0.05 & 0.01 & 0.06 & 0.02 & 0.05 & 0.01 & 0.05 & 0.01 & 0.06 & 0.02 & 0.05 & 0.01 & 0.06 & 0.01 & 0.06 & 0.02 & 0.04 & 0.01 & 0.05       \\
\hline
\end{tabular}}
\caption{Performance Evaluation on Function Name Recovery with Different Architectures and Optimization Levels.}
\label{tbl:nameresult}
\end{table*}
}


\begin{table*}[t!]
\caption{Performance Evaluation on Function Name Recovery with Different Architectures and Optimization Levels.}
\label{tbl:nameresult}
\centering
\footnotesize
\scalebox{0.87}{
\begin{tabular}{lrrrrrrrrrrrr}
\hline
\multicolumn{1}{c}{\multirow{3}{*}{\textbf{Model}}} & \multicolumn{12}{c}{\textbf{Function Name Recovery Performance}}                                                                                                                                                                                                                                                                                                                                                                                 \\ \cline{2-13} 
\multicolumn{1}{c}{}                                & \multicolumn{3}{c|}{\textbf{O0}}                                                                           & \multicolumn{3}{c|}{\textbf{O1}}                                                                           & \multicolumn{3}{c|}{\textbf{O2}}                                                                           & \multicolumn{3}{c}{\textbf{O3}}                                                                           \\ \cline{2-13} 
\multicolumn{1}{c}{}                                & \multicolumn{1}{c}{\textbf{Prec}} & \multicolumn{1}{c}{\textbf{Recall}} & \multicolumn{1}{c|}{\textbf{F1}} & \multicolumn{1}{c}{\textbf{Prec}} & \multicolumn{1}{c}{\textbf{Recall}} & \multicolumn{1}{c|}{\textbf{F1}} & \multicolumn{1}{c}{\textbf{Prec}} & \multicolumn{1}{c}{\textbf{Recall}} & \multicolumn{1}{c|}{\textbf{F1}} & \multicolumn{1}{c}{\textbf{Prec}} & \multicolumn{1}{c}{\textbf{Recall}} & \multicolumn{1}{c}{\textbf{F1}} \\ \hline
\multicolumn{13}{c}{x64}                                                                                                                                                                                                                                                                                                                                                                                                                                                                               \\ \hline
CodeLlama                                           & 0.0513                            & 0.0892                              & \multicolumn{1}{r|}{0.0651}      & 0.0556                            & 0.0924                              & \multicolumn{1}{r|}{0.0695}      & 0.0542                            & 0.0881                              & \multicolumn{1}{r|}{0.0671}      & 0.0512                            & 0.0892                              & 0.0650                          \\
Llama2                                              & 0.0269                            & 0.0492                              & \multicolumn{1}{r|}{0.0348}      & 0.0323                            & 0.0410                              & \multicolumn{1}{r|}{0.0362}      & 0.0317                            & 0.0438                              & \multicolumn{1}{r|}{0.0368}      & 0.0304                            & 0.0425                              & 0.0354                          \\
WizardCoder                                         & 0.0867                            & 0.0943                              & \multicolumn{1}{r|}{0.0903}      & 0.0826                            & 0.0882                              & \multicolumn{1}{r|}{0.0853}      & 0.0904                            & 0.0960                              & \multicolumn{1}{r|}{0.0931}      & 0.0866                            & 0.0955                              & 0.0908                          \\
Deepseek-R1                                         & 0.1272                            & 0.1299                              & \multicolumn{1}{r|}{0.1285}      & 0.1361                            & 0.1376                              & \multicolumn{1}{r|}{0.1368}      & 0.1490                            & 0.1476                              & \multicolumn{1}{r|}{0.1483}      & 0.1460                            & 0.1482                              & 0.1471                          \\
Qwen2.5                                             & 0.1430                            & 0.1353                              & \multicolumn{1}{r|}{0.1391}      & 0.1507                            & 0.1466                              & \multicolumn{1}{r|}{0.1487}      & 0.1472                            & 0.1434                              & \multicolumn{1}{r|}{0.1453}      & 0.1567                            & 0.1546                              & 0.1556                          \\
Qwen3                                               & 0.1313                            & 0.1301                              & \multicolumn{1}{r|}{0.1307}      & 0.1385                            & 0.1377                              & \multicolumn{1}{r|}{0.1381}      & 0.1440                            & 0.1416                              & \multicolumn{1}{r|}{0.1428}      & 0.1515                            & 0.1513                              & 0.1514                          \\
Phi4 Unsloth                                        & 0.1535                            & 0.1583                              & \multicolumn{1}{r|}{0.1558}      & 0.1561                            & 0.1571                              & \multicolumn{1}{r|}{0.1566}      & 0.1556                            & 0.1542                              & \multicolumn{1}{r|}{0.1549}      & 0.1599                            & 0.1636                              & 0.1617                          \\ \hline
\multicolumn{13}{c}{x86}                                                                                                                                                                                                                                                                                                                                                                                                                                                                               \\ \hline
CodeLlama                                           & 0.0975                            & 0.1516                              & \multicolumn{1}{r|}{0.1187}      & 0.1050                            & 0.1642                              & \multicolumn{1}{r|}{0.1281}      & 0.1058                            & 0.1660                              & \multicolumn{1}{r|}{0.1292}      & 0.1047                            & 0.1685                              & 0.1291                          \\
Llama2                                              & 0.0413                            & 0.0788                              & \multicolumn{1}{r|}{0.0542}      & 0.0533                            & 0.0910                              & \multicolumn{1}{r|}{0.0672}      & 0.0480                            & 0.0841                              & \multicolumn{1}{r|}{0.0611}      & 0.0489                            & 0.0912                              & 0.0637                          \\
WizardCoder                                         & 0.1386                            & 0.1395                              & \multicolumn{1}{r|}{0.1390}      & 0.1513                            & 0.1542                              & \multicolumn{1}{r|}{0.1527}      & 0.1429                            & 0.1447                              & \multicolumn{1}{r|}{0.1438}      & 0.1527                            & 0.1549                              & 0.1538                          \\
Deepseek-R1                                         & 0.3486                            & 0.3485                              & \multicolumn{1}{r|}{0.3486}      & 0.3473                            & 0.3502                              & \multicolumn{1}{r|}{0.3487}      & 0.3464                            & 0.3497                              & \multicolumn{1}{r|}{0.3480}      & 0.3544                            & 0.3584                              & 0.3564                          \\
Qwen2.5                                             & 0.3105                            & 0.2972                              & \multicolumn{1}{r|}{0.3037}      & 0.3224                            & 0.3136                              & \multicolumn{1}{r|}{0.3179}      & 0.3258                            & 0.3164                              & \multicolumn{1}{r|}{0.3210}      & 0.3231                            & 0.3195                              & 0.3213                          \\
Qwen3                                               & 0.3763                            & 0.3780                              & \multicolumn{1}{r|}{0.3771}      & 0.3758                            & 0.3691                              & \multicolumn{1}{r|}{0.3724}      & 0.3798                            & 0.3718                              & \multicolumn{1}{r|}{0.3758}      & 0.3814                            & 0.3818                              & 0.3816                          \\
Phi4 Unsloth                                        & 0.2974                            & 0.3040                              & \multicolumn{1}{r|}{0.3007}      & 0.3149                            & 0.3305                              & \multicolumn{1}{r|}{0.3225}      & 0.3199                            & 0.3331                              & \multicolumn{1}{r|}{0.3264}      & 0.3175                            & 0.3350                              & 0.3260                          \\ \hline
\multicolumn{13}{c}{ARM}                                                                                                                                                                                                                                                                                                                                                                                                                                                                               \\ \hline
CodeLlama                                           & 0.0345                            & 0.0570                              & \multicolumn{1}{r|}{0.0429}      & 0.0381                            & 0.0594                              & \multicolumn{1}{r|}{0.0464}      & 0.0428                            & 0.0669                              & \multicolumn{1}{r|}{0.0522}      & 0.0431                            & 0.0697                              & 0.0533                          \\
Llama2                                              & 0.0294                            & 0.0404                              & \multicolumn{1}{r|}{0.0340}      & 0.0286                            & 0.0391                              & \multicolumn{1}{r|}{0.0330}      & 0.0282                            & 0.0398                              & \multicolumn{1}{r|}{0.0330}      & 0.0340                            & 0.0439                              & 0.0383                          \\
WizardCoder                                         & 0.0687                            & 0.0723                              & \multicolumn{1}{r|}{0.0704}      & 0.0678                            & 0.0703                              & \multicolumn{1}{r|}{0.0690}      & 0.0715                            & 0.0754                              & \multicolumn{1}{r|}{0.0734}      & 0.0804                            & 0.0866                              & 0.0834                          \\
Deepseek-R1                                         & 0.0777                            & 0.0802                              & \multicolumn{1}{r|}{0.0790}      & 0.0898                            & 0.0899                              & \multicolumn{1}{r|}{0.0899}      & 0.0923                            & 0.0945                              & \multicolumn{1}{r|}{0.0934}      & 0.0959                            & 0.0998                              & 0.0978                          \\
Qwen2.5                                             & 0.0934                            & 0.0930                              & \multicolumn{1}{r|}{0.0932}      & 0.0918                            & 0.0971                              & \multicolumn{1}{r|}{0.0944}      & 0.1034                            & 0.1115                              & \multicolumn{1}{r|}{0.1073}      & 0.1113                            & 0.1231                              & 0.1169                          \\
Qwen3                                               & 0.0800                            & 0.0777                              & \multicolumn{1}{r|}{0.0788}      & 0.0832                            & 0.0865                              & \multicolumn{1}{r|}{0.0848}      & 0.0859                            & 0.0894                              & \multicolumn{1}{r|}{0.0876}      & 0.0939                            & 0.1014                              & 0.0975                          \\
Phi4 Unsloth                                        & 0.1069                            & 0.1065                              & \multicolumn{1}{r|}{0.1067}      & 0.1097                            & 0.1124                              & \multicolumn{1}{r|}{0.1110}       & 0.1198                            & 0.1235                              & \multicolumn{1}{r|}{0.1216}      & 0.1236                            & 0.1287                              & 0.1261                          \\ \hline
\multicolumn{13}{c}{MIPS}                                                                                                                                                                                                                                                                                                                                                                                                                                                                              \\ \hline
CodeLlama                                           & 0.0180                            & 0.0316                              & \multicolumn{1}{r|}{0.0230}      & 0.0206                            & 0.0313                              & \multicolumn{1}{r|}{0.0248}      & 0.0175                            & 0.0266                              & \multicolumn{1}{r|}{0.0211}      & 0.0175                            & 0.0255                              & 0.0208                          \\
Llama2                                              & 0.0224                            & 0.0282                              & \multicolumn{1}{r|}{0.0250}      & 0.0237                            & 0.0269                              & \multicolumn{1}{r|}{0.0252}      & 0.0270                            & 0.0343                              & \multicolumn{1}{r|}{0.0302}      & 0.0254                            & 0.0313                              & 0.0280                          \\
WizardCoder                                         & 0.0484                            & 0.0419                              & \multicolumn{1}{r|}{0.0449}      & 0.0466                            & 0.0403                              & \multicolumn{1}{r|}{0.0432}      & 0.0512                            & 0.0448                              & \multicolumn{1}{r|}{0.0478}      & 0.0476                            & 0.0400                              & 0.0435                          \\
Deepseek-R1                                         & 0.0470                            & 0.0481                              & \multicolumn{1}{r|}{0.0476}      & 0.0537                            & 0.0508                              & \multicolumn{1}{r|}{0.0522}      & 0.0567                            & 0.0531                              & \multicolumn{1}{r|}{0.0548}      & 0.0509                            & 0.0490                              & 0.0499                          \\
Qwen2.5                                             & 0.0568                            & 0.0531                              & \multicolumn{1}{r|}{0.0549}      & 0.0570                            & 0.0522                              & \multicolumn{1}{r|}{0.0545}      & 0.0557                            & 0.0501                              & \multicolumn{1}{r|}{0.0528}      & 0.0533                            & 0.0485                              & 0.0508                          \\
Qwen3                                               & 0.0507                            & 0.0502                              & \multicolumn{1}{r|}{0.0505}      & 0.0477                            & 0.0471                              & \multicolumn{1}{r|}{0.0474}      & 0.0519                            & 0.0513                              & \multicolumn{1}{r|}{0.0516}      & 0.0495                            & 0.0490                              & 0.0493                          \\
Phi4 Unsloth                                        & 0.0594                            & 0.0598                              & \multicolumn{1}{r|}{0.0596}      & 0.0611                            & 0.0643                              & \multicolumn{1}{r|}{0.0626}      & 0.0675                            & 0.0714                              & \multicolumn{1}{r|}{0.0694}      & 0.0656                            & 0.0693                              & 0.0674                          \\ \hline
\end{tabular}}
\end{table*}

%% file: table/var_result.tex
\ignore{
\begin{table*}[tb!]
\centering
\footnotesize
\scalebox{0.51}{
\begin{tabular}{lllllll|llllll|llllll|llllll} 
\hline
\multicolumn{1}{c}{\multirow{4}{*}{\textbf{Model }}} & \multicolumn{24}{c}{\textbf{Variable Name Recovery Performance }}                                                                                                                                                                                                                                                                                                                                                                                                                                                                                                                                                                                                                                                                                                                                                                                                                                 \\ 
\cline{2-25}
\multicolumn{1}{c}{}                                 & \multicolumn{6}{c|}{\textbf{O0 }}                                                                                                                                                                                      & \multicolumn{6}{c|}{\textbf{O1 }}                                                                                                                                                                                      & \multicolumn{6}{c|}{\textbf{O2 }}                                                                                                                                                                                      & \multicolumn{6}{c}{\textbf{O3 }}                                                                                                                                                                                       \\ 
\cline{2-25}
\multicolumn{1}{c}{}                                 & \multicolumn{2}{c}{\textbf{Prec }}                                    & \multicolumn{2}{c}{\textbf{Recall }}                                  & \multicolumn{2}{c|}{\textbf{F1 }}                                      & \multicolumn{2}{c}{\textbf{Prec }}                                    & \multicolumn{2}{c}{\textbf{Recall }}                                  & \multicolumn{2}{c|}{\textbf{F1 }}                                      & \multicolumn{2}{c}{\textbf{Prec }}                                    & \multicolumn{2}{c}{\textbf{Recall }}                                  & \multicolumn{2}{c|}{\textbf{F1 }}                                      & \multicolumn{2}{c}{\textbf{Prec }}                                    & \multicolumn{2}{c}{\textbf{Recall }}                                  & \multicolumn{2}{c}{\textbf{F1 }}                                       \\
\cline{2-25}
\multicolumn{1}{c}{}                                 & \multicolumn{1}{c}{\textbf{Orig}} & \multicolumn{1}{c}{\textbf{Fine}} & \multicolumn{1}{c}{\textbf{Orig}} & \multicolumn{1}{c}{\textbf{Fine}} & \multicolumn{1}{c}{\textbf{Orig}} & \multicolumn{1}{c|}{\textbf{Fine}} & \multicolumn{1}{c}{\textbf{Orig}} & \multicolumn{1}{c}{\textbf{Fine}} & \multicolumn{1}{c}{\textbf{Orig}} & \multicolumn{1}{c}{\textbf{Fine}} & \multicolumn{1}{c}{\textbf{Orig}} & \multicolumn{1}{c|}{\textbf{Fine}} & \multicolumn{1}{c}{\textbf{Orig}} & \multicolumn{1}{c}{\textbf{Fine}} & \multicolumn{1}{c}{\textbf{Orig}} & \multicolumn{1}{c}{\textbf{Fine}} & \multicolumn{1}{c}{\textbf{Orig}} & \multicolumn{1}{c|}{\textbf{Fine}} & \multicolumn{1}{c}{\textbf{Orig}} & \multicolumn{1}{c}{\textbf{Fine}} & \multicolumn{1}{c}{\textbf{Orig}} & \multicolumn{1}{c}{\textbf{Fine}} & \multicolumn{1}{c}{\textbf{Orig}} & \multicolumn{1}{c}{\textbf{Fine}}  \\ 
\hline
\multicolumn{25}{c}{\textbf{\bf x64 }}                                                                                                                                                                                                                                                                                                                                                                                                                                                                                                                                                                                                                                                                                                                                                                                                                                                                                                                       \\ 
\hline
CodeLlama    & 0.02 & 0.19 & 0.03 & 0.24 & 0.02 & 0.21 & 0.03 & 0.07 & 0.04 & 0.09 & 0.03 & 0.07 & 0.02 & 0.05 & 0.03 & 0.06 & 0.02 & 0.06 & 0.02 & 0.06 & 0.03 & 0.08 & 0.02 & 0.07  \\
Llama2   & 0.01 & 0.22 & 0.02 & 0.26 & 0.02 & 0.24 & 0.02 & 0.05 & 0.02 & 0.06 & 0.02 & 0.05 & 0.01 & 0.04 & 0.02 & 0.05 & 0.01 & 0.05 & 0.02 & 0.06 & 0.02 & 0.07 & 0.02 & 0.06  \\
Deepseek-V2 & 0.03 & 0.27 & 0.04 & 0.45 & 0.04 & 0.34 & 0.08 & 0.05 & 0.09 & 0.08 & 0.08 & 0.06 & 0.05 & 0.05 & 0.06 & 0.07 & 0.05 & 0.06 & 0.06 & 0.06 & 0.06 & 0.09 & 0.06 & 0.07   \\
Deepseek-R1  & 0.02 & 0.15 & 0.03 & 0.23 & 0.03 & 0.18 & 0.03 & 0.06 & 0.03 & 0.08 & 0.03 & 0.07 & 0.03 & 0.06 & 0.04 & 0.08 & 0.03 & 0.07 & 0.02 & 0.06 & 0.03 & 0.07 & 0.02 & 0.07  \\
Qwen   & 0.02 & 0.28 & 0.02 & 0.46 & 0.02 & 0.35 & 0.03 & 0.06 & 0.04 & 0.08 & 0.03 & 0.07 & 0.03 & 0.07 & 0.03 & 0.10 & 0.03 & 0.08 & 0.02 & 0.07 & 0.03 & 0.10 & 0.03 & 0.08  \\
WizardCoder & 0.03 & 0.29 & 0.04 & 0.50 & 0.03 & 0.37 & 0.05 & 0.06 & 0.06 & 0.09 & 0.05 & 0.07 & 0.03 & 0.06 & 0.05 & 0.08 & 0.04 & 0.07 & 0.03 & 0.06 & 0.04 & 0.09 & 0.03 & 0.07     \\ 
\hline
\multicolumn{25}{c}{\bf x86}                                                                                                                                                                                                                                                                                                                                                                                                                                                                                                                                                                                                                                                                                                                                                                                                                                                                                                                                 \\ 
\hline
CodeLlama   & 0.04 & 0.05 & 0.05 & 0.06 & 0.04 & 0.05 & 0.04 & 0.07 & 0.05 & 0.08 & 0.05 & 0.07 & 0.04 & 0.06 & 0.05 & 0.08 & 0.05 & 0.07 & 0.04 & 0.06 & 0.05 & 0.08 & 0.04 & 0.07 \\
Llama2     & 0.03 & 0.04 & 0.03 & 0.03 & 0.03 & 0.04 & 0.03 & 0.04 & 0.04 & 0.03 & 0.03 & 0.04 & 0.03 & 0.03 & 0.03 & 0.03 & 0.03 & 0.03 & 0.02 & 0.03 & 0.03 & 0.03 & 0.03 & 0.03  \\
Deepseek-V2   & 0.05 & 0.08 & 0.05 & 0.09 & 0.05 & 0.08 & 0.05 & 0.08 & 0.06 & 0.09 & 0.06 & 0.09 & 0.05 & 0.07 & 0.05 & 0.08 & 0.05 & 0.08 & 0.06 & 0.07 & 0.06 & 0.09 & 0.06 & 0.08  \\
Deepseek-R1  & 0.04 & 0.11 & 0.03 & 0.13 & 0.03 & 0.12 & 0.04 & 0.10 & 0.03 & 0.12 & 0.04 & 0.11 & 0.04 & 0.09 & 0.03 & 0.11 & 0.03 & 0.10 & 0.05 & 0.09 & 0.05 & 0.11 & 0.05 & 0.10   \\
Qwen   & 0.05 & 0.10 & 0.04 & 0.11 & 0.05 & 0.11 & 0.05 & 0.11 & 0.05 & 0.12 & 0.05 & 0.11 & 0.06 & 0.10 & 0.05 & 0.11 & 0.06 & 0.11 & 0.06 & 0.11 & 0.06 & 0.12 & 0.06 & 0.12  \\
WizardCoder & 0.05 & 0.10 & 0.06 & 0.11 & 0.05 & 0.10 & 0.05 & 0.11 & 0.05 & 0.12 & 0.05 & 0.11 & 0.04 & 0.09 & 0.05 & 0.11 & 0.04 & 0.10 & 0.04 & 0.10 & 0.04 & 0.11 & 0.04 & 0.10  \\ 
\hline
\multicolumn{25}{c}{\bf ARM}                                                                                                                                                                                                                                                                                                                                                                                                                                                                                                                                                                                                                                                                                                                                                                                                                                                                                                                                 \\ 
\hline
CodeLlama  & 0.02 & 0.12 & 0.03 & 0.17 & 0.03 & 0.14 & 0.04 & 0.07 & 0.06 & 0.08 & 0.05 & 0.07 & 0.03 & 0.06 & 0.04 & 0.09 & 0.03 & 0.07 & 0.03 & 0.07 & 0.04 & 0.10 & 0.03 & 0.08  \\
Llama2    & 0.01 & 0.19 & 0.02 & 0.23 & 0.02 & 0.21 & 0.02 & 0.08 & 0.03 & 0.08 & 0.02 & 0.08 & 0.02 & 0.09 & 0.02 & 0.11 & 0.02 & 0.10 & 0.02 & 0.11 & 0.03 & 0.12 & 0.02 & 0.12 \\
Deepseek-V2  & 0.03 & 0.24 & 0.03 & 0.35 & 0.03 & 0.29 & 0.05 & 0.10 & 0.06 & 0.12 & 0.05 & 0.11 & 0.04 & 0.12 & 0.05 & 0.16 & 0.05 & 0.14 & 0.06 & 0.14 & 0.06 & 0.18 & 0.06 & 0.15 \\
Deepseek-R1   & 0.02 & 0.18 & 0.02 & 0.28 & 0.02 & 0.22 & 0.03 & 0.08 & 0.03 & 0.11 & 0.03 & 0.10 & 0.03 & 0.10 & 0.04 & 0.14 & 0.03 & 0.12 & 0.03 & 0.12 & 0.03 & 0.15 & 0.03 & 0.13  \\
Qwen  & 0.02 & 0.27 & 0.03 & 0.40 & 0.02 & 0.32 & 0.04 & 0.12 & 0.03 & 0.15 & 0.03 & 0.13 & 0.03 & 0.16 & 0.03 & 0.20 & 0.03 & 0.18 & 0.04 & 0.19 & 0.04 & 0.25 & 0.04 & 0.21   \\
WizardCoder  & 0.03 & 0.28 & 0.03 & 0.42 & 0.03 & 0.33 & 0.05 & 0.10 & 0.06 & 0.13 & 0.06 & 0.11 & 0.03 & 0.12 & 0.04 & 0.16 & 0.03 & 0.14 & 0.05 & 0.16 & 0.05 & 0.22 & 0.05 & 0.18    \\ 
\hline
\multicolumn{25}{c}{\bf MIPS}                                                                                                                                                                                                                                                                                                                                                                                                                                                                                                                                                                                                                                                                                                                                                                                                                                                                                                                                \\ 
\hline
CodeLlama  & 0.03 & 0.02 & 0.03 & 0.03 & 0.03 & 0.03 & 0.03 & 0.01 & 0.04 & 0.02 & 0.03 & 0.02 & 0.03 & 0.02 & 0.05 & 0.03 & 0.04 & 0.02 & 0.02 & 0.01 & 0.04 & 0.02 & 0.03 & 0.01   \\
Llama2   & 0.02 & 0.02 & 0.03 & 0.02 & 0.03 & 0.02 & 0.02 & 0.01 & 0.03 & 0.01 & 0.03 & 0.01 & 0.02 & 0.01 & 0.03 & 0.01 & 0.02 & 0.01 & 0.02 & 0.01 & 0.03 & 0.02 & 0.02 & 0.01  \\
Deepseek-V2  & 0.03 & 0.03 & 0.02 & 0.04 & 0.02 & 0.03 & 0.02 & 0.01 & 0.02 & 0.02 & 0.02 & 0.02 & 0.02 & 0.02 & 0.03 & 0.03 & 0.03 & 0.03 & 0.02 & 0.02 & 0.02 & 0.03 & 0.02 & 0.02  \\
Deepseek-R1 & 0.02 & 0.05 & 0.02 & 0.06 & 0.02 & 0.06 & 0.02 & 0.06 & 0.02 & 0.08 & 0.02 & 0.07 & 0.02 & 0.06 & 0.02 & 0.08 & 0.02 & 0.07 & 0.01 & 0.06 & 0.01 & 0.08 & 0.01 & 0.07  \\
Qwen & 0.02 & 0.05 & 0.02 & 0.06 & 0.02 & 0.06 & 0.02 & 0.06 & 0.02 & 0.08 & 0.02 & 0.07 & 0.02 & 0.06 & 0.02 & 0.08 & 0.02 & 0.07 & 0.01 & 0.06 & 0.01 & 0.08 & 0.01 & 0.07   \\
WizardCoder & 0.03 & 0.04 & 0.02 & 0.05 & 0.03 & 0.04 & 0.03 & 0.02 & 0.04 & 0.02 & 0.03 & 0.02 & 0.03 & 0.02 & 0.04 & 0.02 & 0.04 & 0.02 & 0.02 & 0.02 & 0.03 & 0.03 & 0.03 & 0.03  \\
\hline
\end{tabular}}
\caption{Performance Evaluation on Variable Name Recovery with Different Architectures and Optimization Levels.}
\label{tbl:varnameresult}
\end{table*}
}

\begin{table*}[t!]
\caption{Performance Evaluation on Variable Name Recovery with Different Architectures and Optimization Levels.}
\label{tbl:varnameresult}
\centering
\footnotesize
\scalebox{0.87}{
\begin{tabular}{lrrrrrrrrrrrr}
\hline
\multicolumn{1}{c}{\multirow{3}{*}{\textbf{Model}}} & \multicolumn{12}{c}{\textbf{Variable Name Recovery Performance}}                                                                                                                                                                                                                                                                                                                                                                                 \\ \cline{2-13} 
\multicolumn{1}{c}{}                                & \multicolumn{3}{c|}{\textbf{O0}}                                                                           & \multicolumn{3}{c|}{\textbf{O1}}                                                                           & \multicolumn{3}{c|}{\textbf{O2}}                                                                           & \multicolumn{3}{c}{\textbf{O3}}                                                                           \\ \cline{2-13} 
\multicolumn{1}{c}{}                                & \multicolumn{1}{c}{\textbf{Prec}} & \multicolumn{1}{c}{\textbf{Recall}} & \multicolumn{1}{c|}{\textbf{F1}} & \multicolumn{1}{c}{\textbf{Prec}} & \multicolumn{1}{c}{\textbf{Recall}} & \multicolumn{1}{c|}{\textbf{F1}} & \multicolumn{1}{c}{\textbf{Prec}} & \multicolumn{1}{c}{\textbf{Recall}} & \multicolumn{1}{c|}{\textbf{F1}} & \multicolumn{1}{c}{\textbf{Prec}} & \multicolumn{1}{c}{\textbf{Recall}} & \multicolumn{1}{c}{\textbf{F1}} \\ \hline
\multicolumn{13}{c}{x64}                                                                                                                                                                                                                                                                                                                                                                                                                                                                               \\ \hline
CodeLlama                                           & 0.0199                            & 0.0312                              & \multicolumn{1}{r|}{0.0243}      & 0.0240                            & 0.0402                              & \multicolumn{1}{r|}{0.0301}      & 0.0290                            & 0.0448                              & \multicolumn{1}{r|}{0.0352}      & 0.0362                            & 0.0517                              & 0.0426                          \\
Llama2                                              & 0.0134                            & 0.0188                              & \multicolumn{1}{r|}{0.0157}      & 0.0161                            & 0.0237                              & \multicolumn{1}{r|}{0.0192}      & 0.0215                            & 0.0318                              & \multicolumn{1}{r|}{0.0256}      & 0.0121                            & 0.0186                              & 0.0147                          \\
WizardCoder                                         & 0.1862                            & 0.2191                              & \multicolumn{1}{r|}{0.2013}      & 0.0336                            & 0.0372                              & \multicolumn{1}{r|}{0.0353}      & 0.0496                            & 0.0558                              & \multicolumn{1}{r|}{0.0526}      & 0.0507                            & 0.0586                              & 0.0543                          \\
Deepseek-R1                                         & 0.1652                            & 0.2411                              & \multicolumn{1}{r|}{0.1961}      & 0.0510                             & 0.0695                              & \multicolumn{1}{r|}{0.0588}      & 0.0481                            & 0.0688                              & \multicolumn{1}{r|}{0.0567}      & 0.0536                            & 0.0756                              & 0.0627                          \\
Qwen2.5                                             & 0.1822                            & 0.2388                              & \multicolumn{1}{r|}{0.2067}      & 0.0414                            & 0.0529                              & \multicolumn{1}{r|}{0.0465}      & 0.0524                            & 0.0698                              & \multicolumn{1}{r|}{0.0599}      & 0.0517                            & 0.0642                              & 0.0572                          \\
Qwen3                                               & 0.0947                            & 0.1167                              & \multicolumn{1}{r|}{0.1046}      & 0.0441                            & 0.0545                              & \multicolumn{1}{r|}{0.0487}      & 0.0488                            & 0.0610                              & \multicolumn{1}{r|}{0.0542}      & 0.0517                            & 0.0622                              & 0.0565                          \\
Phi4 Unsloth                                        & 0.1656                            & 0.2244                              & \multicolumn{1}{r|}{0.1906}      & 0.0507                            & 0.0561                              & \multicolumn{1}{r|}{0.0532}      & 0.0542                            & 0.0716                              & \multicolumn{1}{r|}{0.0617}      & 0.0579                            & 0.0735                              & 0.0648                          \\ \hline
\multicolumn{13}{c}{x86}                                                                                                                                                                                                                                                                                                                                                                                                                                                                               \\ \hline
CodeLlama                                           & 0.0346                            & 0.0433                              & \multicolumn{1}{r|}{0.0385}      & 0.0494                            & 0.0676                              & \multicolumn{1}{r|}{0.0571}      & 0.0491                            & 0.0616                              & \multicolumn{1}{r|}{0.0546}      & 0.0510                            & 0.0625                              & 0.0561                          \\
Llama2                                              & 0.0194                            & 0.0223                              & \multicolumn{1}{r|}{0.0208}      & 0.0331                            & 0.0362                              & \multicolumn{1}{r|}{0.0346}      & 0.0252                            & 0.0276                              & \multicolumn{1}{r|}{0.0264}      & 0.0302                            & 0.0317                              & 0.0310                           \\
WizardCoder                                         & 0.0674                            & 0.0593                              & \multicolumn{1}{r|}{0.0631}      & 0.0708                            & 0.0597                              & \multicolumn{1}{r|}{0.0648}      & 0.0630                            & 0.0509                              & \multicolumn{1}{r|}{0.0563}      & 0.0605                            & 0.0529                              & 0.0564                          \\
Deepseek-R1                                         & 0.0671                            & 0.0866                              & \multicolumn{1}{r|}{0.0756}      & 0.0781                            & 0.1015                              & \multicolumn{1}{r|}{0.0883}      & 0.0744                            & 0.0955                              & \multicolumn{1}{r|}{0.0836}      & 0.0804                            & 0.1040                              & 0.0907                          \\
Qwen2.5                                             & 0.0804                            & 0.0863                              & \multicolumn{1}{r|}{0.0833}      & 0.0813                            & 0.0913                              & \multicolumn{1}{r|}{0.0861}      & 0.0831                            & 0.0966                              & \multicolumn{1}{r|}{0.0894}      & 0.0887                            & 0.0980                              & 0.0932                          \\
Qwen3                                               & 0.0765                            & 0.0872                              & \multicolumn{1}{r|}{0.0815}      & 0.0916                            & 0.1054                              & \multicolumn{1}{r|}{0.0980}      & 0.0893                            & 0.1026                              & \multicolumn{1}{r|}{0.0955}      & 0.0955                            & 0.1089                              & 0.1018                          \\
Phi4 Unsloth                                        & 0.0870                            & 0.1032                              & \multicolumn{1}{r|}{0.0944}      & 0.0922                            & 0.1109                              & \multicolumn{1}{r|}{0.1007}      & 0.0910                            & 0.1096                              & \multicolumn{1}{r|}{0.0994}      & 0.0889                            & 0.1059                              & 0.0966                          \\ \hline
\multicolumn{13}{c}{ARM}                                                                                                                                                                                                                                                                                                                                                                                                                                                                               \\ \hline
CodeLlama                                           & 0.0255                            & 0.0394                              & \multicolumn{1}{r|}{0.0310}      & 0.0261                            & 0.0374                              & \multicolumn{1}{r|}{0.0308}      & 0.0278                            & 0.0405                              & \multicolumn{1}{r|}{0.0329}      & 0.0349                            & 0.0514                              & 0.0416                          \\
Llama2                                              & 0.0138                            & 0.0198                              & \multicolumn{1}{r|}{0.0163}      & 0.0166                            & 0.0208                              & \multicolumn{1}{r|}{0.0185}      & 0.0233                            & 0.0275                              & \multicolumn{1}{r|}{0.0252}      & 0.0142                            & 0.0178                              & 0.0158                          \\
WizardCoder                                         & 0.2105                            & 0.2685                              & \multicolumn{1}{r|}{0.2360}      & 0.1208                            & 0.1366                              & \multicolumn{1}{r|}{0.1282}      & 0.1199                            & 0.1366                              & \multicolumn{1}{r|}{0.1277}      & 0.1327                            & 0.1498                              & 0.1407                          \\
Deepseek-R1                                         & 0.1548                            & 0.2245                              & \multicolumn{1}{r|}{0.1832}      & 0.1052                            & 0.1311                              & \multicolumn{1}{r|}{0.1167}      & 0.1102                            & 0.1419                              & \multicolumn{1}{r|}{0.1240}       & 0.1183                            & 0.1544                              & 0.1339                          \\
Qwen2.5                                             & 0.1836                            & 0.2573                              & \multicolumn{1}{r|}{0.2143}      & 0.1199                            & 0.1544                              & \multicolumn{1}{r|}{0.1350}       & 0.1141                            & 0.1490                              & \multicolumn{1}{r|}{0.1292}      & 0.1247                            & 0.1630                              & 0.1413                          \\
Qwen3                                               & 0.1224                            & 0.1519                              & \multicolumn{1}{r|}{0.1356}      & 0.0882                            & 0.1015                              & \multicolumn{1}{r|}{0.0944}      & 0.0858                            & 0.1018                              & \multicolumn{1}{r|}{0.0931}      & 0.0918                            & 0.1048                              & 0.0979                          \\
Phi4 Unsloth                                        & 0.1817                            & 0.2578                              & \multicolumn{1}{r|}{0.2131}      & 0.1144                            & 0.1459                              & \multicolumn{1}{r|}{0.1283}      & 0.1137                            & 0.1415                              & \multicolumn{1}{r|}{0.1261}      & 0.1263                            & 0.1457                              & 0.1353                          \\ \hline
\multicolumn{13}{c}{MIPS}                                                                                                                                                                                                                                                                                                                                                                                                                                                                              \\ \hline
CodeLlama                                           & 0.0200                            & 0.0232                              & \multicolumn{1}{r|}{0.0214}      & 0.0248                            & 0.0341                              & \multicolumn{1}{r|}{0.0287}      & 0.0158                            & 0.0199                              & \multicolumn{1}{r|}{0.0176}      & 0.0283                            & 0.0330                              & 0.0305                          \\
Llama2                                              & 0.0161                            & 0.0152                              & \multicolumn{1}{r|}{0.0156}      & 0.0180                            & 0.0215                              & \multicolumn{1}{r|}{0.0196}      & 0.0206                            & 0.0220                              & \multicolumn{1}{r|}{0.0213}      & 0.0094                            & 0.0103                              & 0.0098                          \\
WizardCoder                                         & 0.0504                            & 0.0552                              & \multicolumn{1}{r|}{0.0527}      & 0.0473                            & 0.0522                              & \multicolumn{1}{r|}{0.0496}      & 0.0447                            & 0.0499                              & \multicolumn{1}{r|}{0.0472}      & 0.0743                            & 0.0833                              & 0.0785                          \\
Deepseek-R1                                         & 0.0371                            & 0.0455                              & \multicolumn{1}{r|}{0.0409}      & 0.0395                            & 0.0515                              & \multicolumn{1}{r|}{0.0447}      & 0.0412                            & 0.0539                              & \multicolumn{1}{r|}{0.0467}      & 0.0515                            & 0.0616                              & 0.0561                          \\
Qwen2.5                                             & 0.0467                            & 0.0533                              & \multicolumn{1}{r|}{0.0498}      & 0.0412                            & 0.0519                              & \multicolumn{1}{r|}{0.0459}      & 0.0375                            & 0.0464                              & \multicolumn{1}{r|}{0.0415}      & 0.0628                            & 0.0706                              & 0.0665                          \\
Qwen3                                               & 0.0542                            & 0.0602                              & \multicolumn{1}{r|}{0.0571}      & 0.0553                            & 0.0682                              & \multicolumn{1}{r|}{0.0611}      & 0.0491                            & 0.0604                              & \multicolumn{1}{r|}{0.0542}      & 0.0689                            & 0.0759                              & 0.0722                          \\
Phi4 Unsloth                                        & 0.0551                            & 0.0588                              & \multicolumn{1}{r|}{0.0569}      & 0.0608                            & 0.0691                              & \multicolumn{1}{r|}{0.0647}      & 0.0427                            & 0.0511                              & \multicolumn{1}{r|}{0.0465}      & 0.0706                            & 0.0696                              & 0.0701                          \\ \hline
\end{tabular}}
\end{table*}

%% file: table/type_result.tex
\ignore{
\begin{table*}[tb!]
\centering
\footnotesize
\scalebox{0.51}{
\begin{tabular}{lllllll|llllll|llllll|llllll} 
\hline
\multicolumn{1}{c}{\multirow{4}{*}{\textbf{Model }}} & \multicolumn{24}{c}{\textbf{Variable Type Inference Performance }}                                                                                                                                                                                                                                                                                                                                                                                                                                                                                                                                                                                                                                                                                                                                                                                                                                 \\ 
\cline{2-25}
\multicolumn{1}{c}{}                                 & \multicolumn{6}{c|}{\textbf{O0 }}                                                                                                                                                                                      & \multicolumn{6}{c|}{\textbf{O1 }}                                                                                                                                                                                      & \multicolumn{6}{c|}{\textbf{O2 }}                                                                                                                                                                                      & \multicolumn{6}{c}{\textbf{O3 }}                                                                                                                                                                                       \\ 
\cline{2-25}
\multicolumn{1}{c}{}                                 & \multicolumn{2}{c}{\textbf{Prec }}                                    & \multicolumn{2}{c}{\textbf{Recall }}                                  & \multicolumn{2}{c|}{\textbf{F1 }}                                      & \multicolumn{2}{c}{\textbf{Prec }}                                    & \multicolumn{2}{c}{\textbf{Recall }}                                  & \multicolumn{2}{c|}{\textbf{F1 }}                                      & \multicolumn{2}{c}{\textbf{Prec }}                                    & \multicolumn{2}{c}{\textbf{Recall }}                                  & \multicolumn{2}{c|}{\textbf{F1 }}                                      & \multicolumn{2}{c}{\textbf{Prec }}                                    & \multicolumn{2}{c}{\textbf{Recall }}                                  & \multicolumn{2}{c}{\textbf{F1 }}                                       \\
\cline{2-25}
\multicolumn{1}{c}{}                                 & \multicolumn{1}{c}{\textbf{Orig}} & \multicolumn{1}{c}{\textbf{Fine}} & \multicolumn{1}{c}{\textbf{Orig}} & \multicolumn{1}{c}{\textbf{Fine}} & \multicolumn{1}{c}{\textbf{Orig}} & \multicolumn{1}{c|}{\textbf{Fine}} & \multicolumn{1}{c}{\textbf{Orig}} & \multicolumn{1}{c}{\textbf{Fine}} & \multicolumn{1}{c}{\textbf{Orig}} & \multicolumn{1}{c}{\textbf{Fine}} & \multicolumn{1}{c}{\textbf{Orig}} & \multicolumn{1}{c|}{\textbf{Fine}} & \multicolumn{1}{c}{\textbf{Orig}} & \multicolumn{1}{c}{\textbf{Fine}} & \multicolumn{1}{c}{\textbf{Orig}} & \multicolumn{1}{c}{\textbf{Fine}} & \multicolumn{1}{c}{\textbf{Orig}} & \multicolumn{1}{c|}{\textbf{Fine}} & \multicolumn{1}{c}{\textbf{Orig}} & \multicolumn{1}{c}{\textbf{Fine}} & \multicolumn{1}{c}{\textbf{Orig}} & \multicolumn{1}{c}{\textbf{Fine}} & \multicolumn{1}{c}{\textbf{Orig}} & \multicolumn{1}{c}{\textbf{Fine}}  \\ 
\hline
\multicolumn{25}{c}{\textbf{x64 }}                                                                                                                                                                                                                                                                                                                                                                                                                                                                                                                                                                                                                                                                                                                                                                                                                                                                                                                               \\ 
\hline
CodeLlama     &0.03 & 0.06 & 0.02 & 0.03 & 0.02 & 0.04 & 0.02 & 0.09 & 0.01 & 0.03 & 0.01 & 0.04 & 0.02 & 0.07 & 0.01 & 0.02 & 0.02 & 0.04 & 0.02 & 0.06 & 0.01 & 0.02 & 0.02 & 0.03 \\
Llama2        &  0.04 & 0.09 & 0.04 & 0.05 & 0.04 & 0.07 & 0.02 & 0.07 & 0.01 & 0.03 & 0.02 & 0.04 & 0.02 & 0.06 & 0.02 & 0.03 & 0.02 & 0.04 & 0.03 & 0.07 & 0.02 & 0.03 & 0.02 & 0.04  \\
Deepseek-V2   &  0.02 & 0.02 & 0.02 & 0.01 & 0.02 & 0.01 & 0.01 & 0.01 & 0.01 & 0.00 & 0.01 & 0.01 & 0.02 & 0.02 & 0.01 & 0.01 & 0.02 & 0.01 & 0.02 & 0.01 & 0.01 & 0.01 & 0.02 & 0.01  \\
Deepseek-R1 & 0.03 & 0.03 & 0.02 & 0.02 & 0.02 & 0.02 & 0.02 & 0.05 & 0.01 & 0.02 & 0.01 & 0.03 & 0.02 & 0.04 & 0.01 & 0.02 & 0.01 & 0.03 & 0.02 & 0.02 & 0.01 & 0.01 & 0.01 & 0.02    \\
Qwen    & 0.01 & 0.03 & 0.01 & 0.02 & 0.01 & 0.03 & 0.01 & 0.03 & 0.01 & 0.01 & 0.01 & 0.02 & 0.01 & 0.03 & 0.00 & 0.02 & 0.00 & 0.02 & 0.01 & 0.02 & 0.01 & 0.01 & 0.01 & 0.01      \\
WizardCoder  & 0.02 & 0.03 & 0.01 & 0.02 & 0.02 & 0.03 & 0.01 & 0.05 & 0.00 & 0.03 & 0.00 & 0.03 & 0.01 & 0.04 & 0.00 & 0.03 & 0.01 & 0.03 & 0.02 & 0.03 & 0.01 & 0.02 & 0.01 & 0.02                \\ 
\hline
\multicolumn{25}{c}{\bf x86}                                                                                                                                                                                                                                                                                                                                                                                                                                                                                                                                                                                                                                                                                                                                                                                                                                                                                                                                 \\ 
\hline
CodeLlama    & 0.03 & 0.09 & 0.02 & 0.03 & 0.03 & 0.05 & 0.03 & 0.07 & 0.02 & 0.04 & 0.02 & 0.05 & 0.03 & 0.07 & 0.02 & 0.04 & 0.03 & 0.05 & 0.02 & 0.06 & 0.02 & 0.03 & 0.02 & 0.04  \\
Llama2    & 0.05 & 0.09 & 0.04 & 0.05 & 0.04 & 0.06 & 0.03 & 0.07 & 0.03 & 0.04 & 0.03 & 0.05 & 0.04 & 0.08 & 0.03 & 0.05 & 0.04 & 0.06 & 0.03 & 0.07 & 0.02 & 0.04 & 0.03 & 0.05   \\
Deepseek-V2   & 0.03 & 0.02 & 0.02 & 0.01 & 0.02 & 0.01 & 0.03 & 0.02 & 0.03 & 0.01 & 0.03 & 0.01 & 0.03 & 0.02 & 0.03 & 0.02 & 0.03 & 0.02 & 0.03 & 0.02 & 0.03 & 0.01 & 0.03 & 0.01  \\
Deepseek-R1  & 0.03 & 0.05 & 0.01 & 0.03 & 0.01 & 0.04 & 0.02 & 0.05 & 0.01 & 0.04 & 0.01 & 0.04 & 0.03 & 0.06 & 0.01 & 0.04 & 0.02 & 0.05 & 0.03 & 0.06 & 0.01 & 0.04 & 0.01 & 0.04  \\
Qwen  & 0.02 & 0.04 & 0.01 & 0.03 & 0.01 & 0.03 & 0.02 & 0.04 & 0.01 & 0.03 & 0.01 & 0.04 & 0.02 & 0.05 & 0.01 & 0.04 & 0.01 & 0.04 & 0.02 & 0.05 & 0.01 & 0.04 & 0.01 & 0.04  \\
WizardCoder & 0.02 & 0.05 & 0.01 & 0.03 & 0.01 & 0.04 & 0.02 & 0.05 & 0.01 & 0.04 & 0.01 & 0.04 & 0.02 & 0.05 & 0.01 & 0.04 & 0.02 & 0.04 & 0.02 & 0.05 & 0.01 & 0.03 & 0.01 & 0.04   \\ 
\hline
\multicolumn{25}{c}{\bf ARM}                                                                                                                                                                                                                                                                                                                                                                                                                                                                                                                                                                                                                                                                                                                                                                                                                                                                                                                                 \\ 
\hline
CodeLlama  & 0.04 & 0.07 & 0.03 & 0.03 & 0.03 & 0.04 & 0.04 & 0.09 & 0.03 & 0.03 & 0.03 & 0.05 & 0.05 & 0.10 & 0.03 & 0.04 & 0.04 & 0.05 & 0.05 & 0.08 & 0.03 & 0.03 & 0.04 & 0.05   \\
Llama2    & 0.05 & 0.09 & 0.04 & 0.05 & 0.04 & 0.06 & 0.03 & 0.07 & 0.03 & 0.04 & 0.03 & 0.05 & 0.04 & 0.08 & 0.03 & 0.05 & 0.04 & 0.06 & 0.03 & 0.07 & 0.02 & 0.04 & 0.03 & 0.05   \\
Deepseek-V2   & 0.03 & 0.02 & 0.02 & 0.01 & 0.02 & 0.01 & 0.03 & 0.02 & 0.03 & 0.01 & 0.03 & 0.01 & 0.03 & 0.02 & 0.03 & 0.02 & 0.03 & 0.02 & 0.03 & 0.02 & 0.03 & 0.01 & 0.03 & 0.01   \\
Deepseek-R1  & 0.03 & 0.05 & 0.01 & 0.03 & 0.01 & 0.04 & 0.02 & 0.05 & 0.01 & 0.04 & 0.01 & 0.04 & 0.03 & 0.06 & 0.01 & 0.04 & 0.02 & 0.05 & 0.03 & 0.06 & 0.01 & 0.04 & 0.01 & 0.04   \\
Qwen     & 0.02 & 0.04 & 0.01 & 0.03 & 0.01 & 0.03 & 0.02 & 0.04 & 0.01 & 0.03 & 0.01 & 0.04 & 0.02 & 0.05 & 0.01 & 0.04 & 0.01 & 0.04 & 0.02 & 0.05 & 0.01 & 0.04 & 0.01 & 0.04  \\
WizardCoder  & 0.02 & 0.05 & 0.01 & 0.03 & 0.01 & 0.04 & 0.02 & 0.04 & 0.01 & 0.03 & 0.01 & 0.04 & 0.02 & 0.05 & 0.01 & 0.03 & 0.01 & 0.04 & 0.02 & 0.04 & 0.01 & 0.02 & 0.01 & 0.03  \\ 
\hline
\multicolumn{25}{c}{\bf MIPS}                                                                                                                                                                                                                                                                                                                                                                                                                                                                                                                                                                                                                                                                                                                                                                                                                                                                                                                                \\ 
\hline
CodeLlama    & 0.04 & 0.07 & 0.03 & 0.03 & 0.03 & 0.04 & 0.04 & 0.06 & 0.02 & 0.02 & 0.02 & 0.03 & 0.04 & 0.07 & 0.02 & 0.02 & 0.03 & 0.04 & 0.04 & 0.05 & 0.02 & 0.02 & 0.03 & 0.02  \\
Llama2   & 0.08 & 0.10 & 0.05 & 0.05 & 0.06 & 0.07 & 0.06 & 0.11 & 0.04 & 0.05 & 0.05 & 0.07 & 0.07 & 0.11 & 0.04 & 0.05 & 0.06 & 0.07 & 0.06 & 0.09 & 0.03 & 0.04 & 0.04 & 0.06  \\
Deepseek-V2  & 0.04 & 0.03 & 0.04 & 0.02 & 0.04 & 0.02 & 0.04 & 0.02 & 0.03 & 0.01 & 0.03 & 0.02 & 0.03 & 0.02 & 0.02 & 0.01 & 0.03 & 0.01 & 0.03 & 0.03 & 0.02 & 0.01 & 0.03 & 0.02  \\
Deepseek-R1  & 0.05 & 0.03 & 0.02 & 0.02 & 0.03 & 0.02 & 0.06 & 0.03 & 0.02 & 0.02 & 0.03 & 0.02 & 0.05 & 0.04 & 0.02 & 0.03 & 0.02 & 0.03 & 0.05 & 0.02 & 0.02 & 0.01 & 0.03 & 0.02 \\
Qwen   & 0.03 & 0.05 & 0.02 & 0.03 & 0.02 & 0.04 & 0.04 & 0.04 & 0.02 & 0.02 & 0.03 & 0.03 & 0.03 & 0.03 & 0.01 & 0.02 & 0.02 & 0.02 & 0.04 & 0.03 & 0.02 & 0.02 & 0.02 & 0.02  \\
WizardCoder & 0.04 & 0.05 & 0.02 & 0.03 & 0.02 & 0.04 & 0.03 & 0.05 & 0.01 & 0.03 & 0.02 & 0.04 & 0.03 & 0.05 & 0.01 & 0.03 & 0.02 & 0.03 & 0.04 & 0.04 & 0.01 & 0.02 & 0.02 & 0.03 \\
\hline
\end{tabular}}
\caption{Performance Evaluation on Type Inference with Different Architectures and Optimization Levels.}
\label{tbl:typeresult}
\end{table*}}

\begin{table*}[t!]
\caption{Performance Evaluation on Type Inference with Different Architectures and Optimization Levels.}
\label{tbl:typeresult}
\centering
\footnotesize
\scalebox{0.87}{
\begin{tabular}{lrrrrrrrrrrrr}
\hline
\multicolumn{1}{c}{\multirow{3}{*}{\textbf{Model}}} & \multicolumn{12}{c}{\textbf{Type Inference Performance}}                                                                                                                                                                                                                                                                                                                                                                                 \\ \cline{2-13} 
\multicolumn{1}{c}{}                                & \multicolumn{3}{c|}{\textbf{O0}}                                                                           & \multicolumn{3}{c|}{\textbf{O1}}                                                                           & \multicolumn{3}{c|}{\textbf{O2}}                                                                           & \multicolumn{3}{c}{\textbf{O3}}                                                                           \\ \cline{2-13} 
\multicolumn{1}{c}{}                                & \multicolumn{1}{c}{\textbf{Prec}} & \multicolumn{1}{c}{\textbf{Recall}} & \multicolumn{1}{c|}{\textbf{F1}} & \multicolumn{1}{c}{\textbf{Prec}} & \multicolumn{1}{c}{\textbf{Recall}} & \multicolumn{1}{c|}{\textbf{F1}} & \multicolumn{1}{c}{\textbf{Prec}} & \multicolumn{1}{c}{\textbf{Recall}} & \multicolumn{1}{c|}{\textbf{F1}} & \multicolumn{1}{c}{\textbf{Prec}} & \multicolumn{1}{c}{\textbf{Recall}} & \multicolumn{1}{c}{\textbf{F1}} \\ \hline
\multicolumn{13}{c}{x64}                                                                                                                                                                                                                                                                                                                                                                                                                                                                               \\ \hline
CodeLlama                                           & 0.0402                            & 0.0063                              & \multicolumn{1}{r|}{0.0109}      & 0.0392                            & 0.0072                              & \multicolumn{1}{r|}{0.0121}      & 0.0201                            & 0.0028                              & \multicolumn{1}{r|}{0.0049}      & 0.0390                            & 0.0045                              & 0.0081                          \\
Llama2                                              & 0.0214                            & 0.0127                              & \multicolumn{1}{r|}{0.0159}      & 0.0204                            & 0.0112                              & \multicolumn{1}{r|}{0.0145}      & 0.0150                            & 0.0070                              & \multicolumn{1}{r|}{0.0095}      & 0.0139                            & 0.0058                              & 0.0082                          \\
WizardCoder                                         & 0.0576                            & 0.0144                              & \multicolumn{1}{r|}{0.0230}      & 0.0457                            & 0.0095                              & \multicolumn{1}{r|}{0.0157}      & 0.0391                            & 0.0081                              & \multicolumn{1}{r|}{0.0134}      & 0.0402                            & 0.0077                              & 0.0129                          \\
Deepseek-R1                                         & 0.0347                            & 0.0331                              & \multicolumn{1}{r|}{0.0339}      & 0.0369                            & 0.0320                              & \multicolumn{1}{r|}{0.0343}      & 0.0394                            & 0.0335                              & \multicolumn{1}{r|}{0.0362}      & 0.0228                            & 0.0156                              & 0.0185                          \\
Qwen2.5                                             & 0.0198                            & 0.0114                              & \multicolumn{1}{r|}{0.0145}      & 0.0257                            & 0.0130                              & \multicolumn{1}{r|}{0.0173}      & 0.0166                            & 0.0073                              & \multicolumn{1}{r|}{0.0101}      & 0.0217                            & 0.0082                              & 0.0119                          \\
Qwen3                                               & 0.0327                            & 0.0213                              & \multicolumn{1}{r|}{0.0258}      & 0.0443                            & 0.0284                              & \multicolumn{1}{r|}{0.0346}      & 0.0367                            & 0.0223                              & \multicolumn{1}{r|}{0.0278}      & 0.0400                            & 0.0214                              & 0.0279                          \\
Phi4 Unsloth                                        & 0.0474                            & 0.0310                              & \multicolumn{1}{r|}{0.0375}      & 0.0409                            & 0.0256                              & \multicolumn{1}{r|}{0.0315}      & 0.0310                            & 0.0181                              & \multicolumn{1}{r|}{0.0229}      & 0.0370                            & 0.0185                              & 0.0246                          \\ \hline
\multicolumn{13}{c}{x32}                                                                                                                                                                                                                                                                                                                                                                                                                                                                               \\ \hline
CodeLlama                                           & 0.0644                            & 0.0079                              & \multicolumn{1}{r|}{0.0141}      & 0.0611                            & 0.0134                              & \multicolumn{1}{r|}{0.0220}      & 0.0419                            & 0.0091                              & \multicolumn{1}{r|}{0.0149}      & 0.0462                            & 0.0086                              & 0.0145                          \\
Llama2                                              & 0.0312                            & 0.0150                              & \multicolumn{1}{r|}{0.0202}      & 0.0295                            & 0.0226                              & \multicolumn{1}{r|}{0.0256}      & 0.0294                            & 0.0222                              & \multicolumn{1}{r|}{0.0253}      & 0.0337                            & 0.0229                              & 0.0273                          \\
WizardCoder                                         & 0.0813                            & 0.0227                              & \multicolumn{1}{r|}{0.0355}      & 0.0604                            & 0.0176                              & \multicolumn{1}{r|}{0.0273}      & 0.0629                            & 0.0199                              & \multicolumn{1}{r|}{0.0302}      & 0.0602                            & 0.0161                              & 0.0254                          \\
Deepseek-R1                                         & 0.0435                            & 0.0348                              & \multicolumn{1}{r|}{0.0387}      & 0.0594                            & 0.0498                              & \multicolumn{1}{r|}{0.0542}      & 0.0467                            & 0.0425                              & \multicolumn{1}{r|}{0.0445}      & 0.0486                            & 0.0442                              & 0.0463                          \\
Qwen2.5                                             & 0.0363                            & 0.0194                              & \multicolumn{1}{r|}{0.0253}      & 0.0352                            & 0.0219                              & \multicolumn{1}{r|}{0.0270}      & 0.0382                            & 0.0252                              & \multicolumn{1}{r|}{0.0304}      & 0.0359                            & 0.0210                              & 0.0265                          \\
Qwen3                                               & 0.0565                            & 0.0355                              & \multicolumn{1}{r|}{0.0436}      & 0.0477                            & 0.0389                              & \multicolumn{1}{r|}{0.0429}      & 0.0464                            & 0.0409                              & \multicolumn{1}{r|}{0.0435}      & 0.0443                            & 0.0330                              & 0.0378                          \\
Phi4 Unsloth                                        & 0.0647                            & 0.0462                              & \multicolumn{1}{r|}{0.0539}      & 0.0559                            & 0.0412                              & \multicolumn{1}{r|}{0.0475}      & 0.0620                            & 0.0473                              & \multicolumn{1}{r|}{0.0537}      & 0.0575                            & 0.0384                              & 0.0461                          \\ \hline
\multicolumn{13}{c}{ARM}                                                                                                                                                                                                                                                                                                                                                                                                                                                                               \\ \hline
CodeLlama                                           & 0.0346                            & 0.0053                              & \multicolumn{1}{r|}{0.0091}      & 0.0437                            & 0.0069                              & \multicolumn{1}{r|}{0.0120}      & 0.0471                            & 0.0052                              & \multicolumn{1}{r|}{0.0094}      & 0.0237                            & 0.0030                              & 0.0054                          \\
Llama2                                              & 0.0367                            & 0.0255                              & \multicolumn{1}{r|}{0.0301}      & 0.0581                            & 0.0290                              & \multicolumn{1}{r|}{0.0387}      & 0.0653                            & 0.0367                              & \multicolumn{1}{r|}{0.0470}      & 0.0362                            & 0.0158                              & 0.0220                          \\
WizardCoder                                         & 0.0848                            & 0.0206                              & \multicolumn{1}{r|}{0.0332}      & 0.0855                            & 0.0196                              & \multicolumn{1}{r|}{0.0319}      & 0.0697                            & 0.0173                              & \multicolumn{1}{r|}{0.0277}      & 0.0710                            & 0.0123                              & 0.0210                          \\
Deepseek-R1                                         & 0.0535                            & 0.0500                              & \multicolumn{1}{r|}{0.0517}      & 0.0563                            & 0.0407                              & \multicolumn{1}{r|}{0.0472}      & 0.0517                            & 0.0412                              & \multicolumn{1}{r|}{0.0458}      & 0.0389                            & 0.0244                              & 0.0300                          \\
Qwen2.5                                             & 0.0248                            & 0.0150                               & \multicolumn{1}{r|}{0.0187}      & 0.0335                            & 0.0184                              & \multicolumn{1}{r|}{0.0237}      & 0.0471                            & 0.0256                              & \multicolumn{1}{r|}{0.0331}      & 0.0303                            & 0.0125                              & 0.0177                          \\
Qwen3                                               & 0.0264                            & 0.0188                              & \multicolumn{1}{r|}{0.0219}      & 0.0513                            & 0.0360                              & \multicolumn{1}{r|}{0.0423}      & 0.0482                            & 0.0329                              & \multicolumn{1}{r|}{0.0391}      & 0.0400                            & 0.0197                              & 0.0264                          \\
Phi4 Unsloth                                        & 0.0509                            & 0.0358                              & \multicolumn{1}{r|}{0.0420}      & 0.0646                            & 0.0375                              & \multicolumn{1}{r|}{0.0474}      & 0.0650                            & 0.0407                              & \multicolumn{1}{r|}{0.0501}      & 0.0664                            & 0.0318                              & 0.0430                          \\ \hline
\multicolumn{13}{c}{MIPS}                                                                                                                                                                                                                                                                                                                                                                                                                                                                              \\ \hline
CodeLlama                                           & 0.0406                            & 0.0062                              & \multicolumn{1}{r|}{0.0108}      & 0.0431                            & 0.0043                              & \multicolumn{1}{r|}{0.0079}      & 0.0266                            & 0.0023                              & \multicolumn{1}{r|}{0.0043}      & 0.0491                            & 0.0034                              & 0.0063                          \\
Llama2                                              & 0.0464                            & 0.0290                              & \multicolumn{1}{r|}{0.0357}      & 0.0209                            & 0.0068                              & \multicolumn{1}{r|}{0.0102}      & 0.0272                            & 0.0087                              & \multicolumn{1}{r|}{0.0132}      & 0.0219                            & 0.0067                              & 0.0103                          \\
WizardCoder                                         & 0.1068                            & 0.0281                              & \multicolumn{1}{r|}{0.0445}      & 0.0582                            & 0.0087                              & \multicolumn{1}{r|}{0.0151}      & 0.0579                            & 0.0080                              & \multicolumn{1}{r|}{0.0140}      & 0.0885                            & 0.0122                              & 0.0215                          \\
Deepseek-R1                                         & 0.0205                            & 0.0202                              & \multicolumn{1}{r|}{0.0203}      & 0.0257                            & 0.0162                              & \multicolumn{1}{r|}{0.0199}      & 0.0144                            & 0.0082                              & \multicolumn{1}{r|}{0.0105}      & 0.0159                            & 0.0058                              & 0.0085                          \\
Qwen2.5                                             & 0.0228                            & 0.0128                              & \multicolumn{1}{r|}{0.0164}      & 0.0306                            & 0.0119                              & \multicolumn{1}{r|}{0.0171}      & 0.0592                            & 0.0268                              & \multicolumn{1}{r|}{0.0369}      & 0.0298                            & 0.0083                              & 0.0129                          \\
Qwen3                                               & 0.0272                            & 0.0186                              & \multicolumn{1}{r|}{0.0221}      & 0.0463                            & 0.0211                              & \multicolumn{1}{r|}{0.0290}      & 0.0398                            & 0.0188                              & \multicolumn{1}{r|}{0.0255}      & 0.0364                            & 0.0125                              & 0.0186                          \\
Phi4 Unsloth                                        & 0.0396                            & 0.0278                              & \multicolumn{1}{r|}{0.0327}      & 0.0795                            & 0.0468                              & \multicolumn{1}{r|}{0.0589}      & 0.0396                            & 0.0216                              & \multicolumn{1}{r|}{0.0280}      & 0.0287                            & 0.0122                              & 0.0171                          \\ \hline
\end{tabular}}
\end{table*}

%% file: table/asm_result.tex
\ignore{
\begin{table*}[tb!]
\centering
\footnotesize
\scalebox{0.51}{
\begin{tabular}{lllllll|llllll|llllll|llllll} 
\hline
\multicolumn{1}{c}{\multirow{4}{*}{\textbf{Model }}} & \multicolumn{24}{c}{\textbf{Function Name Recovery Performance }}                                                                                                                                                                                                                                                                                                                                                                                                                                                                                                                                                                                                                                                                                                                                                                                                                                 \\ 
\cline{2-25}
\multicolumn{1}{c}{}                                 & \multicolumn{6}{c|}{\textbf{O0 }}                                                                                                                                                                                      & \multicolumn{6}{c|}{\textbf{O1 }}                                                                                                                                                                                      & \multicolumn{6}{c|}{\textbf{O2 }}                                                                                                                                                                                      & \multicolumn{6}{c}{\textbf{O3 }}                                                                                                                                                                                       \\ 
\cline{2-25}
\multicolumn{1}{c}{}                                 & \multicolumn{2}{c}{\textbf{Prec }}                                    & \multicolumn{2}{c}{\textbf{Recall }}                                  & \multicolumn{2}{c|}{\textbf{F1 }}                                      & \multicolumn{2}{c}{\textbf{Prec }}                                    & \multicolumn{2}{c}{\textbf{Recall }}                                  & \multicolumn{2}{c|}{\textbf{F1 }}                                      & \multicolumn{2}{c}{\textbf{Prec }}                                    & \multicolumn{2}{c}{\textbf{Recall }}                                  & \multicolumn{2}{c|}{\textbf{F1 }}                                      & \multicolumn{2}{c}{\textbf{Prec }}                                    & \multicolumn{2}{c}{\textbf{Recall }}                                  & \multicolumn{2}{c}{\textbf{F1 }}                                       \\
\cline{2-25}
\multicolumn{1}{c}{}                                 & \multicolumn{1}{c}{\textbf{Orig}} & \multicolumn{1}{c}{\textbf{Fine}} & \multicolumn{1}{c}{\textbf{Orig}} & \multicolumn{1}{c}{\textbf{Fine}} & \multicolumn{1}{c}{\textbf{Orig}} & \multicolumn{1}{c|}{\textbf{Fine}} & \multicolumn{1}{c}{\textbf{Orig}} & \multicolumn{1}{c}{\textbf{Fine}} & \multicolumn{1}{c}{\textbf{Orig}} & \multicolumn{1}{c}{\textbf{Fine}} & \multicolumn{1}{c}{\textbf{Orig}} & \multicolumn{1}{c|}{\textbf{Fine}} & \multicolumn{1}{c}{\textbf{Orig}} & \multicolumn{1}{c}{\textbf{Fine}} & \multicolumn{1}{c}{\textbf{Orig}} & \multicolumn{1}{c}{\textbf{Fine}} & \multicolumn{1}{c}{\textbf{Orig}} & \multicolumn{1}{c|}{\textbf{Fine}} & \multicolumn{1}{c}{\textbf{Orig}} & \multicolumn{1}{c}{\textbf{Fine}} & \multicolumn{1}{c}{\textbf{Orig}} & \multicolumn{1}{c}{\textbf{Fine}} & \multicolumn{1}{c}{\textbf{Orig}} & \multicolumn{1}{c}{\textbf{Fine}}  \\ 
\hline
\multicolumn{25}{c}{\textbf{x64 }}               \\ 
\hline
CodeLlama    & 0.02 & 0.13 & 0.04 & 0.09 & 0.02 & 0.10 & 0.02 & 0.15 & 0.04 & 0.10 & 0.03 & 0.12 & 0.02 & 0.13 & 0.04 & 0.09 & 0.02 & 0.11 & 0.02 & 0.13 & 0.05 & 0.10 & 0.02 & 0.11  \\
Llama2   & 0.01 & 0.12 & 0.03 & 0.09 & 0.01 & 0.11 & 0.01 & 0.12 & 0.03 & 0.10 & 0.01 & 0.11 & 0.01 & 0.11 & 0.03 & 0.10 & 0.01 & 0.10 & 0.01 & 0.11 & 0.02 & 0.10 & 0.01 & 0.10  \\
Deepseek-V2 & 0.02 & 0.03 & 0.03 & 0.04 & 0.02 & 0.03 & 0.02 & 0.05 & 0.03 & 0.05 & 0.02 & 0.05 & 0.02 & 0.05 & 0.02 & 0.06 & 0.02 & 0.05 & 0.02 & 0.04 & 0.03 & 0.05 & 0.02 & 0.05  \\
Deepseek-R1 & 0.01 & 0.06 & 0.03 & 0.10 & 0.01 & 0.07 & 0.01 & 0.05 & 0.04 & 0.09 & 0.02 & 0.06 & 0.01 & 0.04 & 0.04 & 0.07 & 0.02 & 0.05 & 0.01 & 0.05 & 0.03 & 0.08 & 0.02 & 0.06  \\
Qwen   & 0.02 & 0.13 & 0.03 & 0.09 & 0.03 & 0.11 & 0.03 & 0.14 & 0.04 & 0.10 & 0.03 & 0.12 & 0.02 & 0.13 & 0.03 & 0.10 & 0.03 & 0.11 & 0.02 & 0.13 & 0.03 & 0.10 & 0.03 & 0.11 \\
WizardCoder & 0.00 & 0.11 & 0.01 & 0.09 & 0.00 & 0.10 & 0.00 & 0.14 & 0.02 & 0.11 & 0.01 & 0.12 & 0.00 & 0.13 & 0.02 & 0.10 & 0.01 & 0.12 & 0.00 & 0.13 & 0.01 & 0.10 & 0.01 & 0.12   \\ 
\hline
\multicolumn{25}{c}{\bf x86}                                                                                                                                                                                                                                                                                                                                                                                                                                                                                                                                                                                                                                                                                                                                                                                                                                                                                                                                 \\ 
\hline
CodeLlama  & 0.02 & 0.12 & 0.04 & 0.09 & 0.02 & 0.10 & 0.02 & 0.11 & 0.04 & 0.08 & 0.02 & 0.10 & 0.01 & 0.12 & 0.03 & 0.08 & 0.02 & 0.10 & 0.02 & 0.13 & 0.04 & 0.09 & 0.02 & 0.11  \\
Llama2   & 0.01 & 0.10 & 0.02 & 0.09 & 0.01 & 0.10 & 0.01 & 0.08 & 0.03 & 0.08 & 0.01 & 0.08 & 0.01 & 0.08 & 0.03 & 0.08 & 0.02 & 0.08 & 0.01 & 0.08 & 0.02 & 0.08 & 0.01 & 0.08  \\
Deepseek-V2  & 0.01 & 0.03 & 0.02 & 0.04 & 0.02 & 0.04 & 0.02 & 0.05 & 0.03 & 0.05 & 0.02 & 0.05 & 0.01 & 0.05 & 0.02 & 0.05 & 0.02 & 0.05 & 0.02 & 0.04 & 0.02 & 0.05 & 0.02 & 0.05 \\
Deepseek-R1   & 0.01 & 0.06 & 0.04 & 0.09 & 0.02 & 0.07 & 0.01 & 0.06 & 0.03 & 0.09 & 0.02 & 0.07 & 0.01 & 0.05 & 0.03 & 0.08 & 0.02 & 0.06 & 0.01 & 0.05 & 0.03 & 0.09 & 0.02 & 0.07  \\
Qwen  & 0.02 & 0.11 & 0.03 & 0.09 & 0.02 & 0.10 & 0.03 & 0.11 & 0.04 & 0.09 & 0.03 & 0.10 & 0.02 & 0.10 & 0.03 & 0.08 & 0.03 & 0.09 & 0.03 & 0.10 & 0.04 & 0.08 & 0.03 & 0.09   \\
WizardCoder & 0.00 & 0.12 & 0.02 & 0.10 & 0.01 & 0.10 & 0.00 & 0.12 & 0.02 & 0.10 & 0.01 & 0.11 & 0.00 & 0.12 & 0.02 & 0.10 & 0.01 & 0.11 & 0.00 & 0.13 & 0.01 & 0.11 & 0.00 & 0.12   \\ 
\hline
\multicolumn{25}{c}{\bf ARM}                                                                                                                                                                                                                                                                                                                                                                                                                                                                                                                                                                                                                                                                                                                                                                                                                                                                                                                                 \\ 
\hline
CodeLlama & 0.02 & 0.10 & 0.03 & 0.10 & 0.02 & 0.10 & 0.02 & 0.08 & 0.03 & 0.10 & 0.02 & 0.09 & 0.01 & 0.07 & 0.02 & 0.09 & 0.01 & 0.08 & 0.01 & 0.09 & 0.02 & 0.12 & 0.02 & 0.10  \\
Llama2   & 0.02 & 0.08 & 0.05 & 0.07 & 0.03 & 0.07 & 0.02 & 0.08 & 0.03 & 0.07 & 0.02 & 0.08 & 0.02 & 0.10 & 0.03 & 0.08 & 0.02 & 0.09 & 0.02 & 0.12 & 0.04 & 0.10 & 0.02 & 0.11   \\
Deepseek-V2 & 0.01 & 0.05 & 0.02 & 0.05 & 0.01 & 0.05 & 0.02 & 0.04 & 0.02 & 0.04 & 0.02 & 0.04 & 0.01 & 0.03 & 0.02 & 0.04 & 0.01 & 0.03 & 0.01 & 0.04 & 0.02 & 0.04 & 0.02 & 0.04 \\
Deepseek-R1  & 0.02 & 0.05 & 0.04 & 0.08 & 0.02 & 0.07 & 0.01 & 0.05 & 0.04 & 0.08 & 0.02 & 0.07 & 0.01 & 0.05 & 0.03 & 0.07 & 0.02 & 0.06 & 0.01 & 0.05 & 0.03 & 0.07 & 0.02 & 0.06 \\
Qwen  & 0.02 & 0.10 & 0.03 & 0.09 & 0.03 & 0.09 & 0.03 & 0.10 & 0.03 & 0.09 & 0.03 & 0.10 & 0.02 & 0.10 & 0.02 & 0.09 & 0.02 & 0.10 & 0.02 & 0.11 & 0.03 & 0.10 & 0.02 & 0.10  \\
WizardCoder & 0.01 & 0.07 & 0.02 & 0.08 & 0.01 & 0.08 & 0.01 & 0.09 & 0.02 & 0.11 & 0.01 & 0.10 & 0.01 & 0.09 & 0.02 & 0.10 & 0.01 & 0.09 & 0.01 & 0.11 & 0.02 & 0.12 & 0.01 & 0.11   \\ 
\hline
\multicolumn{25}{c}{\bf MIPS}                                                                                                                                                                                                                                                                                                                                                                                                                                                                                                                                                                                                                                                                                                                                                                                                                                                                                                                                \\ 
\hline
CodeLlama & 0.02 & 0.05 & 0.03 & 0.07 & 0.02 & 0.06 & 0.01 & 0.04 & 0.02 & 0.06 & 0.01 & 0.05 & 0.01 & 0.04 & 0.02 & 0.06 & 0.01 & 0.05 & 0.01 & 0.04 & 0.02 & 0.06 & 0.01 & 0.05  \\
Llama2  & 0.01 & 0.03 & 0.03 & 0.03 & 0.02 & 0.03 & 0.02 & 0.04 & 0.04 & 0.04 & 0.02 & 0.04 & 0.01 & 0.04 & 0.03 & 0.03 & 0.02 & 0.03 & 0.02 & 0.04 & 0.04 & 0.04 & 0.02 & 0.04 \\
Deepseek-V2  & 0.01 & 0.02 & 0.02 & 0.03 & 0.01 & 0.03 & 0.01 & 0.03 & 0.01 & 0.04 & 0.01 & 0.03 & 0.01 & 0.02 & 0.02 & 0.03 & 0.01 & 0.03 & 0.01 & 0.03 & 0.01 & 0.04 & 0.01 & 0.03  \\
Deepseek-R1 & 0.01 & 0.04 & 0.03 & 0.05 & 0.02 & 0.05 & 0.01 & 0.04 & 0.03 & 0.04 & 0.01 & 0.04 & 0.01 & 0.04 & 0.02 & 0.04 & 0.01 & 0.04 & 0.01 & 0.04 & 0.02 & 0.04 & 0.01 & 0.04 \\
Qwen  & 0.02 & 0.07 & 0.03 & 0.07 & 0.03 & 0.07 & 0.02 & 0.08 & 0.03 & 0.08 & 0.02 & 0.08 & 0.02 & 0.07 & 0.02 & 0.08 & 0.02 & 0.08 & 0.02 & 0.07 & 0.02 & 0.07 & 0.02 & 0.07  \\
WizardCoder & 0.01 & 0.06 & 0.02 & 0.06 & 0.01 & 0.06 & 0.00 & 0.07 & 0.01 & 0.07 & 0.00 & 0.07 & 0.00 & 0.07 & 0.01 & 0.07 & 0.00 & 0.07 & 0.00 & 0.08 & 0.01 & 0.07 & 0.00 & 0.08 \\
\hline
\end{tabular}}
\caption{Performance Evaluation on Function Name Recovery with Different Architectures and Optimization Levels via Assembly Code.}
\label{tbl:asmresult}
\end{table*}
}

\ignore{
\begin{table}
\centering
\footnotesize
\scalebox{0.85}{
\begin{tabular}{llllllllll} 
\hline
\multicolumn{1}{c}{\multirow{3}{*}{\textbf{Arch}}} & \multicolumn{1}{c}{\multirow{3}{*}{\textbf{Model }}} & \multicolumn{8}{c}{\textbf{Function Name Recovery Performance }}                                                                                                                                                                                                                               \\ 
\cline{3-10}
\multicolumn{1}{c}{}                               & \multicolumn{1}{c}{}                                 & \multicolumn{2}{c}{\textbf{O0 }}                                      & \multicolumn{2}{c}{\textbf{O1 }}                                      & \multicolumn{2}{c}{\textbf{O2 }}                                      & \multicolumn{2}{c}{\textbf{O3 }}                                       \\ 
\cline{3-10}
\multicolumn{1}{c}{}                               & \multicolumn{1}{c}{}                                 & \multicolumn{1}{c}{\textbf{Orig}} & \multicolumn{1}{c}{\textbf{Fine}} & \multicolumn{1}{c}{\textbf{Orig}} & \multicolumn{1}{c}{\textbf{Fine}} & \multicolumn{1}{c}{\textbf{Orig}} & \multicolumn{1}{c}{\textbf{Fine}} & \multicolumn{1}{c}{\textbf{Orig}} & \multicolumn{1}{c}{\textbf{Fine}}  \\ 
\hline
\multirow{9}{*}{x64}                               & CodeLlama                                            & 0.00                              & 0.00                              & 0.00                              & 0.00                              & 0.00                              & 0.00                              & 0.00                              & 0.00                               \\
                                                   & Llama2                                               & 0.00                              & 0.00                              & 0.00                              & 0.00                              & 0.00                              & 0.00                              & 0.00                              & 0.00                               \\
                                                   & Deepseek-V2                                          & 0.00                              & 0.00                              & 0.00                              & 0.00                              & 0.00                              & 0.00                              & 0.00                              & 0.00                               \\
                                                   & Deepseek-R1                                          & 0.00                              & 0.00                              & 0.00                              & 0.00                              & 0.00                              & 0.00                              & 0.00                              & 0.00                               \\
                                                   & Qwen2.5                                              & 0.00                              & 0.00                              & 0.00                              & 0.00                              & 0.00                              & 0.00                              & 0.00                              & 0.00                               \\
                                                   & Qwen 3                                               & 0.00                              & 0.00                              & 0.00                              & 0.00                              & 0.00                              & 0.00                              & 0.00                              & 0.00                               \\
                                                   & WizardCoder                                          & 0.00                              & 0.00                              & 0.00                              & 0.00                              & 0.00                              & 0.00                              & 0.00                              & 0.00                               \\
                                                   & Phi4                                                 & 0.00                              & 0.00                              & 0.00                              & 0.00                              & 0.00                              & 0.00                              & 0.00                              & 0.00                               \\
                                                   & Gemma3                                               & 0.00                              & 0.00                              & 0.00                              & 0.00                              & 0.00                              & 0.00                              & 0.00                              & 0.00                               \\ 
\hline
\end{tabular}}
\caption{Performance Evaluation on Function Name Recovery with Different Architectures and Optimization Levels via Assembly Code.}
\label{tbl:asmresult}
\end{table}}

\begin{table*}[t!]
\caption{Performance Evaluation on Function Name Recovery with Different Architectures and Optimization Levels via Assembly Code.}
\label{tbl:asmresult}
\centering
\footnotesize
\scalebox{0.87}{
\begin{tabular}{lrrrrrrrrrrrr}
\hline
\multicolumn{1}{c}{\multirow{3}{*}{\textbf{Model}}} & \multicolumn{12}{c}{\textbf{Function Name Recovery Performance}}                                                                                                                                                                                                                                                                                                                                                                                 \\ \cline{2-13} 
\multicolumn{1}{c}{}                                & \multicolumn{3}{c|}{\textbf{O0}}                                                                           & \multicolumn{3}{c|}{\textbf{O1}}                                                                           & \multicolumn{3}{c|}{\textbf{O2}}                                                                           & \multicolumn{3}{c}{\textbf{O3}}                                                                           \\ \cline{2-13} 
\multicolumn{1}{c}{}                                & \multicolumn{1}{c}{\textbf{Prec}} & \multicolumn{1}{c}{\textbf{Recall}} & \multicolumn{1}{c|}{\textbf{F1}} & \multicolumn{1}{c}{\textbf{Prec}} & \multicolumn{1}{c}{\textbf{Recall}} & \multicolumn{1}{c|}{\textbf{F1}} & \multicolumn{1}{c}{\textbf{Prec}} & \multicolumn{1}{c}{\textbf{Recall}} & \multicolumn{1}{c|}{\textbf{F1}} & \multicolumn{1}{c}{\textbf{Prec}} & \multicolumn{1}{c}{\textbf{Recall}} & \multicolumn{1}{c}{\textbf{F1}} \\ \hline
\multicolumn{13}{c}{x64}                                                                                                                                                                                                                                                                                                                                                                                                                                                                               \\ \hline
CodeLlama                                           & 0.0849                            & 0.0770                              & \multicolumn{1}{r|}{0.0808}      & 0.0870                            & 0.0675                              & \multicolumn{1}{r|}{0.0760}      & 0.1182                            & 0.1046                              & \multicolumn{1}{r|}{0.1110}      & 0.1219                            & 0.1079                              & 0.1145                          \\
Llama2                                              & 0.0838                            & 0.0933                              & \multicolumn{1}{r|}{0.0883}      & 0.0750                            & 0.0872                              & \multicolumn{1}{r|}{0.0806}      & 0.1211                            & 0.1307                              & \multicolumn{1}{r|}{0.1257}      & 0.1297                            & 0.1414                              & 0.1353                          \\
WizardCoder                                         & 0.1393                            & 0.1085                              & \multicolumn{1}{r|}{0.1220}      & 0.1266                            & 0.1044                              & \multicolumn{1}{r|}{0.1144}      & 0.1546                            & 0.1286                              & \multicolumn{1}{r|}{0.1404}      & 0.1508                            & 0.1257                              & 0.1371                          \\
Deepseek-R1                                         & 0.0886                            & 0.0742                              & \multicolumn{1}{r|}{0.0807}      & 0.1207                            & 0.1126                              & \multicolumn{1}{r|}{0.1165}      & 0.1526                            & 0.1563                              & \multicolumn{1}{r|}{0.1545}      & 0.1423                            & 0.1474                              & 0.1448                          \\
Qwen2.5                                             & 0.0868                            & 0.0754                              & \multicolumn{1}{r|}{0.0807}      & 0.1246                            & 0.1058                              & \multicolumn{1}{r|}{0.1145}      & 0.1575                            & 0.1347                              & \multicolumn{1}{r|}{0.1452}      & 0.1509                            & 0.1280                              & 0.1385                          \\
Qwen3                                               & 0.1253                            & 0.1180                              & \multicolumn{1}{r|}{0.1215}      & 0.1349                            & 0.1294                              & \multicolumn{1}{r|}{0.1321}      & 0.1719                            & 0.1635                              & \multicolumn{1}{r|}{0.1676}      & 0.1734                            & 0.1723                              & 0.1729                          \\
Phi4 Unsloth                                        & 0.1308                            & 0.1285                              & \multicolumn{1}{r|}{0.1296}      & 0.1648                            & 0.1624                              & \multicolumn{1}{r|}{0.1636}      & 0.1950                            & 0.1890                              & \multicolumn{1}{r|}{0.1919}      & 0.1977                            & 0.1926                              & 0.1951                          \\ \hline
\multicolumn{13}{c}{x86}                                                                                                                                                                                                                                                                                                                                                                                                                                                                               \\ \hline
CodeLlama                                           & 0.0909                            & 0.0900                              & \multicolumn{1}{r|}{0.0905}      & 0.0889                            & 0.0807                              & \multicolumn{1}{r|}{0.0846}      & 0.0909                            & 0.0825                              & \multicolumn{1}{r|}{0.0865}      & 0.0948                            & 0.0864                              & 0.0904                          \\
Llama2                                              & 0.0592                            & 0.0848                              & \multicolumn{1}{r|}{0.0697}      & 0.0564                            & 0.0667                              & \multicolumn{1}{r|}{0.0611}      & 0.0521                            & 0.0643                              & \multicolumn{1}{r|}{0.0576}      & 0.0525                            & 0.0533                              & 0.0529                          \\
WizardCoder                                         & 0.1274                            & 0.1244                              & \multicolumn{1}{r|}{0.1259}      & 0.1282                            & 0.1176                              & \multicolumn{1}{r|}{0.1227}      & 0.1308                            & 0.1173                              & \multicolumn{1}{r|}{0.1237}      & 0.1438                            & 0.1295                              & 0.1363                          \\
Deepseek-R1                                         & 0.1140                            & 0.1047                              & \multicolumn{1}{r|}{0.1091}      & 0.0891                            & 0.0861                              & \multicolumn{1}{r|}{0.0876}      & 0.0947                            & 0.0939                              & \multicolumn{1}{r|}{0.0943}      & 0.1013                            & 0.1005                              & 0.1009                          \\
Qwen2.5                                             & 0.1144                            & 0.0929                              & \multicolumn{1}{r|}{0.1025}      & 0.1026                            & 0.0802                              & \multicolumn{1}{r|}{0.0900}      & 0.0833                            & 0.0669                              & \multicolumn{1}{r|}{0.0742}      & 0.1034                            & 0.0819                              & 0.0914                          \\
Qwen3                                               & 0.1456                            & 0.1349                              & \multicolumn{1}{r|}{0.1400}      & 0.1448                            & 0.1268                              & \multicolumn{1}{r|}{0.1353}      & 0.1425                            & 0.1248                              & \multicolumn{1}{r|}{0.1331}      & 0.1475                            & 0.1302                              & 0.1383                          \\
Phi4 Unsloth                                        & 0.1733                            & 0.1593                              & \multicolumn{1}{r|}{0.1660}      & 0.1457                            & 0.1390                              & \multicolumn{1}{r|}{0.1423}      & 0.1594                            & 0.1482                              & \multicolumn{1}{r|}{0.1536}      & 0.1757                            & 0.1638                              & 0.1695                          \\ \hline
\multicolumn{13}{c}{ARM}                                                                                                                                                                                                                                                                                                                                                                                                                                                                               \\ \hline
CodeLlama                                           & 0.0844                            & 0.0801                              & \multicolumn{1}{r|}{0.0822}      & 0.0926                            & 0.0796                              & \multicolumn{1}{r|}{0.0856}      & 0.094                             & 0.0892                              & \multicolumn{1}{r|}{0.0916}      & 0.1040                            & 0.0977                              & 0.1008                          \\
Llama2                                              & 0.0458                            & 0.1010                              & \multicolumn{1}{r|}{0.0630}      & 0.0705                            & 0.1377                              & \multicolumn{1}{r|}{0.0932}      & 0.0735                            & 0.1475                              & \multicolumn{1}{r|}{0.0981}      & 0.0835                            & 0.1671                              & 0.1113                          \\
WizardCoder                                         & 0.0756                            & 0.0794                              & \multicolumn{1}{r|}{0.0774}      & 0.0720                            & 0.0720                              & \multicolumn{1}{r|}{0.0720}      & 0.0770                            & 0.0813                              & \multicolumn{1}{r|}{0.0791}      & 0.0767                            & 0.0837                              & 0.0800                          \\
Deepseek-R1                                         & 0.0705                            & 0.0901                              & \multicolumn{1}{r|}{0.0791}      & 0.0752                            & 0.1161                              & \multicolumn{1}{r|}{0.0913}      & 0.0771                            & 0.1145                              & \multicolumn{1}{r|}{0.0922}      & 0.083                             & 0.1294                              & 0.1011                          \\
Qwen2.5                                             & 0.0683                            & 0.0725                              & \multicolumn{1}{r|}{0.0703}      & 0.0805                            & 0.0881                              & \multicolumn{1}{r|}{0.0841}      & 0.0776                            & 0.0860                              & \multicolumn{1}{r|}{0.0816}      & 0.0822                            & 0.0841                              & 0.0831                          \\
Qwen3                                               & 0.0752                            & 0.0797                              & \multicolumn{1}{r|}{0.0774}      & 0.0921                            & 0.0961                              & \multicolumn{1}{r|}{0.0941}      & 0.0973                            & 0.1032                              & \multicolumn{1}{r|}{0.1002}      & 0.1080                            & 0.1176                              & 0.1126                          \\
Phi4 Unsloth                                        & 0.0994                            & 0.1190                              & \multicolumn{1}{r|}{0.1083}      & 0.1240                            & 0.1462                              & \multicolumn{1}{r|}{0.1342}      & 0.1299                            & 0.1495                              & \multicolumn{1}{r|}{0.1390}      & 0.1381                            & 0.1652                              & 0.1504                          \\ \hline
\multicolumn{13}{c}{MIPS}                                                                                                                                                                                                                                                                                                                                                                                                                                                                              \\ \hline
CodeLlama                                           & 0.0794                            & 0.0630                              & \multicolumn{1}{r|}{0.0702}      & 0.0777                            & 0.0540                              & \multicolumn{1}{r|}{0.0637}      & 0.0873                            & 0.0638                              & \multicolumn{1}{r|}{0.0737}      & 0.0715                            & 0.0560                              & 0.0628                          \\
Llama2                                              & 0.0315                            & 0.0577                              & \multicolumn{1}{r|}{0.0407}      & 0.0447                            & 0.0706                              & \multicolumn{1}{r|}{0.0547}      & 0.0424                            & 0.0706                              & \multicolumn{1}{r|}{0.0530}      & 0.0307                            & 0.0563                              & 0.0398                          \\
WizardCoder                                         & 0.0693                            & 0.0681                              & \multicolumn{1}{r|}{0.0687}      & 0.0745                            & 0.0679                              & \multicolumn{1}{r|}{0.0711}      & 0.0733                            & 0.0662                              & \multicolumn{1}{r|}{0.0696}      & 0.0595                            & 0.0570                              & 0.0582                          \\
Deepseek-R1                                         & 0.0630                            & 0.0745                              & \multicolumn{1}{r|}{0.0683}      & 0.0719                            & 0.0816                              & \multicolumn{1}{r|}{0.0764}      & 0.0626                            & 0.0684                              & \multicolumn{1}{r|}{0.0654}      & 0.0707                            & 0.0805                              & 0.0753                          \\
Qwen2.5                                             & 0.0794                            & 0.0732                              & \multicolumn{1}{r|}{0.0762}      & 0.0834                            & 0.0728                              & \multicolumn{1}{r|}{0.0778}      & 0.0828                            & 0.0691                              & \multicolumn{1}{r|}{0.0753}      & 0.0805                            & 0.0711                              & 0.0755                          \\
Qwen3                                               & 0.0792                            & 0.0858                              & \multicolumn{1}{r|}{0.0824}      & 0.0861                            & 0.0831                              & \multicolumn{1}{r|}{0.0846}      & 0.0893                            & 0.0847                              & \multicolumn{1}{r|}{0.0869}      & 0.0818                            & 0.0820                              & 0.0819                          \\
Phi4 Unsloth                                        & 0.0906                            & 0.0964                              & \multicolumn{1}{r|}{0.0934}      & 0.1066                            & 0.1081                              & \multicolumn{1}{r|}{0.1073}      & 0.1099                            & 0.1065                              & \multicolumn{1}{r|}{0.1082}      & 0.1084                            & 0.1101                              & 0.1093                          \\ \hline
\end{tabular}}
\end{table*}

%% file: table/real_world.tex
\ignore{
\begin{table*}[tb!]
\centering
\footnotesize
\scalebox{0.51}{
\begin{tabular}{lllllll|llllll|llllll} 
\hline
\multicolumn{1}{c}{\multirow{3}{*}{\textbf{Model }}} & \multicolumn{6}{c|}{\textbf{Function Name Recovery}}                                                                                                                                                                                                          & \multicolumn{6}{c|}{\textbf{Variable Name Recovery}}                                                                                                                                                                                                                            & \multicolumn{6}{c}{\textbf{Type Inference}}                                                                                                                                                                                                                                     \\ 
\cline{2-19}
\multicolumn{1}{c}{}                                 & \multicolumn{2}{c}{\textbf{Prec }}                                     & \multicolumn{2}{c}{\textbf{Recall }}                                                     & \multicolumn{2}{c|}{\textbf{F1 }}                                                         & \multicolumn{2}{c}{\textbf{Prec }}                                                       & \multicolumn{2}{c}{\textbf{Recall }}                                                     & \multicolumn{2}{c|}{\textbf{F1 }}                                                         & \multicolumn{2}{c}{\textbf{Prec }}                                                       & \multicolumn{2}{c}{\textbf{Recall }}                                                     & \multicolumn{2}{c}{\textbf{F1 }}                                                          \\ 
\cline{2-19}
\multicolumn{1}{c}{}                                 & \multicolumn{1}{c}{\textbf{Ghi}} & \multicolumn{1}{c}{\textbf{IDA}} & \multicolumn{1}{c}{\textbf{\textbf{Ghi}}} & \multicolumn{1}{c}{\textbf{\textbf{IDA}}} & \multicolumn{1}{c}{\textbf{\textbf{Ghi}}} & \multicolumn{1}{c|}{\textbf{\textbf{IDA}}} & \multicolumn{1}{c}{\textbf{\textbf{Ghi}}} & \multicolumn{1}{c}{\textbf{\textbf{IDA}}} & \multicolumn{1}{c}{\textbf{\textbf{Ghi}}} & \multicolumn{1}{c}{\textbf{\textbf{IDA}}} & \multicolumn{1}{c}{\textbf{\textbf{Ghi}}} & \multicolumn{1}{c|}{\textbf{\textbf{IDA}}} & \multicolumn{1}{c}{\textbf{\textbf{Ghi}}} & \multicolumn{1}{c}{\textbf{\textbf{IDA}}} & \multicolumn{1}{c}{\textbf{\textbf{Ghi}}} & \multicolumn{1}{c}{\textbf{\textbf{IDA}}} & \multicolumn{1}{c}{\textbf{\textbf{Ghi}}} & \multicolumn{1}{c}{\textbf{\textbf{IDA}}}  \\ 
\hline
CodeLlama                                              & 0.01 & 0.02 & 0.03 & 0.04 & 0.02 & 0.03 & 0.02 & 0.03 & 0.03 & 0.04 & 0.02 & 0.03 & 0.05 & 0.06 & 0.01 & 0.01 & 0.02 & 0.02                                        \\
Llama2                                                  & 0.01 & 0.01 & 0.02 & 0.03 & 0.01 & 0.02 &  0.02 & 0.04 & 0.03 & 0.05 & 0.02 & 0.04 & 0.03 & 0.11 & 0.01 & 0.02 & 0.01 & 0.03                                       \\
Deepseek-V2                                               & 0.01 & 0.01 & 0.01 & 0.01 & 0.01 & 0.01 & 0.02 & 0.03 & 0.03 & 0.04 & 0.03 & 0.03 &  0.02 & 0.01 & 0.01 & 0.00 & 0.01 & 0.00                                        \\
Deepseek-R1                                            & 0.02 & 0.01 & 0.03 & 0.02 & 0.02 & 0.02 & 0.01 & 0.00 & 0.01 & 0.00 & 0.01 & 0.00 & 0.03 & 0.03 & 0.00 & 0.00 & 0.01 & 0.00                                        \\
Qwen                                            & 0.01 & 0.01 & 0.03 & 0.02 & 0.02 & 0.01 &  0.02 & 0.03 & 0.04 & 0.03 & 0.03 & 0.03 & 0.04 & 0.05 & 0.01 & 0.01 & 0.02 & 0.01                                      \\
WizardCoder                                          & 0.01 & 0.01 & 0.02 & 0.02 & 0.01 & 0.01 & 0.02 & 0.04 & 0.04 & 0.06 & 0.03 & 0.05 &  0.05 & 0.09 & 0.01 & 0.02 & 0.02 & 0.03                                     \\
\hline
\end{tabular}}
\caption{Performance Evaluation on Every Task from Real World Firmware with Two Distinct Decompilers.}
\label{tbl:realworld}
\end{table*}}

\begin{table*}[t!]
\caption{Performance Evaluation on Every Task from Real World Firmware with Two Distinct Decompilers.}
\label{tbl:realworld}
\centering
\footnotesize
\scalebox{0.87}{
\begin{tabular}{lllllll|llllll|llllll}
\hline
\multicolumn{1}{c}{\multirow{3}{*}{\textbf{Model}}} & \multicolumn{6}{c|}{\textbf{Function Name Recovery}}                                                                                                                                                             & \multicolumn{6}{c|}{\textbf{Variable Name Recovery}}                                                                                                                                                             & \multicolumn{6}{c}{\textbf{Type Inference}}                                                                                                                                                                     \\ \cline{2-19} 
\multicolumn{1}{c}{}                                & \multicolumn{2}{c}{\textbf{Prec}}                                   & \multicolumn{2}{c}{\textbf{Recall}}                                 & \multicolumn{2}{c|}{\textbf{F1}}                                     & \multicolumn{2}{c}{\textbf{Prec}}                                   & \multicolumn{2}{c}{\textbf{Recall}}                                 & \multicolumn{2}{c|}{\textbf{F1}}                                     & \multicolumn{2}{c}{\textbf{Prec}}                                   & \multicolumn{2}{c}{\textbf{Recall}}                                 & \multicolumn{2}{c}{\textbf{F1}}                                     \\ \cline{2-19} 
\multicolumn{1}{c}{}                                & \multicolumn{1}{c}{\textbf{Ghi}} & \multicolumn{1}{c}{\textbf{IDA}} & \multicolumn{1}{c}{\textbf{Ghi}} & \multicolumn{1}{c}{\textbf{IDA}} & \multicolumn{1}{c}{\textbf{Ghi}} & \multicolumn{1}{c|}{\textbf{IDA}} & \multicolumn{1}{c}{\textbf{Ghi}} & \multicolumn{1}{c}{\textbf{IDA}} & \multicolumn{1}{c}{\textbf{Ghi}} & \multicolumn{1}{c}{\textbf{IDA}} & \multicolumn{1}{c}{\textbf{Ghi}} & \multicolumn{1}{c|}{\textbf{IDA}} & \multicolumn{1}{c}{\textbf{Ghi}} & \multicolumn{1}{c}{\textbf{IDA}} & \multicolumn{1}{c}{\textbf{Ghi}} & \multicolumn{1}{c}{\textbf{IDA}} & \multicolumn{1}{c}{\textbf{Ghi}} & \multicolumn{1}{c}{\textbf{IDA}} \\ \hline
CodeLlama                                           & 0.0219                           & 0.0312                           & 0.0370                           & 0.0635                           & 0.0276                           & 0.0419                            & 0.0144                           & 0.0365                           & 0.0252                           & 0.0564                           & 0.0183                           & 0.0443                            & 0.0072                           & 0.0567                           & 0.0026                           & 0.0232                           & 0.0038                           & 0.0330                           \\
Llama2                                              & 0.0224                           & 0.0265                           & 0.0342                           & 0.0366                           & 0.0271                           & 0.0307                            & 0.0170                           & 0.0142                           & 0.0283                           & 0.0132                           & 0.0213                           & 0.0137                            & 0.0068                           & 0.0093                           & 0.0087                           & 0.0073                           & 0.0076                           & 0.0082                           \\
WizardCoder                                         & 0.0283                           & 0.0471                           & 0.0313                           & 0.0383                           & 0.0297                           & 0.0422                            & 0.1634                           & 0.0184                           & 0.2732                           & 0.0223                           & 0.2045                           & 0.0202                            & 0.0401                           & 0.0115                           & 0.0219                           & 0.0082                           & 0.0283                           & 0.0096                           \\
Deepseek-R1                                         & 0.0351                           & 0.0249                           & 0.0358                           & 0.0402                           & 0.0354                           & 0.0307                            & 0.1436                           & 0.0074                           & 0.2466                           & 0.0108                           & 0.1815                           & 0.0087                            & 0.0097                           & 0.0111                           & 0.0096                           & 0.0091                           & 0.0097                           & 0.0100                           \\
Qwen2.5                                             & 0.0356                           & 0.0490                           & 0.0409                           & 0.0518                           & 0.0380                           & 0.0504                            & 0.1484                           & 0.0163                           & 0.2528                           & 0.0199                           & 0.1870                           & 0.0179                            & 0.0080                           & 0.0077                           & 0.0087                           & 0.0059                           & 0.0084                           & 0.0067                           \\
Qwen3                                               & 0.0321                           & 0.0557                           & 0.0361                           & 0.0788                           & 0.0340                           & 0.0653                            & 0.1284                           & 0.0297                           & 0.2071                           & 0.0380                           & 0.1585                           & 0.0333                            & 0.0080                           & 0.0280                           & 0.0087                           & 0.0273                           & 0.0084                           & 0.0277                           \\
Phi4 Unsloth                                        & 0.0489                           & 0.0694                           & 0.0533                           & 0.1010                           & 0.0510                           & 0.0823                            & 0.1357                           & 0.0202                           & 0.2319                           & 0.0282                           & 0.1712                           & 0.0235                            & 0.0150                           & 0.0262                           & 0.0140                           & 0.0219                           & 0.0145                           & 0.0238                           \\ \hline
\end{tabular}}
\end{table*}

%% file: table/ablation.tex
\ignore{
\begin{table*}
\centering
\footnotesize
\scalebox{0.72}{
\begin{tabular}{lclll|lll|lll|lll} 
\hline
\multicolumn{1}{c}{\multirow{3}{*}{\textbf{Model}}} & \multirow{3}{*}{\begin{tabular}[c]{@{}c@{}}\textbf{Code}\\\textbf{Word}\\\textbf{Net}\end{tabular}} & \multicolumn{12}{c}{\textbf{Name Recovery Performance}}                                                                                                                                                                                                                                                                                                                                                                                           \\ 
\cline{3-14}
\multicolumn{1}{c}{}                                &                                                                                                     & \multicolumn{3}{c|}{\textbf{O0}}                                                                           & \multicolumn{3}{c|}{\textbf{O1}}                                                                           & \multicolumn{3}{c|}{\textbf{O2}}                                                                           & \multicolumn{3}{c}{\textbf{O3}}                                                                            \\ 
\cline{3-14}
\multicolumn{1}{c}{}                                &                                                                                                     & \multicolumn{1}{c}{\textbf{Prec}} & \multicolumn{1}{c}{\textbf{Recall}} & \multicolumn{1}{c|}{\textbf{F1}} & \multicolumn{1}{c}{\textbf{Prec}} & \multicolumn{1}{c}{\textbf{Recall}} & \multicolumn{1}{c|}{\textbf{F1}} & \multicolumn{1}{c}{\textbf{Prec}} & \multicolumn{1}{c}{\textbf{Recall}} & \multicolumn{1}{c|}{\textbf{F1}} & \multicolumn{1}{c}{\textbf{Prec}} & \multicolumn{1}{c}{\textbf{Recall}} & \multicolumn{1}{c}{\textbf{F1}}  \\ 
\hline
\multicolumn{14}{c}{Function Name}                                                                                                                                                                                                                                                                                                                                                                                                                                                                                                                                                                            \\ 
\hline
\multirow{2}{*}{CodeGen}                            & \checkmark                                                                                                   & 0.08                              & 0.07                                & 0.08                             & 0.09                              & 0.08                                & 0.08                             & 0.09                              & 0.08                                & 0.08                             & 0.08                              & 0.07                                & 0.08                             \\
                                                    & \ding{55}                                                                                                   & 0.01(\textcolor{red}{$\downarrow$}87.5\%)                    & 0.01(\textcolor{red}{$\downarrow$}85.7\%)                      & 0.01(\textcolor{red}{$\downarrow$}87.5\%)                   & 0.01(\textcolor{red}{$\downarrow$}88.9\%)                     & 0.01(\textcolor{red}{$\downarrow$}87.5\%)                       & 0.01(\textcolor{red}{$\downarrow$}87.5\%)                    & 0.01(\textcolor{red}{$\downarrow$}88.9\%)                    & 0.01(\textcolor{red}{$\downarrow$}87.5\%)                      & 0.01(\textcolor{red}{$\downarrow$}87.5\%)                   & 0.01(\textcolor{red}{$\downarrow$}87.5\%)                    & 0.01(\textcolor{red}{$\downarrow$}85.7\%)                      & 0.01(\textcolor{red}{$\downarrow$}87.5\%)                   \\
\multirow{2}{*}{CodeGen2}                           & \checkmark                                                                                                   & 0.05                              & 0.05                                & 0.05                             & 0.06                              & 0.06                                & 0.06                             & 0.07                              & 0.07                                & 0.07                             & 0.07                              & 0.07                                & 0.07                             \\
                                                    & \ding{55}                                                                                                   & 0.01(\textcolor{red}{$\downarrow$}80.0\%)                     & 0.01(\textcolor{red}{$\downarrow$}80.0\%)                       & 0.01(\textcolor{red}{$\downarrow$}80.0\%)                    & 0.02(\textcolor{red}{$\downarrow$}66.7\%)                     & 0.01(\textcolor{red}{$\downarrow$}83.3\%)                       & 0.02(\textcolor{red}{$\downarrow$}66.7\%)                    & 0.02(\textcolor{red}{$\downarrow$}71.4\%)                     & 0.02(\textcolor{red}{$\downarrow$}71.4\%)                       & 0.02(\textcolor{red}{$\downarrow$}71.4\%)                    & 0.02(\textcolor{red}{$\downarrow$}71.4\%)                     & 0.02(\textcolor{red}{$\downarrow$}71.4\%)                       & 0.02(\textcolor{red}{$\downarrow$}71.4\%)                    \\
\multirow{2}{*}{Llama2}                             & \checkmark                                                                                                   & 0.13                              & 0.04                                & 0.06                             & 0.12                              & 0.03                                & 0.05                             & 0.14                              & 0.04                                & 0.06                             & 0.13                              & 0.04                                & 0.06                             \\
                                                    & \ding{55}                                                                                                   & 0.02(\textcolor{red}{$\downarrow$}84.6\%)                     & 0.01(\textcolor{red}{$\downarrow$}75.0\%)                      & 0.01(\textcolor{red}{$\downarrow$}83.3\%)                    & 0.02(\textcolor{red}{$\downarrow$}83.3\%)                     & 0.0(\textcolor{red}{$\downarrow$}100.0\%)                       & 0.01(\textcolor{red}{$\downarrow$}80.0\%)                    & 0.02(\textcolor{red}{$\downarrow$}85.7\%)                     & 0.01(\textcolor{red}{$\downarrow$}75.0\%)                      & 0.01(\textcolor{red}{$\downarrow$}83.3\%)                    & 0.03(\textcolor{red}{$\downarrow$}76.9\%)                     & 0.01(\textcolor{red}{$\downarrow$}75.0\%)                       & 0.01(\textcolor{red}{$\downarrow$}83.3\%)                    \\
\multirow{2}{*}{CodeLlama}                          & \checkmark                                                                                                   & 0.33                              & 0.32                                & 0.33                             & 0.37                              & 0.36                                & 0.37                             & 0.35                              & 0.34                                & 0.35                             & 0.34                              & 0.34                                & 0.34                             \\
                                                    & \ding{55}                                                                                                   & 0.11(\textcolor{red}{$\downarrow$}66.7\%)                     & 0.11(\textcolor{red}{$\downarrow$}65.6\%)                       & 0.11(\textcolor{red}{$\downarrow$}66.7\%)                    & 0.12(\textcolor{red}{$\downarrow$}67.5\%)                     & 0.12(\textcolor{red}{$\downarrow$}75.0\%)                       & 0.12(\textcolor{red}{$\downarrow$}67.5\%)                    & 0.12(\textcolor{red}{$\downarrow$}65.7\%)                     & 0.12(\textcolor{red}{$\downarrow$}64.7\%)                       & 0.12(\textcolor{red}{$\downarrow$}65.7\%)                    & 0.13(\textcolor{red}{$\downarrow$}61.7\%)                     & 0.13(\textcolor{red}{$\downarrow$}61.7\%)                       & 0.13(\textcolor{red}{$\downarrow$}61.7\%)                    \\
\multirow{2}{*}{CodeT5Plus}                         & \checkmark                                                                                                   & 0.01                              & 0.01                                & 0.01                             & 0.02                              & 0.02                                & 0.02                             & 0.02                              & 0.02                                & 0.02                             & 0.01                              & 0.01                                & 0.01                             \\
                                                    & \ding{55}                                                                                                   & 0.0(\textcolor{red}{$\downarrow$}100.0\%)                     & 0.0(\textcolor{red}{$\downarrow$}100.0\%)                       & 0.0(\textcolor{red}{$\downarrow$}100.0\%)                    & 0.0(\textcolor{red}{$\downarrow$}100.0\%)                     & 0.0(\textcolor{red}{$\downarrow$}100.0\%)                       & 0.0(\textcolor{red}{$\downarrow$}100.0\%)                    & 0.0(\textcolor{red}{$\downarrow$}100.0\%)                     & 0.0(\textcolor{red}{$\downarrow$}100.0\%)                       & 0.0(\textcolor{red}{$\downarrow$}100.0\%)                    & 0.0(\textcolor{red}{$\downarrow$}100.0\%)                     & 0.0(\textcolor{red}{$\downarrow$}100.0\%)                       & 0.0(\textcolor{red}{$\downarrow$}100.0\%)                    \\
\multirow{2}{*}{StarCoder}                          & \checkmark                                                                                                   & 0.26                              & 0.25                                & 0.26                             & 0.3                               & 0.28                                & 0.29                             & 0.28                              & 0.26                                & 0.27                             & 0.29                              & 0.27                                & 0.28                             \\
                                                    & \ding{55}                                                                                                   & 0.08(\textcolor{red}{$\downarrow$}69.2\%)                     & 0.08(\textcolor{red}{$\downarrow$}68.0\%)                       & 0.08(\textcolor{red}{$\downarrow$}69.2\%)                    & 0.09(\textcolor{red}{$\downarrow$}70.0\%)                     & 0.08(\textcolor{red}{$\downarrow$}71.4\%)                       & 0.08(\textcolor{red}{$\downarrow$}72.4\%)                    & 0.09(\textcolor{red}{$\downarrow$}67.8\%)                     & 0.08(\textcolor{red}{$\downarrow$}69.2\%)                       & 0.08(\textcolor{red}{$\downarrow$}70.3\%)                    & 0.09(\textcolor{red}{$\downarrow$}68.9\%)                     & 0.09(\textcolor{red}{$\downarrow$}66.7\%)                       & 0.09(\textcolor{red}{$\downarrow$}67.8\%)                    \\
\multirow{2}{*}{WizardCoder}                        & \checkmark                                                                                                   & 0.28                              & 0.28                                & 0.28                             & 0.32                              & 0.31                                & 0.32                             & 0.31                              & 0.3                                 & 0.3                              & 0.31                              & 0.3                                 & 0.3                              \\
                                                    & \ding{55}                                                                                                   & 0.09(\textcolor{red}{$\downarrow$}67.8\%)                     & 0.09(\textcolor{red}{$\downarrow$}67.8\%)                       & 0.09(\textcolor{red}{$\downarrow$}67.8\%)                    & 0.09(\textcolor{red}{$\downarrow$}71.8\%)                     & 0.09(\textcolor{red}{$\downarrow$}70.9\%)                       & 0.09(\textcolor{red}{$\downarrow$}71.8\%)                    & 0.09(\textcolor{red}{$\downarrow$}70.9\%)                     & 0.09(\textcolor{red}{$\downarrow$}67.7\%)                       & 0.09(\textcolor{red}{$\downarrow$}67.7\%)                    & 0.1(\textcolor{red}{$\downarrow$}67.7\%)                      & 0.1(\textcolor{red}{$\downarrow$}66.7\%)                        & 0.1(\textcolor{red}{$\downarrow$}66.7\%)                     \\ 
\hline
\multicolumn{14}{c}{Variable Name}                                                                                                                                                                                                                                                                                                                                                                                                                                                                                                                                                                            \\ 
\hline
\multirow{2}{*}{CodeGen}                            & \checkmark                                                                                                   & 0.02                              & 0.02                                & 0.02                             & 0.02                              & 0.02                                & 0.02                             & 0.01                              & 0.01                                & 0.01                             & 0.02                              & 0.02                                & 0.02                             \\
                                                    & \ding{55}                                                                                                   & 0.0(\textcolor{red}{$\downarrow$}100.0\%)                     & 0.0(\textcolor{red}{$\downarrow$}100.0\%)                       & 0.0(\textcolor{red}{$\downarrow$}100.0\%)                    & 0.01(\textcolor{red}{$\downarrow$}50.0\%)                     & 0.01(\textcolor{red}{$\downarrow$}50.0\%)                       & 0.01(\textcolor{red}{$\downarrow$}50.0\%)                    & 0.0(\textcolor{red}{$\downarrow$}100.0\%)                     & 0.0(\textcolor{red}{$\downarrow$}100.0\%)                       & 0.0(\textcolor{red}{$\downarrow$}100.0\%)                    & 0.0(\textcolor{red}{$\downarrow$}100.0\%)                     & 0.0(\textcolor{red}{$\downarrow$}100.0\%)                       & 0.0(\textcolor{red}{$\downarrow$}100.0\%)                    \\
\multirow{2}{*}{CodeGen2}                           & \checkmark                                                                                                   & 0.02                              & 0.02                                & 0.02                             & 0.02                              & 0.02                                & 0.02                             & 0.02                              & 0.02                                & 0.02                             & 0.02                              & 0.02                                & 0.02                             \\
                                                    & \ding{55}                                                                                                   & 0.01(\textcolor{red}{$\downarrow$}50.0\%)                     & 0.01(\textcolor{red}{$\downarrow$}50.0\%)                       & 0.01(\textcolor{red}{$\downarrow$}50.0\%)                    & 0.01(\textcolor{red}{$\downarrow$}50.0\%)                     & 0.01(\textcolor{red}{$\downarrow$}50.0\%)                       & 0.01(\textcolor{red}{$\downarrow$}50.0\%)                    & 0.01(\textcolor{red}{$\downarrow$}50.0\%)                     & 0.01(\textcolor{red}{$\downarrow$}50.0\%)                       & 0.01(\textcolor{red}{$\downarrow$}50.0\%)                    & 0.01(\textcolor{red}{$\downarrow$}50.0\%)                     & 0.01(\textcolor{red}{$\downarrow$}50.0\%)                       & 0.01(\textcolor{red}{$\downarrow$}50.0\%)                    \\
\multirow{2}{*}{Llama2}                             & \checkmark                                                                                                   & 0.04                              & 0.01                                & 0.02                             & 0.04                              & 0.01                                & 0.02                             & 0.04                              & 0.01                                & 0.02                             & 0.05                              & 0.01                                & 0.02                             \\
                                                    & X                                                                                                   & 0.01(\textcolor{red}{$\downarrow$}75.0\%)                     & 0.0(\textcolor{red}{$\downarrow$}100.0\%)                       & 0.01(\textcolor{red}{$\downarrow$}50.0\%)                    & 0.02(\textcolor{red}{$\downarrow$}50.0\%)                     & 0.0(\textcolor{red}{$\downarrow$}100.0\%)                       & 0.01(\textcolor{red}{$\downarrow$}50.0\%)                    & 0.02(\textcolor{red}{$\downarrow$}50.0\%)                     & 0.0(\textcolor{red}{$\downarrow$}100.0\%)                       & 0.01(\textcolor{red}{$\downarrow$}50.0\%)                    & 0.02(\textcolor{red}{$\downarrow$}60.0\%)                     & 0.01(0.0\%)                         & 0.01(\textcolor{red}{$\downarrow$}50.0\%)                    \\
\multirow{2}{*}{CodeLlama}                          & \checkmark                                                                                                   & 0.11                              & 0.11                                & 0.11                             & 0.14                              & 0.14                                & 0.14                             & 0.14                              & 0.14                                & 0.14                             & 0.14                              & 0.14                                & 0.14                             \\
                                                    & \ding{55}                                                                                                   & 0.06(\textcolor{red}{$\downarrow$}45.5\%)                     & 0.06(\textcolor{red}{$\downarrow$}45.5\%)                       & 0.06(\textcolor{red}{$\downarrow$}45.5\%)                    & 0.09(\textcolor{red}{$\downarrow$}35.7\%)                     & 0.09(\textcolor{red}{$\downarrow$}35.7\%)                       & 0.09(\textcolor{red}{$\downarrow$}35.7\%)                    & 0.1(\textcolor{red}{$\downarrow$}28.6\%)                      & 0.09(\textcolor{red}{$\downarrow$}35.7\%)                       & 0.1(\textcolor{red}{$\downarrow$}28.6\%)                     & 0.09(\textcolor{red}{$\downarrow$}35.7\%)                     & 0.09(\textcolor{red}{$\downarrow$}35.7\%)                       & 0.09(\textcolor{red}{$\downarrow$}35.7\%)                    \\
\multirow{2}{*}{CodeT5Plus}                         & \checkmark                                                                                                   & 0.01                              & 0.01                                & 0.01                             & 0.01                              & 0.01                                & 0.01                             & 0.01                              & 0.01                                & 0.01                             & 0.01                              & 0.01                                & 0.01                             \\
                                                    & \ding{55}                                                                                                   & 0.0(\textcolor{red}{$\downarrow$}100.0\%)                     & 0.0(\textcolor{red}{$\downarrow$}100.0\%)                       & 0.0(\textcolor{red}{$\downarrow$}100.0\%)                    & 0.0(\textcolor{red}{$\downarrow$}100.0\%)                     & 0.0(\textcolor{red}{$\downarrow$}100.0\%)                       & 0.0(\textcolor{red}{$\downarrow$}100.0\%)                    & 0.0(\textcolor{red}{$\downarrow$}100.0\%)                     & 0.0(\textcolor{red}{$\downarrow$}100.0\%)                       & 0.0(\textcolor{red}{$\downarrow$}100.0\%)                    & 0.0(\textcolor{red}{$\downarrow$}100.0\%)                     & 0.0(\textcolor{red}{$\downarrow$}100.0\%)                       & 0.0(\textcolor{red}{$\downarrow$}100.0\%)                    \\
\multirow{2}{*}{StarCoder}                          & \checkmark                                                                                                   & 0.07                              & 0.06                                & 0.06                             & 0.09                              & 0.09                                & 0.09                             & 0.09                              & 0.08                                & 0.08                             & 0.09                              & 0.08                                & 0.08                             \\
                                                    & \ding{55}                                                                                                   & 0.03(\textcolor{red}{$\downarrow$}57.1\%)                     & 0.02(\textcolor{red}{$\downarrow$}66.7\%)                       & 0.02(\textcolor{red}{$\downarrow$}66.7\%)                    & 0.04(\textcolor{red}{$\downarrow$}55.6\%)                     & 0.04(\textcolor{red}{$\downarrow$}55.6\%)                       & 0.04(\textcolor{red}{$\downarrow$}55.6\%)                    & 0.05(\textcolor{red}{$\downarrow$}44.4\%)                     & 0.04(\textcolor{red}{$\downarrow$}50.0\%)                       & 0.04(\textcolor{red}{$\downarrow$}50.0\%)                    & 0.04(\textcolor{red}{$\downarrow$}55.6\%)                     & 0.04(\textcolor{red}{$\downarrow$}50.0\%)                       & 0.04(\textcolor{red}{$\downarrow$}50.0\%)                    \\
\multirow{2}{*}{WizardCoder}                        & \checkmark                                                                                                   & 0.1                               & 0.09                                & 0.09                             & 0.13                              & 0.12                                & 0.12                             & 0.12                              & 0.11                                & 0.12                             & 0.12                              & 0.11                                & 0.12                             \\
                                                    & \ding{55}                                                                                                   & 0.03(\textcolor{red}{$\downarrow$}70.0\%)                     & 0.03(\textcolor{red}{$\downarrow$}66.7\%)                       & 0.03(\textcolor{red}{$\downarrow$}66.7\%)                    & 0.05(\textcolor{red}{$\downarrow$}61.5\%)                     & 0.05(\textcolor{red}{$\downarrow$}58.3\%)                       & 0.05(\textcolor{red}{$\downarrow$}58.3\%)                    & 0.06(\textcolor{red}{$\downarrow$}50.0\%)                     & 0.05(\textcolor{red}{$\downarrow$}54.5\%)                       & 0.05(\textcolor{red}{$\downarrow$}58.3\%)                    & 0.05(\textcolor{red}{$\downarrow$}58.3\%)                     & 0.05(\textcolor{red}{$\downarrow$}54.5\%)                       & 0.05(\textcolor{red}{$\downarrow$}58.3\%)                    \\
\hline
\end{tabular}}
\caption{Performance Evaluation on Name Recovery with and without CodeWordNet in X64 Architecture.}
\label{tbl:ablation}
\end{table*}}

\ignore{
\begin{figure*}[!htb]
\centering
    
   \begin{subfigure}{0.98\textwidth}
     \centering
     \includegraphics[width=0.24\linewidth]{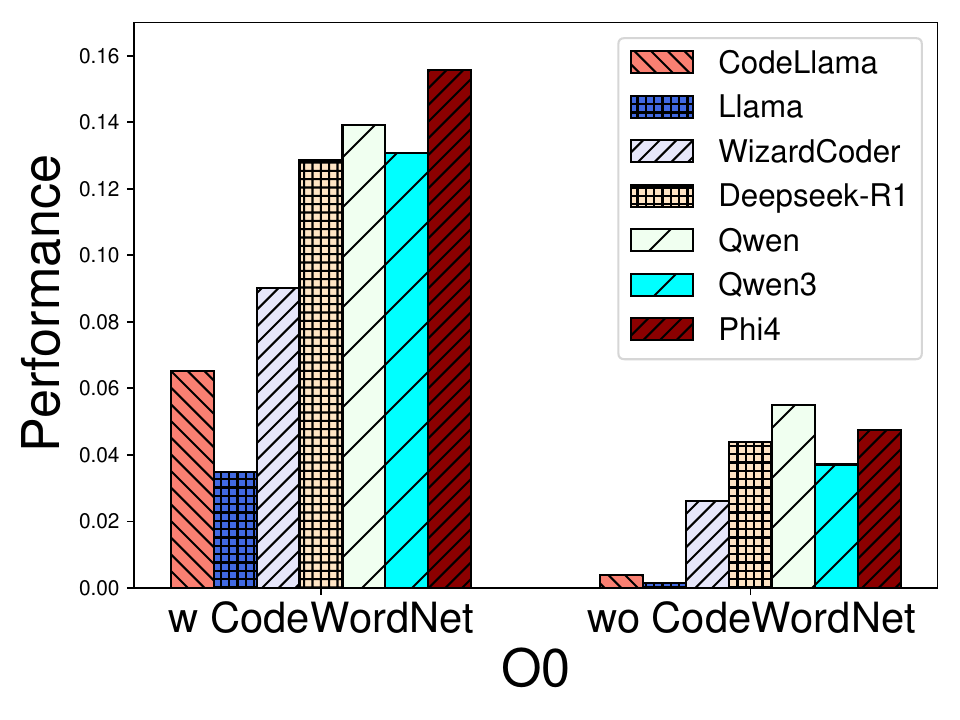}
     \includegraphics[width=0.24\linewidth]{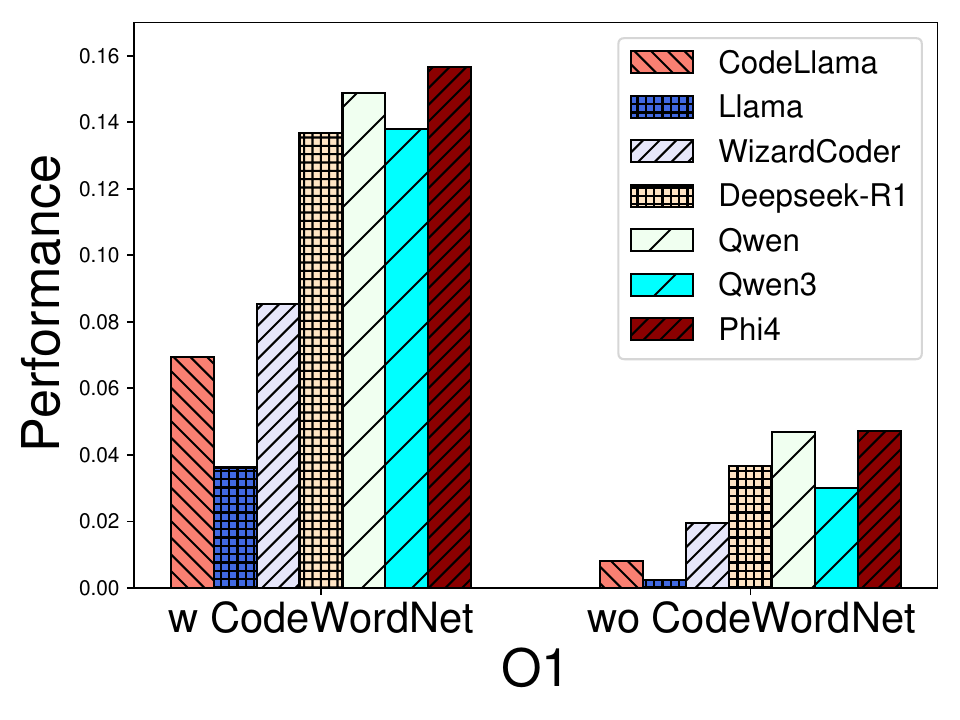}
     \includegraphics[width=0.24\linewidth]{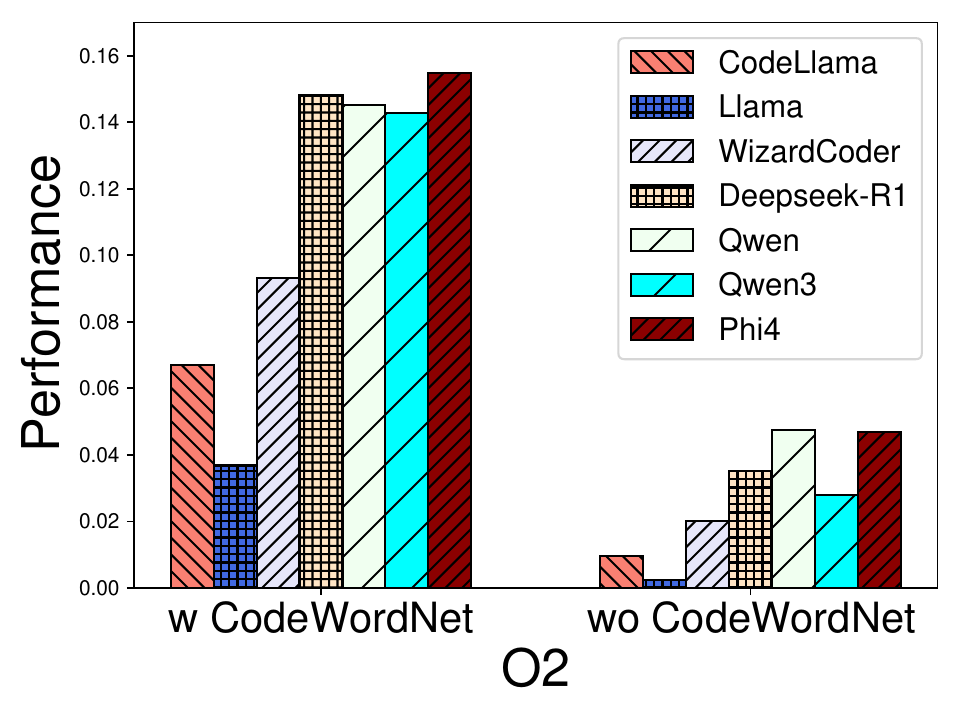}
     \includegraphics[width=0.24\linewidth]{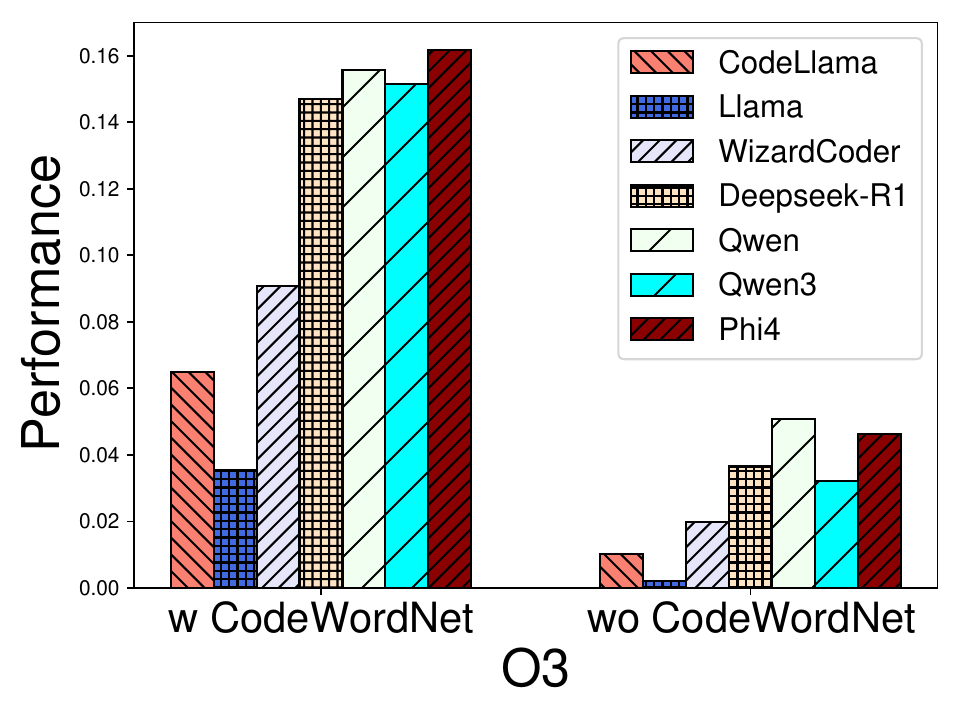}
     \caption{Function Name Recovery}
     \label{fig:ablationfunc}
   \end{subfigure}
   \begin{subfigure}{0.98\textwidth}
     \centering
     \includegraphics[width=0.24\linewidth]{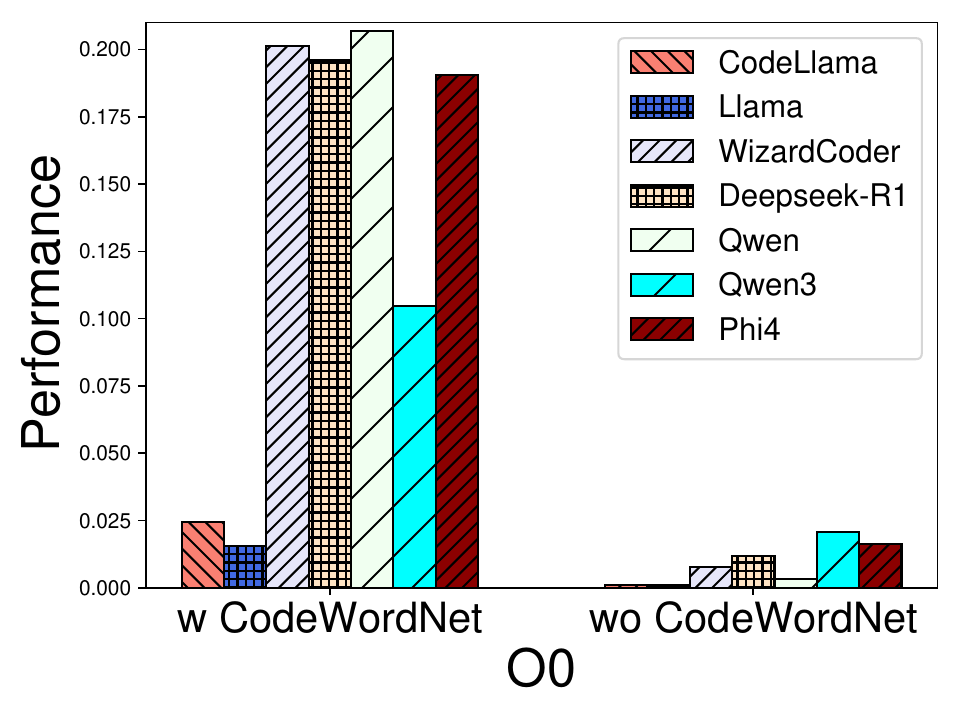}
     \includegraphics[width=0.24\linewidth]{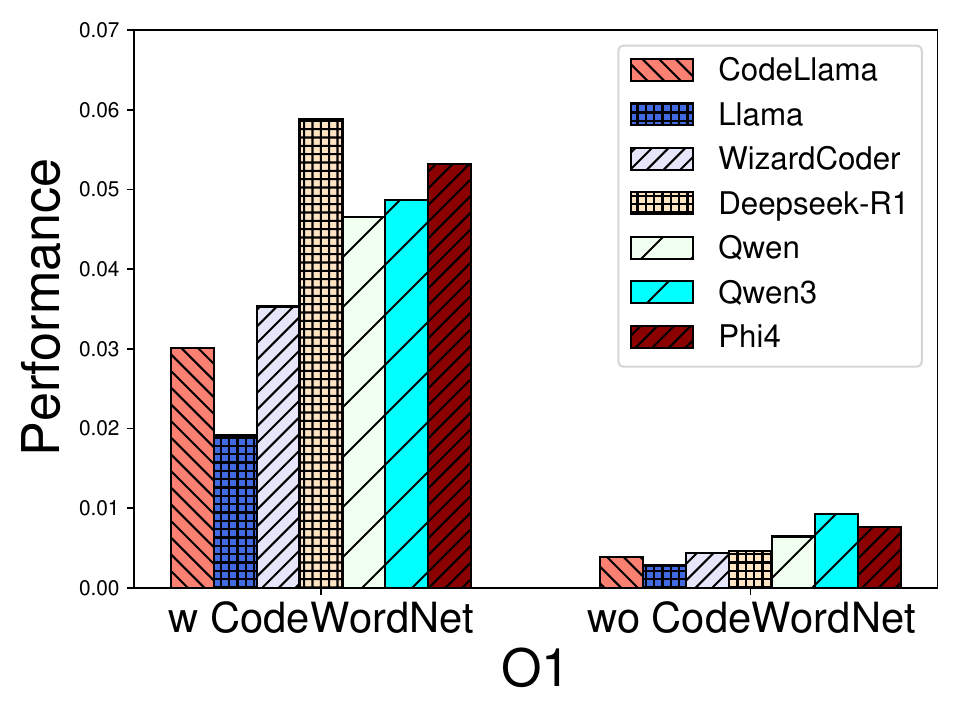}
     \includegraphics[width=0.24\linewidth]{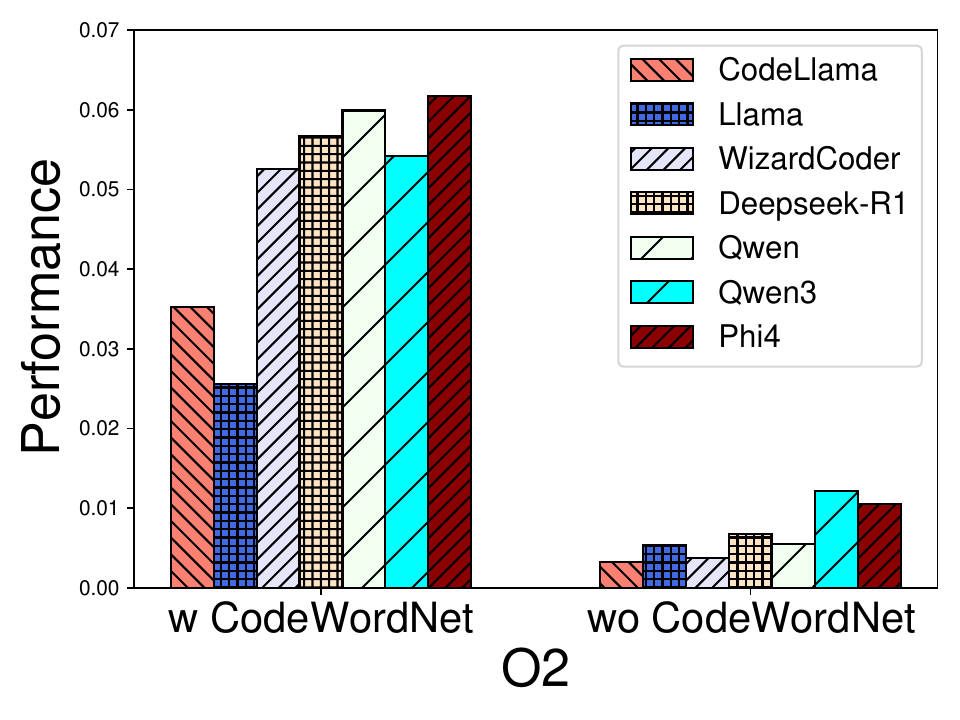}
     \includegraphics[width=0.24\linewidth]{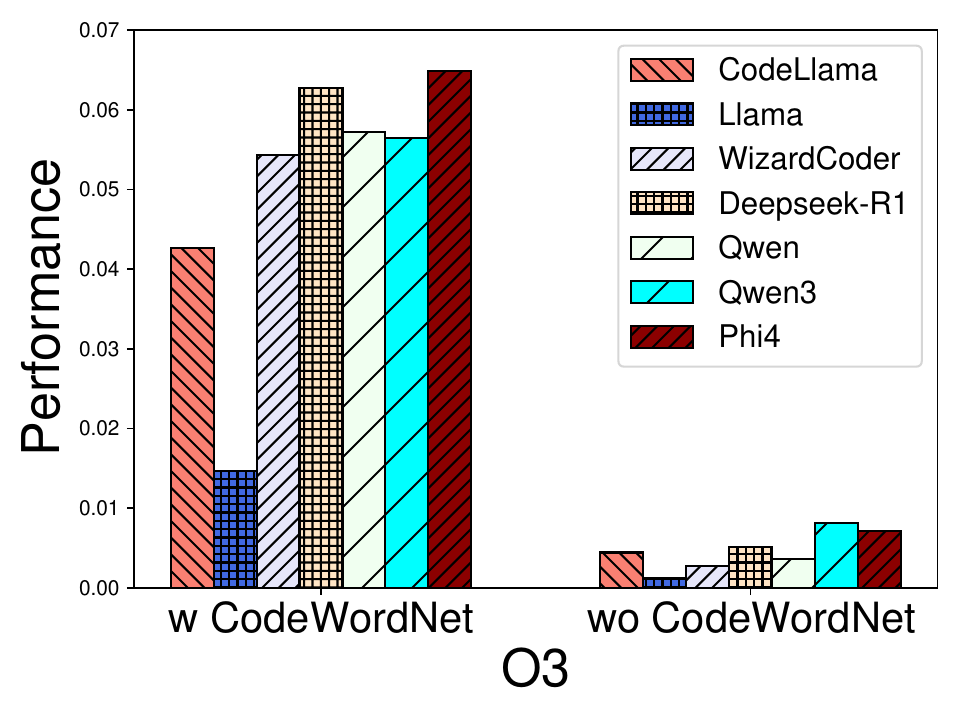}
     \caption{Variable Name Recovery}
     \label{fig:ablationvar}
   \end{subfigure}
   \caption{F1 Score on Function and Variable Name Recovery with and without CodeWordNet in x64 Architecture via Decompiled Code.}
   \label{fig:ablation}
\end{figure*}}

\begin{figure*}[!tb]
\centering
    
   \begin{subfigure}{0.99\textwidth}
     \centering
     \includegraphics[width=0.24\linewidth]{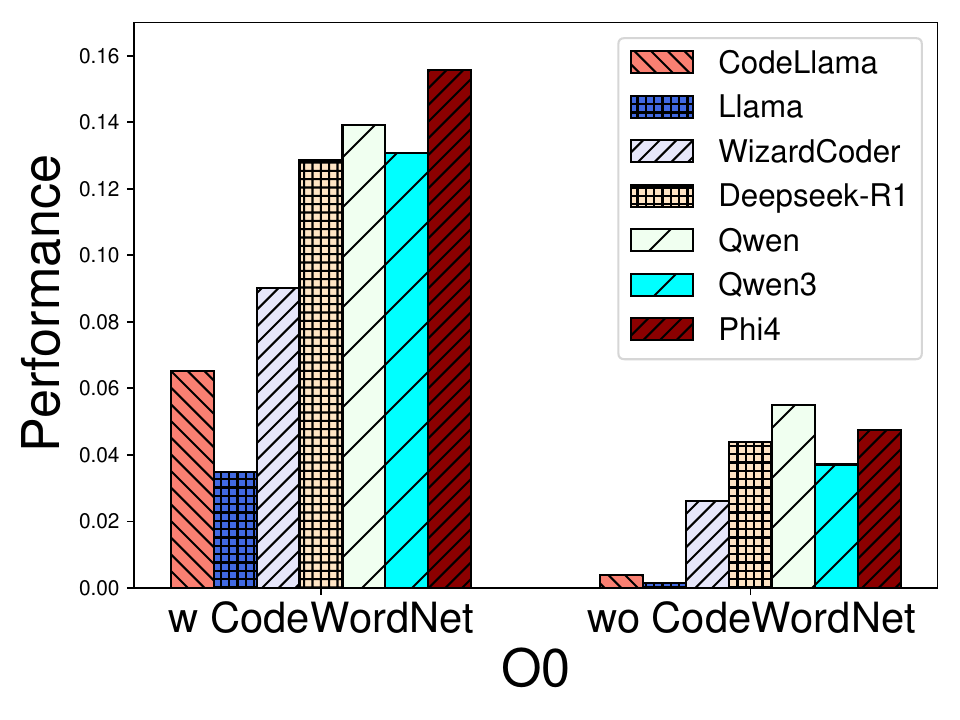}
     \includegraphics[width=0.24\linewidth]{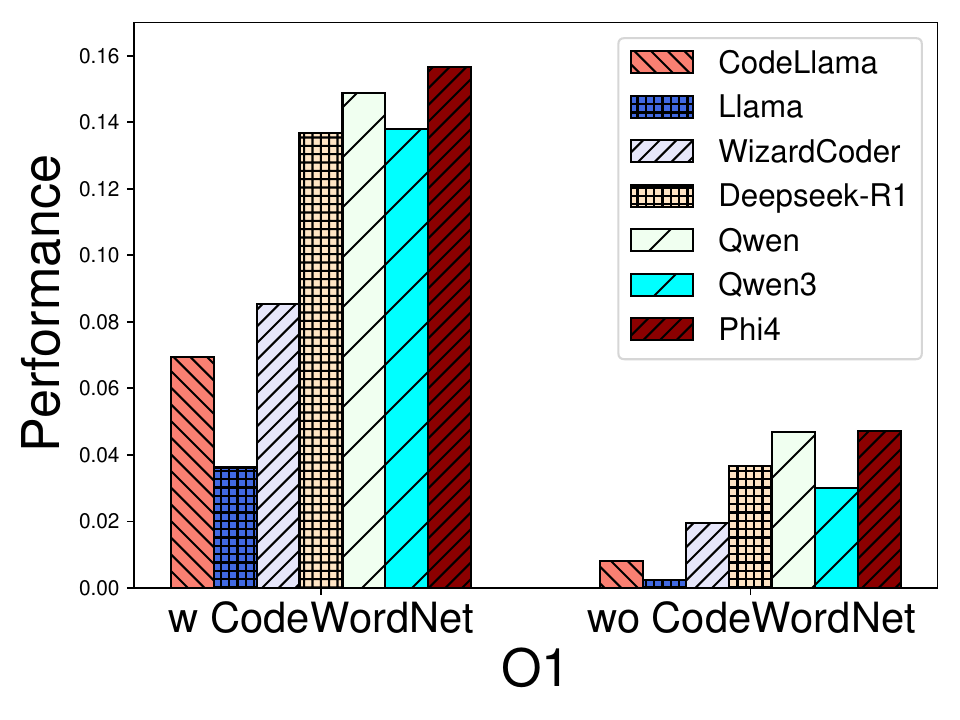}
     \includegraphics[width=0.24\linewidth]{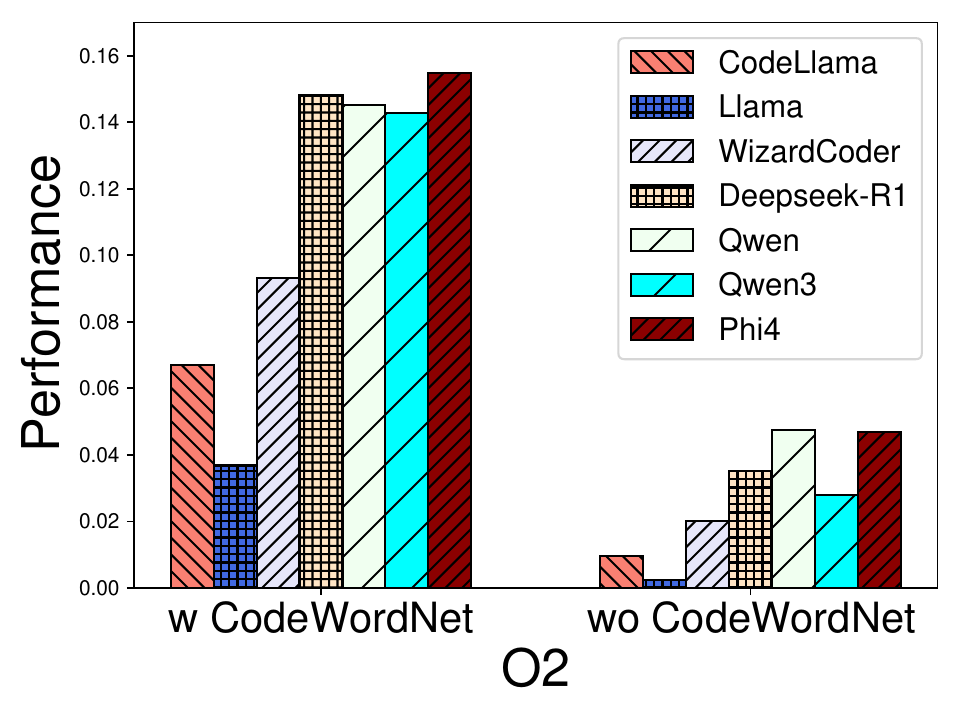}
     \includegraphics[width=0.24\linewidth]{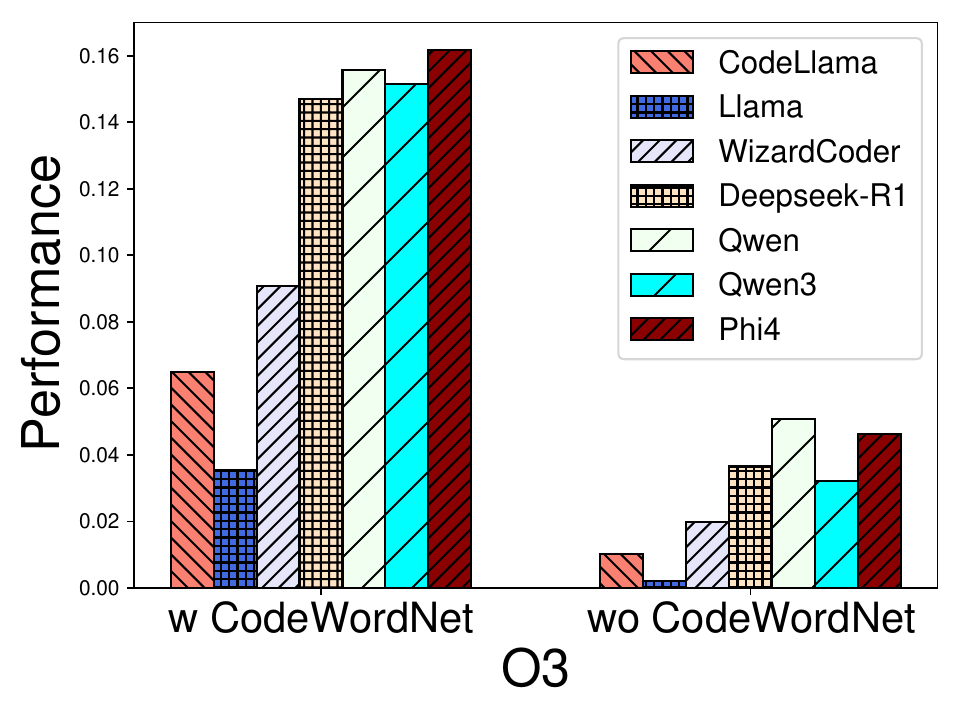}
     \vspace{-0.1in}
     \caption{Function Name Recovery}
     \label{fig:ablationfunc}
   \end{subfigure}
   \begin{subfigure}{0.99\textwidth}
     \centering
     \includegraphics[width=0.24\linewidth]{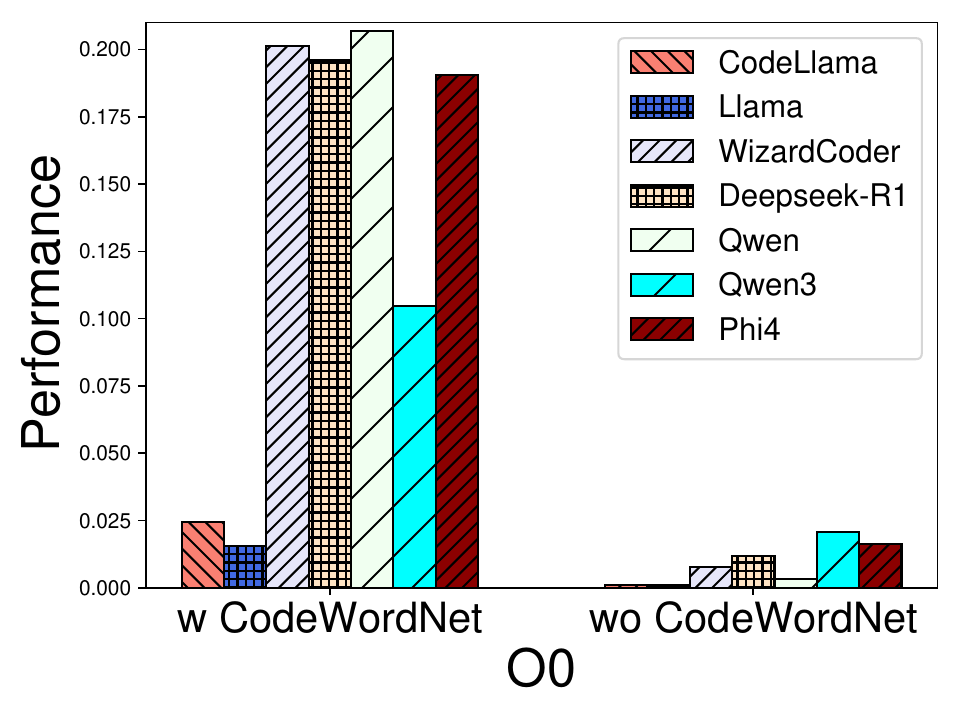}
     \includegraphics[width=0.24\linewidth]{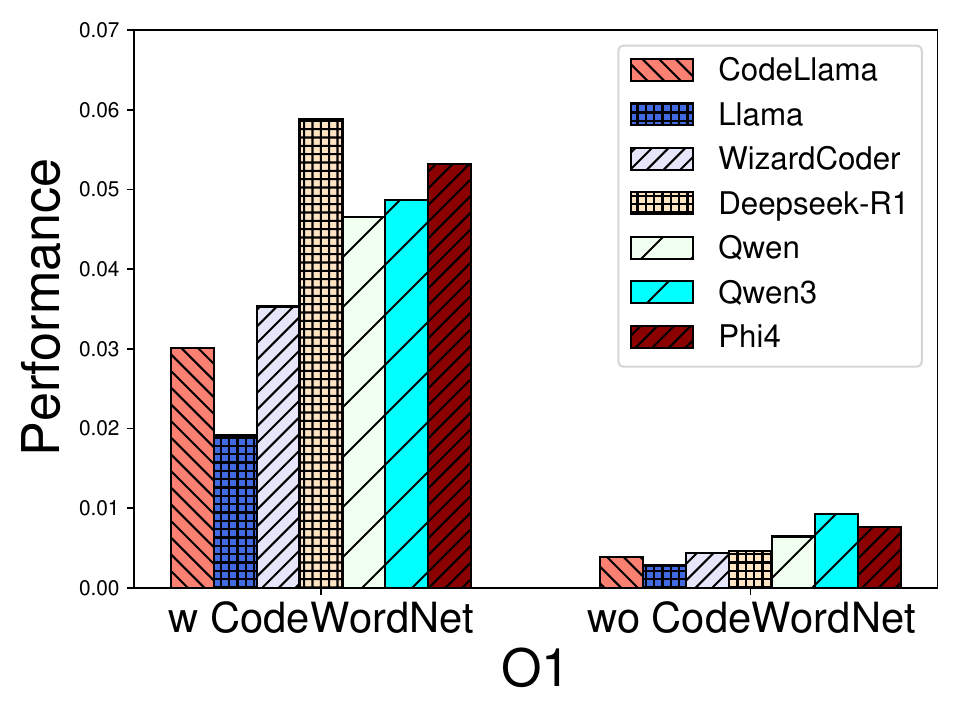}
     \includegraphics[width=0.24\linewidth]{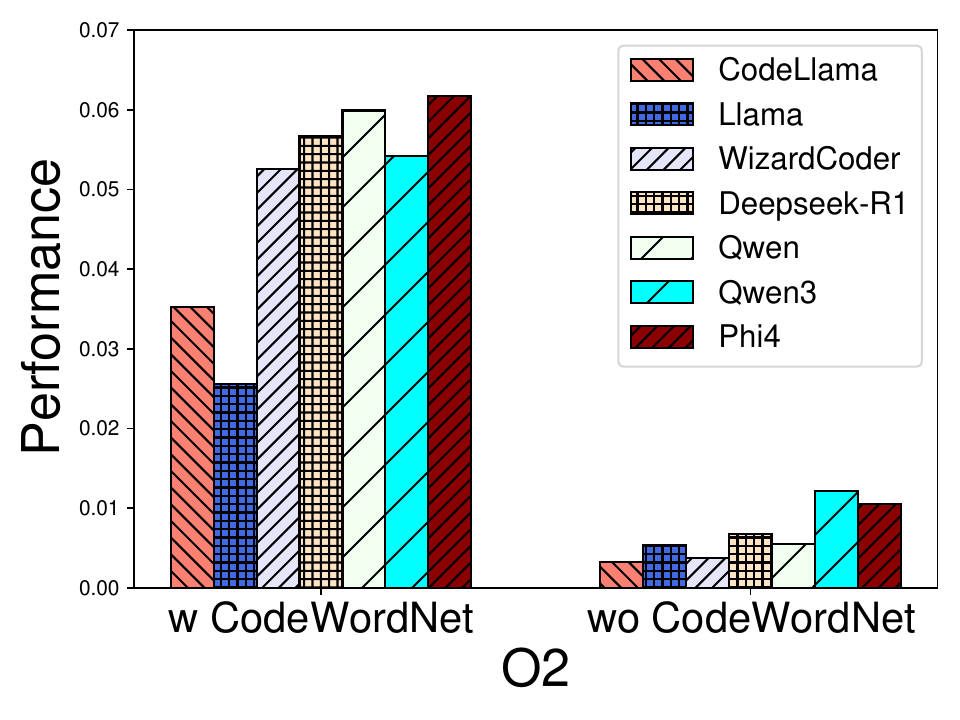}
     \includegraphics[width=0.24\linewidth]{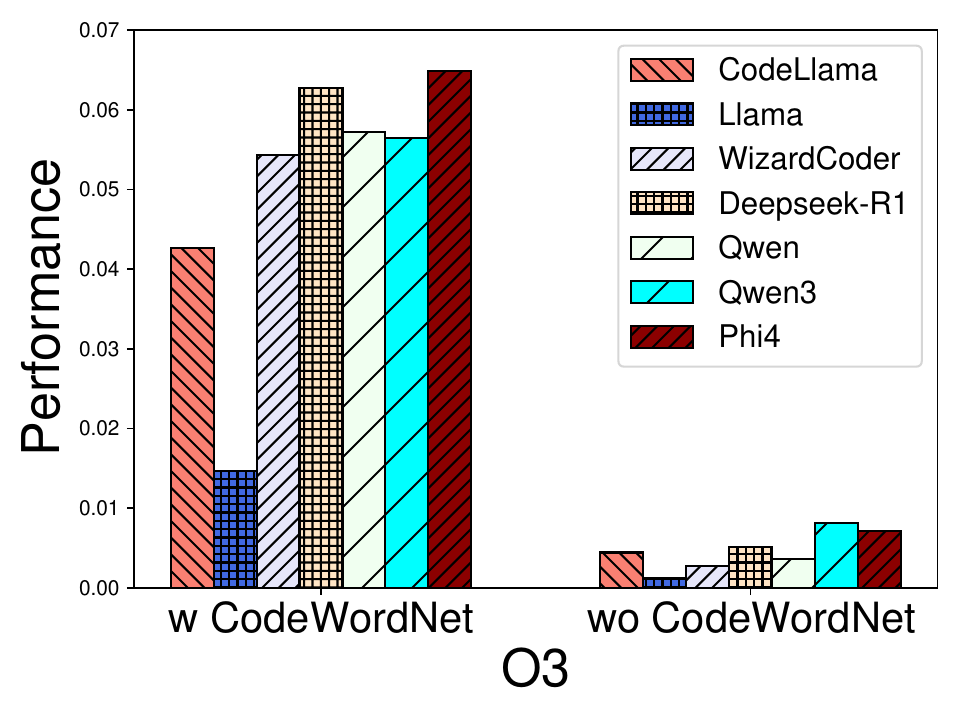}
     \vspace{-0.1in}
     \caption{Variable Name Recovery}
     \vspace{-0.1in}
     \label{fig:ablationvar}
   \end{subfigure}
   \caption{F1 Score on Function and Variable Name Recovery with and without CodeWordNet in X64 Architecture via Decompiled Code.}
   \label{fig:ablation}
\end{figure*}

%% file: table/diff_size.tex
\ignore{
\begin{table*}
\centering
\footnotesize
\scalebox{0.72}{
\begin{tabular}{lllll|lll|lll|lll} 
\hline
\multicolumn{1}{c}{\multirow{3}{*}{\textbf{Model}}} & \multirow{3}{*}{\textbf{Size}} & \multicolumn{12}{c}{\textbf{Performance}}                                                                                                                                                                                                                                                                                                                                                                                                         \\ 
\cline{3-14}
\multicolumn{1}{c}{}                                &                                & \multicolumn{3}{c|}{\textbf{O0}}                                                                           & \multicolumn{3}{c|}{\textbf{O1}}                                                                           & \multicolumn{3}{c|}{\textbf{O2}}                                                                           & \multicolumn{3}{c}{\textbf{O3}}                                                                            \\ 
\cline{3-14}
\multicolumn{1}{c}{}                                &                                & \multicolumn{1}{c}{\textbf{Prec}} & \multicolumn{1}{c}{\textbf{Recall}} & \multicolumn{1}{c|}{\textbf{F1}} & \multicolumn{1}{c}{\textbf{Prec}} & \multicolumn{1}{c}{\textbf{Recall}} & \multicolumn{1}{c|}{\textbf{F1}} & \multicolumn{1}{c}{\textbf{Prec}} & \multicolumn{1}{c}{\textbf{Recall}} & \multicolumn{1}{c|}{\textbf{F1}} & \multicolumn{1}{c}{\textbf{Prec}} & \multicolumn{1}{c}{\textbf{Recall}} & \multicolumn{1}{c}{\textbf{F1}}  \\ 
\hline
\multicolumn{14}{c}{Function Name Recovery}                                                                                                                                                                                                                                                                                                                                                                                                                                                                                              \\ 
\hline
\multirow{3}{*}{CodeLlama}                          & 7B                             & 0.23                              & 0.16                                & 0.19                             & 0.28                              & 0.18                                & 0.22                             & 0.26                              & 0.17                                & 0.2                              & 0.24                              & 0.17                                & 0.2                              \\
                                                    & 13B                            & 0.33(\textcolor{blue}{$\uparrow$}43.5\%)                      & 0.32(\textcolor{blue}{$\uparrow$}100.0\%)                       & 0.33(\textcolor{blue}{$\uparrow$}73.7\%)                     & 0.37(\textcolor{blue}{$\uparrow$}32.1\%)                      & 0.36(\textcolor{blue}{$\uparrow$}100.0\%)                       & 0.37(\textcolor{blue}{$\uparrow$}68.2\%)                     & 0.35(\textcolor{blue}{$\uparrow$}34.6\%)                      & 0.34(\textcolor{blue}{$\uparrow$}100.0\%)                       & 0.35(\textcolor{blue}{$\uparrow$}75.0\%)                     & 0.34(\textcolor{blue}{$\uparrow$}41.7\%)                      & 0.34(\textcolor{blue}{$\uparrow$}100.0\%)                       & 0.34(\textcolor{blue}{$\uparrow$}70.0\%)                     \\
                                                    & 34B                            & 0.24(\textcolor{blue}{$\uparrow$}4.3\%)                       & 0.16(-)                         & 0.19(-)                      & 0.27(\textcolor{red}{$\downarrow$}3.6\%)                      & 0.15(\textcolor{red}{$\downarrow$}16.7\%)                       & 0.19(\textcolor{red}{$\downarrow$}13.6\%)                    & 0.28(\textcolor{blue}{$\uparrow$}7.7\%)                       & 0.14(\textcolor{red}{$\downarrow$}17.6\%)                       & 0.19(\textcolor{red}{$\downarrow$}5.0\%)                     & 0.26(\textcolor{blue}{$\uparrow$}8.3\%)                       & 0.14(\textcolor{red}{$\downarrow$}17.6\%)                       & 0.18(\textcolor{red}{$\downarrow$}10.0\%)                    \\ 
\hline
\multicolumn{14}{c}{Variable Name Recovery}                                                                                                                                                                                                                                                                                                                                                                                                                                                                                              \\ 
\hline
\multirow{3}{*}{CodeLlama}                          & 7B                             & 0.06                              & 0.04                                & 0.05                             & 0.06                              & 0.04                                & 0.05                             & 0.06                              & 0.04                                & 0.05                             & 0.07                              & 0.04                                & 0.05                             \\
                                                    & 13B                            & 0.11(\textcolor{blue}{$\uparrow$}83.3\%)                      & 0.11(\textcolor{blue}{$\uparrow$}175.0\%)                       & 0.11(\textcolor{blue}{$\uparrow$}120.0\%)                    & 0.14(\textcolor{blue}{$\uparrow$}133.3\%)                     & 0.14(\textcolor{blue}{$\uparrow$}250.0\%)                       & 0.14(\textcolor{blue}{$\uparrow$}180.0\%)                    & 0.14(\textcolor{blue}{$\uparrow$}133.3\%)                     & 0.14(\textcolor{blue}{$\uparrow$}250.0\%)                       & 0.14(\textcolor{blue}{$\uparrow$}180.0\%)                    & 0.14(\textcolor{blue}{$\uparrow$}100.0\%)                     & 0.14(\textcolor{blue}{$\uparrow$}250.0\%)                       & 0.14(\textcolor{blue}{$\uparrow$}180.0\%)                    \\
                                                    & 34B                            & 0.05(\textcolor{red}{$\downarrow$}16.7\%)                     & 0.04(-)                         & 0.04(\textcolor{red}{$\downarrow$}20.0\%)                    & 0.08(\textcolor{blue}{$\uparrow$}33.3\%)                      & 0.04(-)                         & 0.06(\textcolor{blue}{$\uparrow$}20.0\%)                     & 0.08(\textcolor{blue}{$\uparrow$}33.3\%)                      & 0.05(\textcolor{blue}{$\uparrow$}25.0\%)                        & 0.06(\textcolor{blue}{$\uparrow$}20.0\%)                     & 0.08(\textcolor{blue}{$\uparrow$}14.3\%)                      & 0.05(\textcolor{blue}{$\uparrow$}25.0\%)                        & 0.06(\textcolor{blue}{$\uparrow$}20.0\%)                     \\ 
\hline
\multicolumn{14}{c}{Type Inference}                                                                                                                                                                                                                                                                                                                                                                                                                                                                                                      \\ 
\hline
\multirow{3}{*}{CodeLlama}                          & 7B                             & 0.08                              & 0.05                                & 0.06                             & 0.1                               & 0.06                                & 0.07                             & 0.1                               & 0.05                                & 0.07                             & 0.1                               & 0.06                                & 0.07                             \\
                                                    & 13B                            & 0.13(\textcolor{blue}{$\uparrow$}62.5\%)                      & 0.11(\textcolor{blue}{$\uparrow$}120.0\%)                       & 0.12(\textcolor{blue}{$\uparrow$}100.0\%)                    & 0.15(\textcolor{blue}{$\uparrow$}50.0\%)                      & 0.12(\textcolor{blue}{$\uparrow$}100.0\%)                       & 0.13(\textcolor{blue}{$\uparrow$}85.7\%)                     & 0.15(\textcolor{blue}{$\uparrow$}50.0\%)                      & 0.12(\textcolor{blue}{$\uparrow$}140.0\%)                       & 0.14(\textcolor{blue}{$\uparrow$}100.0\%)                    & 0.15\textcolor{blue}{$\uparrow$}50.0\%)                      & 0.12(\textcolor{blue}{$\uparrow$}100.0\%)                       & 0.13(\textcolor{blue}{$\uparrow$}85.7\%)                     \\
                                                    & 34B                            & 0.09(\textcolor{blue}{$\uparrow$}12.5\%)                      & 0.06(\textcolor{blue}{$\uparrow$}20.0\%)                        & 0.07(\textcolor{blue}{$\uparrow$}16.7\%)                     & 0.11(\textcolor{blue}{$\uparrow$}10.0\%)                      & 0.05(\textcolor{red}{$\downarrow$}16.7\%)                       & 0.07(-)                      & 0.11(10.0\%)                      & 0.05(-)                         & 0.07(-)                      & 0.11(\textcolor{blue}{$\uparrow$}10.0\%)                      & 0.05(\textcolor{red}{$\downarrow$}16.7\%)                       & 0.07(-)                      \\
\hline
\end{tabular}}
\caption{Performance Evaluation with Different Size of CodeLlama Models in X64 Architecture.}
\label{tbl:diffsize}
\end{table*}}

\ignore{
    \begin{subfigure}{0.95\textwidth}
     \includegraphics[width=0.33\linewidth]{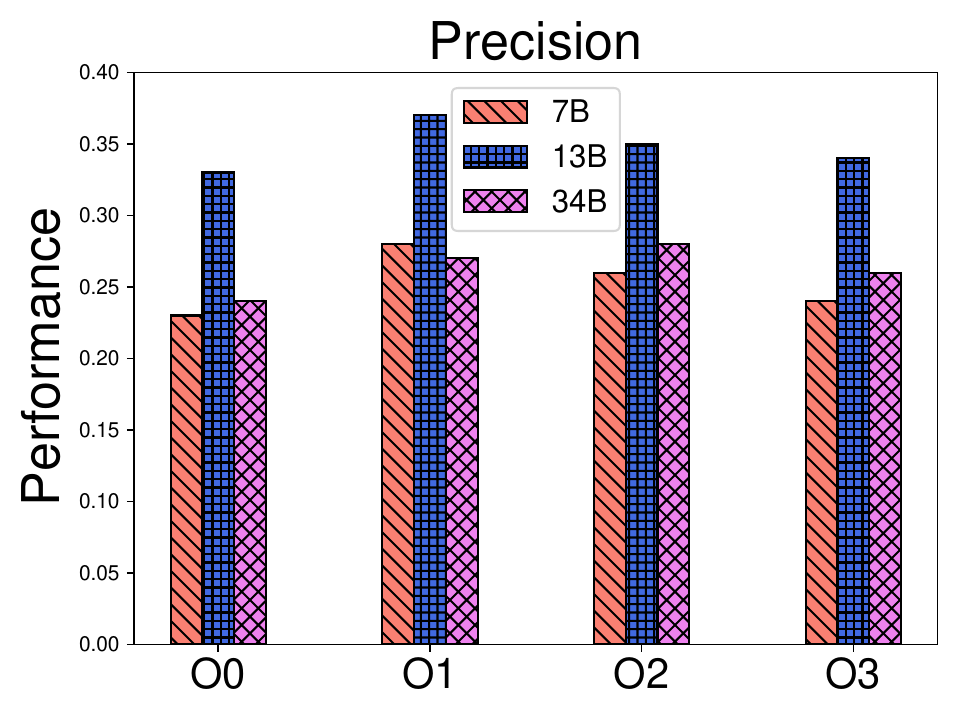}
     \includegraphics[width=0.33\linewidth]{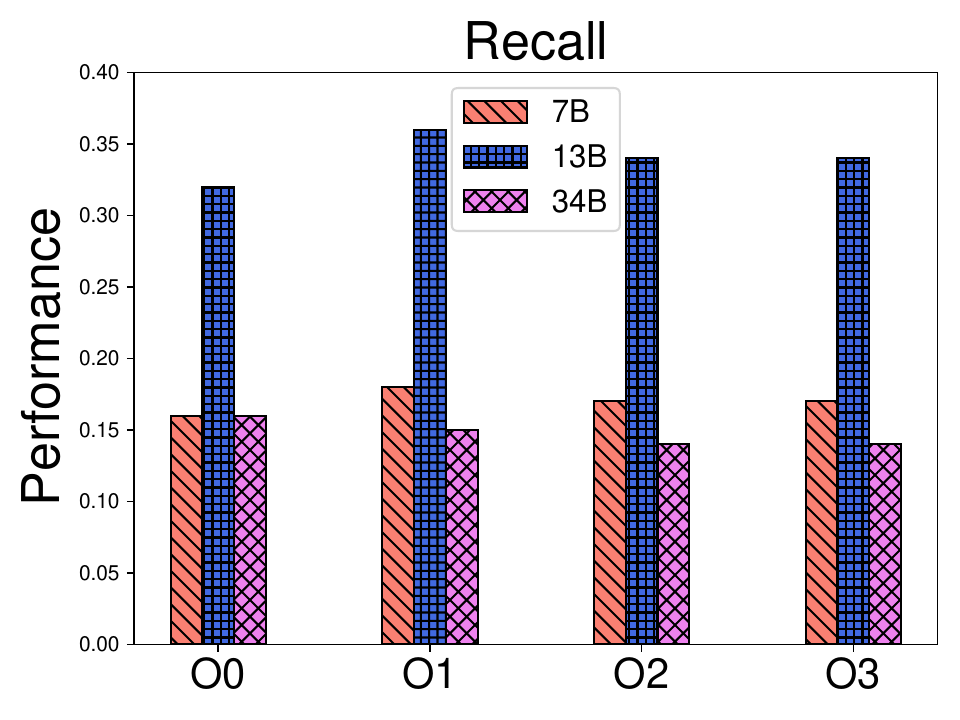}
     \includegraphics[width=0.33\linewidth]{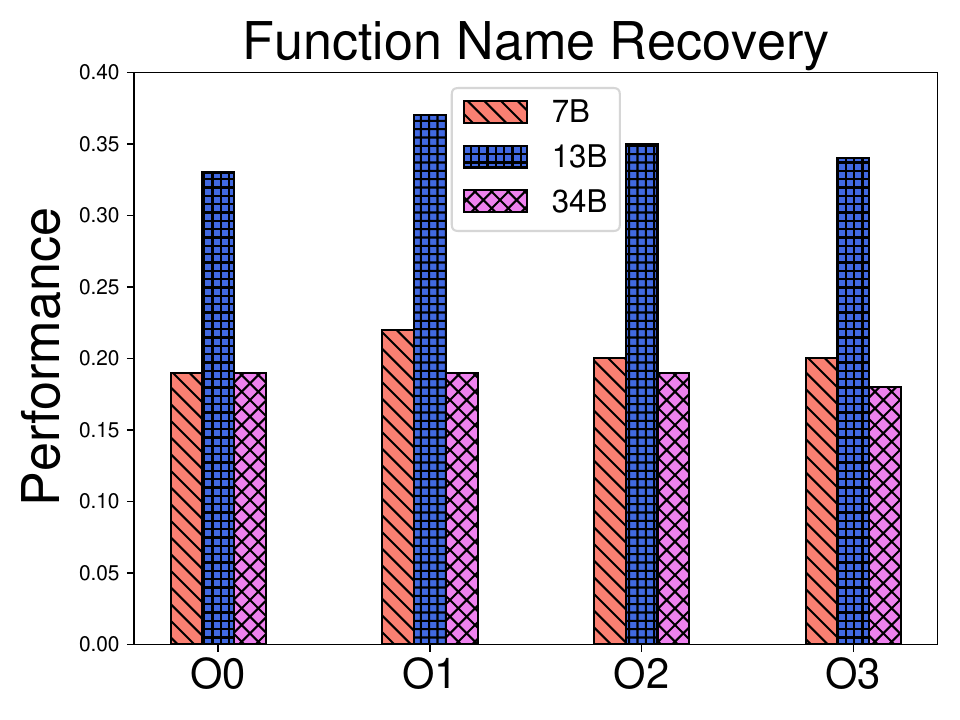}
     \caption{Function name recovery}
   \end{subfigure}
   \begin{subfigure}{0.95\textwidth}
     \includegraphics[width=0.33\linewidth]{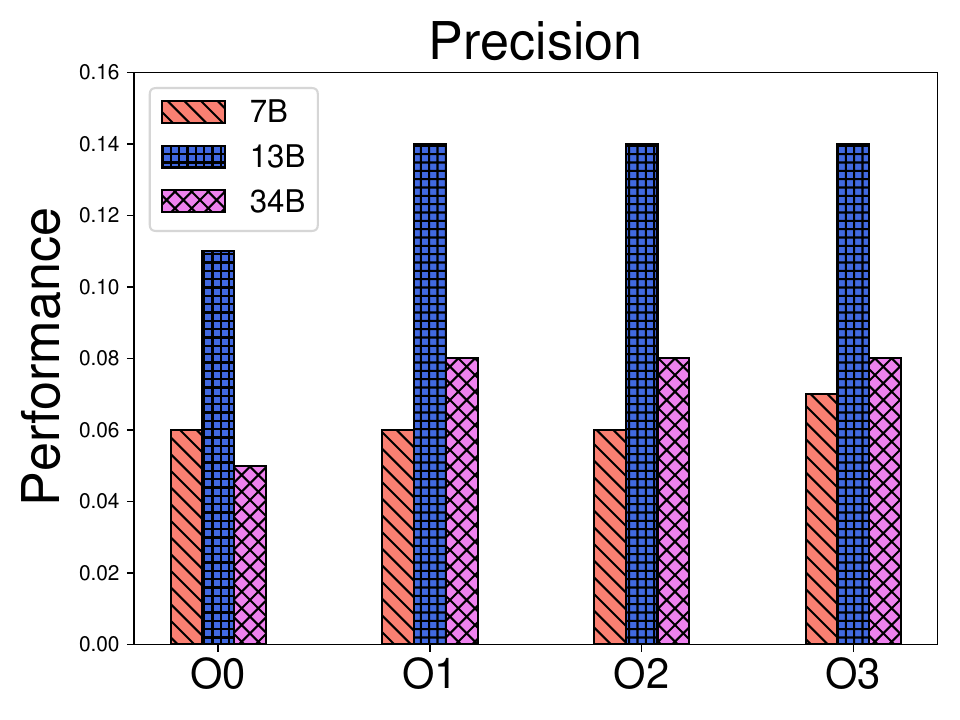}
     \includegraphics[width=0.33\linewidth]{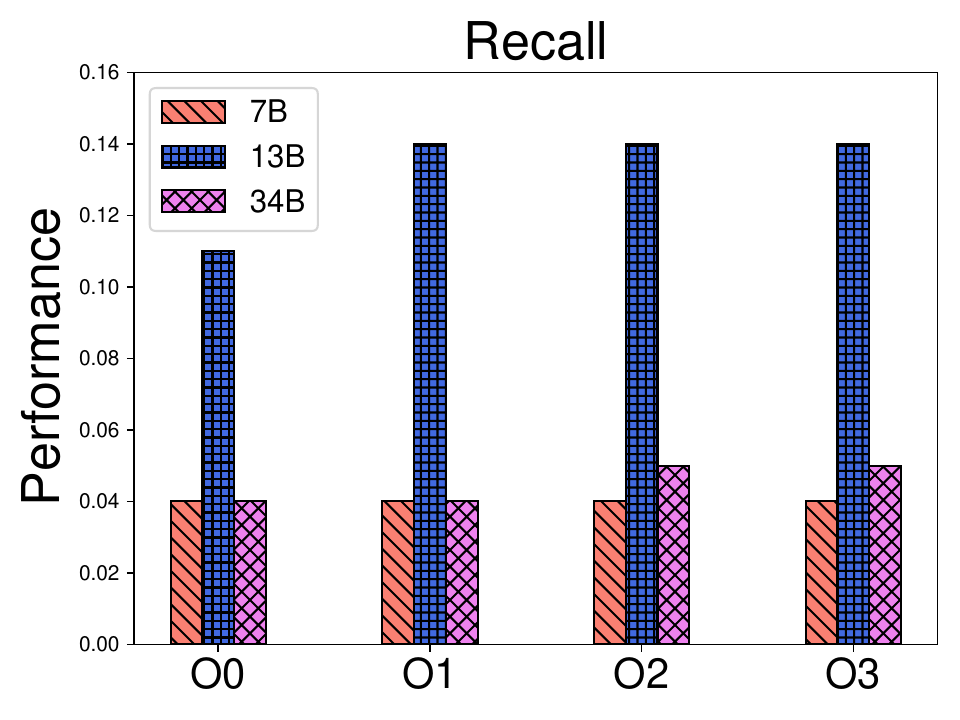}
    \includegraphics[width=0.33\linewidth]{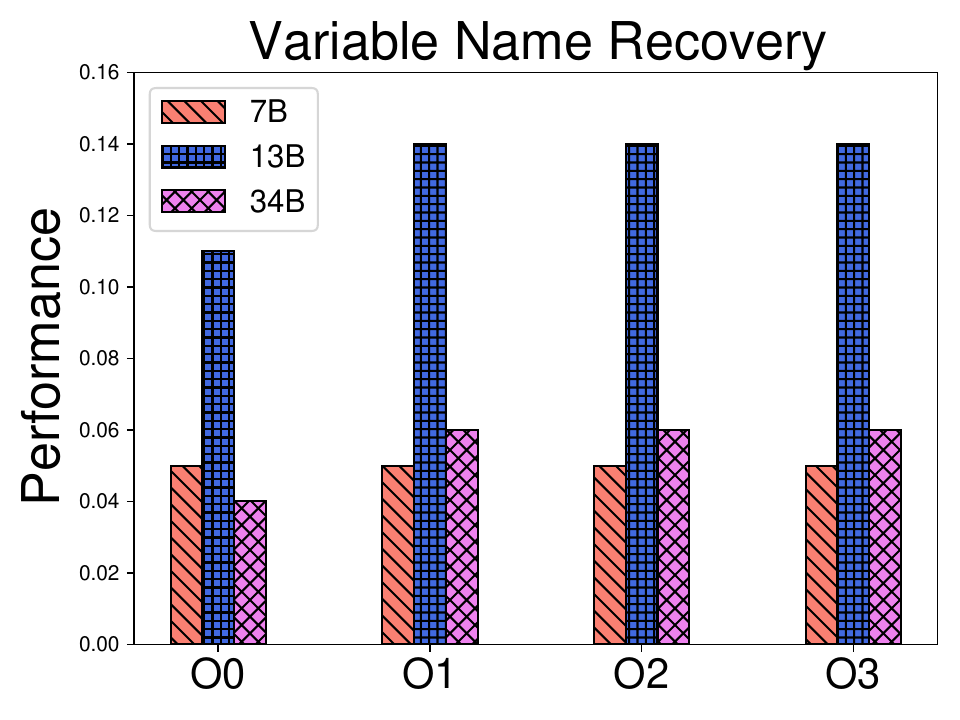}
     \caption{Variable name recovery}
   \end{subfigure} 
}

\begin{figure*}[!tb]
\centering
   \begin{subfigure}{0.99\textwidth}
   \centering
       \includegraphics[width=0.24\linewidth]{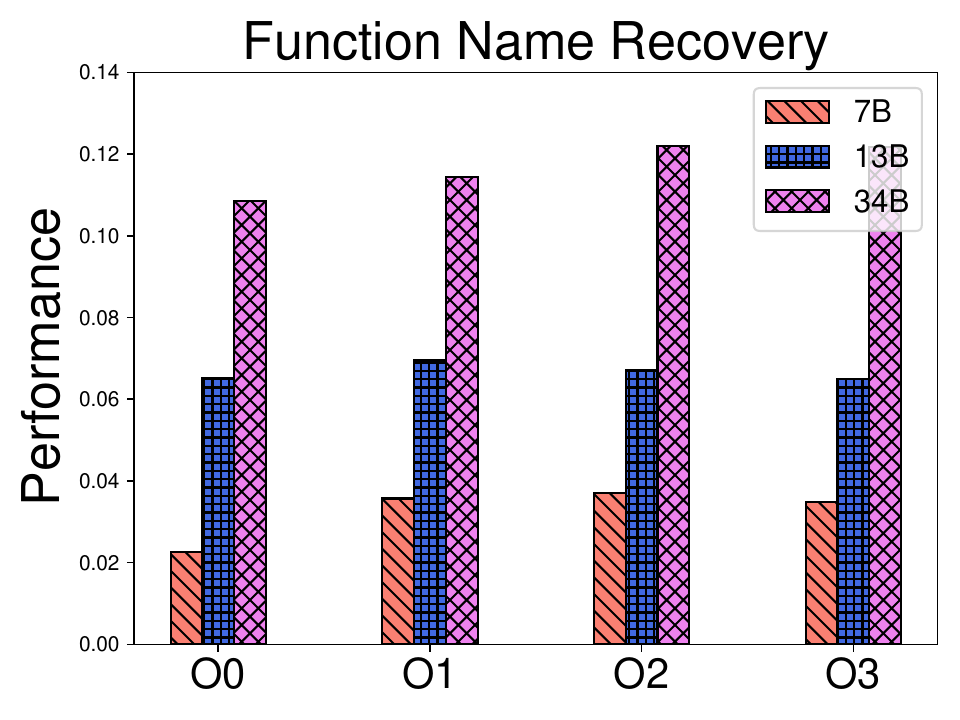}
       \hspace{0.2in}
     \includegraphics[width=0.24\linewidth]{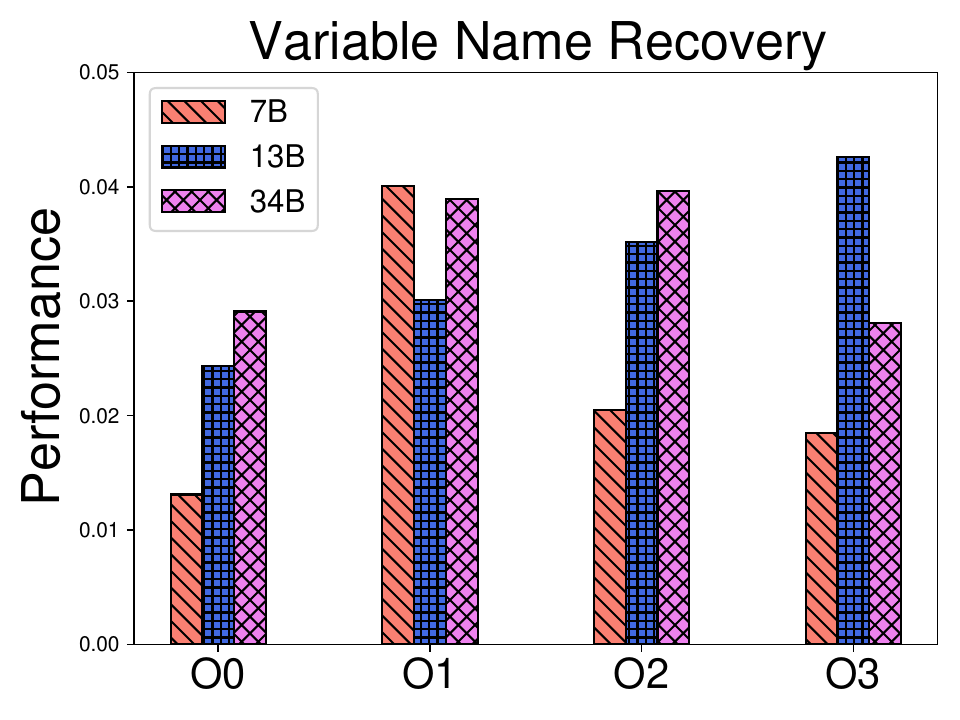}
     \hspace{0.2in}
    \includegraphics[width=0.24\linewidth]{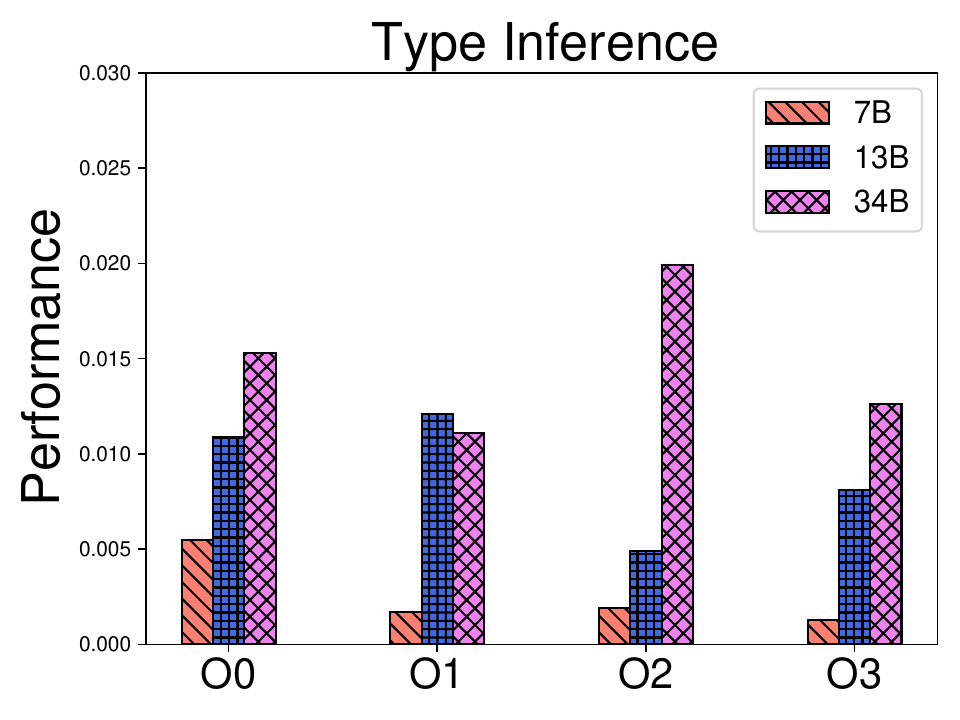}
   \end{subfigure}
   \vspace{-0.15in}
   \caption{F1 Score of Name Recovery and Type Inference with Different Sizes of CodeLlama in x64 Architecture.}
   \label{fig:diffsize}
\end{figure*}

%% file: table/finetune.tex
\begin{table}[tb!]
\caption{Time Spent on the Fine-tuning Process}
\label{tbl:finetime}
\footnotesize
\centering
\scalebox{0.87}{
\begin{tabular}{lr} 
\hline
\textbf{Model} & \textbf{Fine-tuning Time (s/200 steps)}  \\
\hline
CodeLlama      & $5,580$                                  \\
Llama2         & $5,640$                                  \\
WizardCoder    & $4,620$                                  \\
Deepseek-R1    & $9,540$                                  \\
Qwen2.5           & $6,000$                                  \\
Qwen3           & $6,060$                                  \\
Phi4         & $5,880$                                  \\
\hline
\end{tabular}}
\end{table}

%% file: paper/08_future_work.tex
In this section, we discuss the current \sys and outline potential directions for future development. 
%

\parag{Expanding the Scope of Tasks}
While our current evaluation focuses on name recovery and type inference, binary analysis encompasses a much broader array of challenges. Future iterations of \sys will integrate benchmarks for memory corruption detection, control-flow integrity (CFI) analysis, and binary similarity detection. By expanding the task set, we aim to provide a more holistic view of LLM capabilities in security-critical contexts.

\parag{Model Generalization and Data Leakage}
A common concern in LLM-based studies is data leakage from the pre-training or fine-tuning phases. We argue that the \sys pipeline significantly mitigates this risk. By transforming source code through both decompilation and normalization, the resulting input format differs substantially from the original source found in public repositories. This makes it highly unlikely that the model relies on memorized code during evaluation. Moreover, we plan to extend our analysis to a wider variety of models to determine if these performance trends are universal across different architectures.

\parag{Decompiler Dependencies and Limitations}
\sys relies on decompiler outputs for generating its universal representations. However, decompilers are inherently imperfect. We have observed instances where tools like Ghidra introduce artifacts not present in the source, such as stack canaries, or omit vital type annotations. These errors can propagate through the preprocessing stage and affect symbol replacement. To improve robustness, we are investigating ``decompiler-agnostic'' input formats and plan to incorporate additional tools like IDA Pro and angr to identify which decompilers are most ``LLM-friendly''.

\parag{Advanced Deduplication and Maintenance}
Current deduplication in \sys relies on direct comparison of assembly and decompiled code. However, functions that are semantically identical can appear different due to compiler optimizations or varying memory addresses. We intend to implement more sophisticated, semantics-driven deduplication techniques to further reduce the risk of data leakage. Finally, to ensure the long-term utility of the benchmark, we will maintain \sys with regular updates, including real-world firmware and binaries subjected to advanced obfuscation.

\ignore{
\parag{Reverse Engineering Tasks} 
Our current work primarily focuses on two core tasks: name recovery and type inference. 
However, many other reverse engineering challenges remain unexplored, such as memory corruption detection, control-flow integrity analysis, and binary similarity detection. 
In future iterations, we plan to extend \sys to cover a broader set of tasks, thereby enabling more comprehensive evaluation of LLMs in reverse engineering.

\parag{Further Analysis}
Since we apply \sys to a single model, we expand it to multiple models to verify whether every LLM has trouble in performing the given tasks. 
This can provide us insights about understanding capabilities of current LLMs in various reverse engineering tasks. 

\parag{Data Leakage}
It is reasonable to argue that \sys may suffer from data leakage from pre-training or fine-tuning process.
However, since \sys converts the source code through two processes, decompilation and normalization, which results in different formats from the original code.
That is, the functions of \sys are very unlikely utilized during the early stage, which prevent data leakage. 

%
%

%
%
%

\parag{Decompilers} 
Like other reverse engineering benchmarks, \sys relies heavily on decompiler output for generating assembly and decompiled code representations. 
However, decompilers are imperfect and occasionally fail to reconstruct variables or types accurately. 
For instance, we observed cases where decompilers introduced undefined variables not present in the original source code (e.g., stack canaries), or omitted certain type annotations, which then affected symbol replacement during preprocessing. 
These limitations highlight the need for complementary approaches that either refine decompiler outputs or generate input formats independent of a single decompiler’s quality.

In practice, we also encountered instances where Ghidra failed to generate decompiled code because the decompilation process became stuck. 
Moreover, discrepancies can arise across different versions of decompilers: results may vary, and scripts designed for one version may no longer function in another. 
To mitigate these issues, we plan to continuously verify and update our scripts to ensure compatibility and correctness over time.

Furthermore, we generate our universal format through two of the most popular decompilers, Ghidra and IDA Pro.
However, since there are more decompliers, such as angr~\cite{wang2017angr}, it could be interesting to find out which decompiler is LLM-friendly. 

\parag{Dataset Maintenance and Expansion}
Datasets must evolve continuously to remain useful. 
We plan to maintain \sys with regular updates, including new projects, real-world firmware, and binaries compiled under additional compilation strategies such as obfuscation. 
Such extensions will broaden \sys’s applicability to security-critical domains, including malware detection, firmware analysis, and obfuscation-resilient reverse engineering.

\parag{Deduplication}
Currently, \sys removes duplicate functions by comparing decompiled and assembly code. 
However, semantically identical functions can still appear different due to factors such as memory address differences, compiler optimizations, or naming variations. 
More sophisticated deduplication techniques are therefore required, including normalization of decompiled code and function metadata. This will reduce the risk of data leakage between training and test sets.

Additionally, some binaries include calls to well-known external library functions. 
In such cases, Ghidra represents them as external calls, and their decompiled code often consists of only one or two instructions. 
We measure the lines of the decompiled code to detect these external functions.
Since this method can 
}


%% file: paper/10_conclusion.tex
In recent years, numerous reverse engineering approaches have been proposed, leveraging techniques ranging from traditional machine learning to modern LLMs. 
In particular, LLM-based methods often rely on fine-tuning existing models for specialized tasks. 
However, the field currently lacks a standardized benchmark for evaluating both pre-trained and fine-tuned LLMs on reverse engineering tasks, largely due to the absence of comprehensive datasets.
To address this gap, we introduce \sys, a large-scale benchmark dataset comprising 96 projects compiled across multiple architectures and optimization levels. 
\sys enables systematic measurement of LLM performance on a variety of reverse engineering tasks. 
A key feature of the benchmark is its transformation of executable binaries into a universal, human-readable format, supported by a knowledge base that preserves byte-level stack layout information. 
Importantly, \sys does not oversimplify reverse engineering challenges, ensuring that evaluation remains realistic and reflective of real-world conditions.
We evaluate various LLMs with \sys, and it demonstrates the difficulties of LLM for given tasks. 

%
%
%
%

%% file: paper.bbl

\begin{thebibliography}{41}


\ifx \showCODEN    \undefined \def \showCODEN     #1{\unskip}     \fi
\ifx \showISBNx    \undefined \def \showISBNx     #1{\unskip}     \fi
\ifx \showISBNxiii \undefined \def \showISBNxiii  #1{\unskip}     \fi
\ifx \showISSN     \undefined \def \showISSN      #1{\unskip}     \fi
\ifx \showLCCN     \undefined \def \showLCCN      #1{\unskip}     \fi
\ifx \shownote     \undefined \def \shownote      #1{#1}          \fi
\ifx \showarticletitle \undefined \def \showarticletitle #1{#1}   \fi
\ifx \showURL      \undefined \def \showURL       {\relax}        \fi
\providecommand\bibfield[2]{#2}
\providecommand\bibinfo[2]{#2}
\providecommand\natexlab[1]{#1}
\providecommand\showeprint[2][]{arXiv:#2}

\bibitem[pop(2023)]%
        {popularllm}
 \bibinfo{year}{2023}\natexlab{}.
\newblock \bibinfo{title}{{Large Language Model for Software Engineering}}.
\newblock
\newblock
\shownote{\url{https://github.com/gai4se/LLM4SE?tab=readme-ov-file\#model-list}}.


\bibitem[tra(2023)]%
        {transformerAPI}
 \bibinfo{year}{2023}\natexlab{}.
\newblock \bibinfo{title}{{Python Transformers}}.
\newblock
\newblock
\shownote{\url{https://pypi.org/project/transformers/}}.


\bibitem[AI(2024)]%
        {unsloth}
\bibfield{author}{\bibinfo{person}{Unsloth AI}.} \bibinfo{year}{2024}\natexlab{}.
\newblock \bibinfo{title}{Unsloth Phi-4}.
\newblock
\urldef\tempurl%
\url{https://huggingface.co/unsloth/phi-4}
\showURL{%
\tempurl}


\bibitem[Banerjee et~al\mbox{.}(2021)]%
        {banerjee2021variable}
\bibfield{author}{\bibinfo{person}{Pratyay Banerjee}, \bibinfo{person}{Kuntal~Kumar Pal}, \bibinfo{person}{Fish Wang}, {and} \bibinfo{person}{Chitta Baral}.} \bibinfo{year}{2021}\natexlab{}.
\newblock \showarticletitle{Variable name recovery in decompiled binary code using constrained masked language modeling}.
\newblock \bibinfo{journal}{\emph{arXiv preprint arXiv:2103.12801}} (\bibinfo{year}{2021}).
\newblock


\bibitem[Chen et~al\mbox{.}(2022)]%
        {chen2022augmenting}
\bibfield{author}{\bibinfo{person}{Qibin Chen}, \bibinfo{person}{Jeremy Lacomis}, \bibinfo{person}{Edward~J Schwartz}, \bibinfo{person}{Claire Le~Goues}, \bibinfo{person}{Graham Neubig}, {and} \bibinfo{person}{Bogdan Vasilescu}.} \bibinfo{year}{2022}\natexlab{}.
\newblock \showarticletitle{Augmenting decompiler output with learned variable names and types}. In \bibinfo{booktitle}{\emph{31st USENIX Security Symposium (USENIX Security 22)}}. \bibinfo{pages}{4327--4343}.
\newblock


\bibitem[David et~al\mbox{.}(2020)]%
        {david2020neural}
\bibfield{author}{\bibinfo{person}{Yaniv David}, \bibinfo{person}{Uri Alon}, {and} \bibinfo{person}{Eran Yahav}.} \bibinfo{year}{2020}\natexlab{}.
\newblock \showarticletitle{Neural reverse engineering of stripped binaries using augmented control flow graphs}.
\newblock \bibinfo{journal}{\emph{Proceedings of the ACM on Programming Languages}} \bibinfo{volume}{4}, \bibinfo{number}{OOPSLA} (\bibinfo{year}{2020}), \bibinfo{pages}{1--28}.
\newblock


\bibitem[Dettmers et~al\mbox{.}(2023)]%
        {dettmers2023qlora}
\bibfield{author}{\bibinfo{person}{Tim Dettmers}, \bibinfo{person}{Artidoro Pagnoni}, \bibinfo{person}{Ari Holtzman}, {and} \bibinfo{person}{Luke Zettlemoyer}.} \bibinfo{year}{2023}\natexlab{}.
\newblock \showarticletitle{Qlora: Efficient finetuning of quantized llms}.
\newblock \bibinfo{journal}{\emph{Advances in neural information processing systems}}  \bibinfo{volume}{36} (\bibinfo{year}{2023}), \bibinfo{pages}{10088--10115}.
\newblock


\bibitem[Duan et~al\mbox{.}(2020)]%
        {duan2020deepbindiff}
\bibfield{author}{\bibinfo{person}{Yue Duan}, \bibinfo{person}{Xuezixiang Li}, \bibinfo{person}{Jinghan Wang}, {and} \bibinfo{person}{Heng Yin}.} \bibinfo{year}{2020}\natexlab{}.
\newblock \showarticletitle{Deepbindiff: Learning program-wide code representations for binary diffing}. In \bibinfo{booktitle}{\emph{Network and distributed system security symposium}}.
\newblock


\bibitem[Feitelson et~al\mbox{.}(2020)]%
        {feitelson2020developers}
\bibfield{author}{\bibinfo{person}{Dror~G Feitelson}, \bibinfo{person}{Ayelet Mizrahi}, \bibinfo{person}{Nofar Noy}, \bibinfo{person}{Aviad~Ben Shabat}, \bibinfo{person}{Or Eliyahu}, {and} \bibinfo{person}{Roy Sheffer}.} \bibinfo{year}{2020}\natexlab{}.
\newblock \showarticletitle{How developers choose names}.
\newblock \bibinfo{journal}{\emph{IEEE Transactions on Software Engineering}} \bibinfo{volume}{48}, \bibinfo{number}{1} (\bibinfo{year}{2020}), \bibinfo{pages}{37--52}.
\newblock


\bibitem[Feng et~al\mbox{.}(2020)]%
        {feng2020p2im}
\bibfield{author}{\bibinfo{person}{Bo Feng}, \bibinfo{person}{Alejandro Mera}, {and} \bibinfo{person}{Long Lu}.} \bibinfo{year}{2020}\natexlab{}.
\newblock \showarticletitle{$\{$P2IM$\}$: Scalable and hardware-independent firmware testing via automatic peripheral interface modeling}. In \bibinfo{booktitle}{\emph{29th USENIX Security Symposium (USENIX Security 20)}}. \bibinfo{pages}{1237--1254}.
\newblock


\bibitem[Guo et~al\mbox{.}(2025)]%
        {guo2025deepseek}
\bibfield{author}{\bibinfo{person}{Daya Guo}, \bibinfo{person}{Dejian Yang}, \bibinfo{person}{Haowei Zhang}, \bibinfo{person}{Junxiao Song}, \bibinfo{person}{Ruoyu Zhang}, \bibinfo{person}{Runxin Xu}, \bibinfo{person}{Qihao Zhu}, \bibinfo{person}{Shirong Ma}, \bibinfo{person}{Peiyi Wang}, \bibinfo{person}{Xiao Bi}, {et~al\mbox{.}}} \bibinfo{year}{2025}\natexlab{}.
\newblock \showarticletitle{Deepseek-r1: Incentivizing reasoning capability in llms via reinforcement learning}.
\newblock \bibinfo{journal}{\emph{arXiv preprint arXiv:2501.12948}} (\bibinfo{year}{2025}).
\newblock


\bibitem[He et~al\mbox{.}(2018)]%
        {he2018debin}
\bibfield{author}{\bibinfo{person}{Jingxuan He}, \bibinfo{person}{Pesho Ivanov}, \bibinfo{person}{Petar Tsankov}, \bibinfo{person}{Veselin Raychev}, {and} \bibinfo{person}{Martin Vechev}.} \bibinfo{year}{2018}\natexlab{}.
\newblock \showarticletitle{Debin: Predicting debug information in stripped binaries}. In \bibinfo{booktitle}{\emph{Proceedings of the 2018 ACM SIGSAC Conference on Computer and Communications Security}}. \bibinfo{pages}{1667--1680}.
\newblock


\bibitem[Hindle et~al\mbox{.}(2016)]%
        {hindle2016naturalness}
\bibfield{author}{\bibinfo{person}{Abram Hindle}, \bibinfo{person}{Earl~T Barr}, \bibinfo{person}{Mark Gabel}, \bibinfo{person}{Zhendong Su}, {and} \bibinfo{person}{Premkumar Devanbu}.} \bibinfo{year}{2016}\natexlab{}.
\newblock \showarticletitle{On the naturalness of software}.
\newblock \bibinfo{journal}{\emph{Commun. ACM}} \bibinfo{volume}{59}, \bibinfo{number}{5} (\bibinfo{year}{2016}), \bibinfo{pages}{122--131}.
\newblock


\bibitem[Hu et~al\mbox{.}(2022)]%
        {hu2022lora}
\bibfield{author}{\bibinfo{person}{Edward~J Hu}, \bibinfo{person}{Yelong Shen}, \bibinfo{person}{Phillip Wallis}, \bibinfo{person}{Zeyuan Allen-Zhu}, \bibinfo{person}{Yuanzhi Li}, \bibinfo{person}{Shean Wang}, \bibinfo{person}{Lu Wang}, \bibinfo{person}{Weizhu Chen}, {et~al\mbox{.}}} \bibinfo{year}{2022}\natexlab{}.
\newblock \showarticletitle{Lora: Low-rank adaptation of large language models.}
\newblock \bibinfo{journal}{\emph{ICLR}} \bibinfo{volume}{1}, \bibinfo{number}{2} (\bibinfo{year}{2022}), \bibinfo{pages}{3}.
\newblock


\bibitem[Huang et~al\mbox{.}(2019)]%
        {huang2019clinicalbert}
\bibfield{author}{\bibinfo{person}{Kexin Huang}, \bibinfo{person}{Jaan Altosaar}, {and} \bibinfo{person}{Rajesh Ranganath}.} \bibinfo{year}{2019}\natexlab{}.
\newblock \showarticletitle{Clinicalbert: Modeling clinical notes and predicting hospital readmission}.
\newblock \bibinfo{journal}{\emph{arXiv preprint arXiv:1904.05342}} (\bibinfo{year}{2019}).
\newblock


\bibitem[Jiang et~al\mbox{.}(2025)]%
        {jiang2025beyond}
\bibfield{author}{\bibinfo{person}{Linxi Jiang}, \bibinfo{person}{Xin Jin}, {and} \bibinfo{person}{Zhiqiang Lin}.} \bibinfo{year}{2025}\natexlab{}.
\newblock \showarticletitle{Beyond Classification: Inferring Function Names in Stripped Binaries via Domain Adapted LLMs.}. In \bibinfo{booktitle}{\emph{NDSS}}.
\newblock


\bibitem[Jin et~al\mbox{.}(2023)]%
        {jin2023binary}
\bibfield{author}{\bibinfo{person}{Xin Jin}, \bibinfo{person}{Jonathan Larson}, \bibinfo{person}{Weiwei Yang}, {and} \bibinfo{person}{Zhiqiang Lin}.} \bibinfo{year}{2023}\natexlab{}.
\newblock \showarticletitle{Binary code summarization: Benchmarking chatgpt/gpt-4 and other large language models}.
\newblock \bibinfo{journal}{\emph{arXiv preprint arXiv:2312.09601}} (\bibinfo{year}{2023}).
\newblock


\bibitem[Jin et~al\mbox{.}(2022)]%
        {jin2022symlm}
\bibfield{author}{\bibinfo{person}{Xin Jin}, \bibinfo{person}{Kexin Pei}, \bibinfo{person}{Jun~Yeon Won}, {and} \bibinfo{person}{Zhiqiang Lin}.} \bibinfo{year}{2022}\natexlab{}.
\newblock \showarticletitle{Symlm: Predicting function names in stripped binaries via context-sensitive execution-aware code embeddings}. In \bibinfo{booktitle}{\emph{Proceedings of the 2022 ACM SIGSAC Conference on Computer and Communications Security}}. \bibinfo{pages}{1631--1645}.
\newblock


\bibitem[Kim et~al\mbox{.}(2023)]%
        {kim2023transformer}
\bibfield{author}{\bibinfo{person}{Hyunjin Kim}, \bibinfo{person}{Jinyeong Bak}, \bibinfo{person}{Kyunghyun Cho}, {and} \bibinfo{person}{Hyungjoon Koo}.} \bibinfo{year}{2023}\natexlab{}.
\newblock \showarticletitle{A Transformer-based Function Symbol Name Inference Model from an Assembly Language for Binary Reversing}. In \bibinfo{booktitle}{\emph{Proceedings of the 2023 ACM Asia Conference on Computer and Communications Security}}. \bibinfo{pages}{951--965}.
\newblock


\bibitem[Lacomis et~al\mbox{.}(2019)]%
        {lacomis2019dire}
\bibfield{author}{\bibinfo{person}{Jeremy Lacomis}, \bibinfo{person}{Pengcheng Yin}, \bibinfo{person}{Edward Schwartz}, \bibinfo{person}{Miltiadis Allamanis}, \bibinfo{person}{Claire Le~Goues}, \bibinfo{person}{Graham Neubig}, {and} \bibinfo{person}{Bogdan Vasilescu}.} \bibinfo{year}{2019}\natexlab{}.
\newblock \showarticletitle{Dire: A neural approach to decompiled identifier naming}. In \bibinfo{booktitle}{\emph{2019 34th IEEE/ACM International Conference on Automated Software Engineering (ASE)}}. IEEE, \bibinfo{pages}{628--639}.
\newblock


\bibitem[Lee et~al\mbox{.}(2020)]%
        {lee2020biobert}
\bibfield{author}{\bibinfo{person}{Jinhyuk Lee}, \bibinfo{person}{Wonjin Yoon}, \bibinfo{person}{Sungdong Kim}, \bibinfo{person}{Donghyeon Kim}, \bibinfo{person}{Sunkyu Kim}, \bibinfo{person}{Chan~Ho So}, {and} \bibinfo{person}{Jaewoo Kang}.} \bibinfo{year}{2020}\natexlab{}.
\newblock \showarticletitle{BioBERT: a pre-trained biomedical language representation model for biomedical text mining}.
\newblock \bibinfo{journal}{\emph{Bioinformatics}} \bibinfo{volume}{36}, \bibinfo{number}{4} (\bibinfo{year}{2020}), \bibinfo{pages}{1234--1240}.
\newblock


\bibitem[Li et~al\mbox{.}(2023)]%
        {li2023starcoder}
\bibfield{author}{\bibinfo{person}{Raymond Li}, \bibinfo{person}{Loubna~Ben Allal}, \bibinfo{person}{Yangtian Zi}, \bibinfo{person}{Niklas Muennighoff}, \bibinfo{person}{Denis Kocetkov}, \bibinfo{person}{Chenghao Mou}, \bibinfo{person}{Marc Marone}, \bibinfo{person}{Christopher Akiki}, \bibinfo{person}{Jia Li}, \bibinfo{person}{Jenny Chim}, {et~al\mbox{.}}} \bibinfo{year}{2023}\natexlab{}.
\newblock \showarticletitle{StarCoder: may the source be with you!}
\newblock \bibinfo{journal}{\emph{arXiv preprint arXiv:2305.06161}} (\bibinfo{year}{2023}).
\newblock


\bibitem[Luo et~al\mbox{.}(2023)]%
        {luo2023wizardcoder}
\bibfield{author}{\bibinfo{person}{Ziyang Luo}, \bibinfo{person}{Can Xu}, \bibinfo{person}{Pu Zhao}, \bibinfo{person}{Qingfeng Sun}, \bibinfo{person}{Xiubo Geng}, \bibinfo{person}{Wenxiang Hu}, \bibinfo{person}{Chongyang Tao}, \bibinfo{person}{Jing Ma}, \bibinfo{person}{Qingwei Lin}, {and} \bibinfo{person}{Daxin Jiang}.} \bibinfo{year}{2023}\natexlab{}.
\newblock \showarticletitle{WizardCoder: Empowering Code Large Language Models with Evol-Instruct}.
\newblock \bibinfo{journal}{\emph{arXiv preprint arXiv:2306.08568}} (\bibinfo{year}{2023}).
\newblock


\bibitem[NSA({[n.\,d.]})]%
        {ghidra}
\bibfield{author}{\bibinfo{person}{NSA}.} \bibinfo{year}{[n.\,d.]}\natexlab{}.
\newblock \bibinfo{title}{Ghidra Decompiler}.
\newblock \bibinfo{howpublished}{\url{https://ghidra-sre.org/}}.
\newblock
\newblock
\shownote{Accessed: 2025-04-01}.


\bibitem[OpenAI({[n.\,d.]})]%
        {chatGPT}
\bibfield{author}{\bibinfo{person}{OpenAI}.} \bibinfo{year}{[n.\,d.]}\natexlab{}.
\newblock \bibinfo{title}{chatGPT}.
\newblock \bibinfo{howpublished}{\url{https://chat.openai.com/}}.
\newblock
\newblock
\shownote{Accessed: 2026-04-10}.


\bibitem[Pal et~al\mbox{.}(2024)]%
        {pal2024len}
\bibfield{author}{\bibinfo{person}{Kuntal~Kumar Pal}, \bibinfo{person}{Ati~Priya Bajaj}, \bibinfo{person}{Pratyay Banerjee}, \bibinfo{person}{Audrey Dutcher}, \bibinfo{person}{Mutsumi Nakamura}, \bibinfo{person}{Zion~Leonahenahe Basque}, \bibinfo{person}{Himanshu Gupta}, \bibinfo{person}{Saurabh~Arjun Sawant}, \bibinfo{person}{Ujjwala Anantheswaran}, \bibinfo{person}{Yan Shoshitaishvili}, {et~al\mbox{.}}} \bibinfo{year}{2024}\natexlab{}.
\newblock \showarticletitle{“Len or index or count, anything but v1”: Predicting Variable Names in Decompilation Output with Transfer Learning}. In \bibinfo{booktitle}{\emph{2024 IEEE Symposium on Security and Privacy (SP)}}. IEEE Computer Society, \bibinfo{pages}{152--152}.
\newblock


\bibitem[Pei et~al\mbox{.}(2021)]%
        {pei2021stateformer}
\bibfield{author}{\bibinfo{person}{Kexin Pei}, \bibinfo{person}{Jonas Guan}, \bibinfo{person}{Matthew Broughton}, \bibinfo{person}{Zhongtian Chen}, \bibinfo{person}{Songchen Yao}, \bibinfo{person}{David Williams-King}, \bibinfo{person}{Vikas Ummadisetty}, \bibinfo{person}{Junfeng Yang}, \bibinfo{person}{Baishakhi Ray}, {and} \bibinfo{person}{Suman Jana}.} \bibinfo{year}{2021}\natexlab{}.
\newblock \showarticletitle{StateFormer: Fine-grained type recovery from binaries using generative state modeling}. In \bibinfo{booktitle}{\emph{Proceedings of the 29th ACM Joint Meeting on European Software Engineering Conference and Symposium on the Foundations of Software Engineering}}. \bibinfo{pages}{690--702}.
\newblock


\bibitem[Pei et~al\mbox{.}(2020)]%
        {pei2020trex}
\bibfield{author}{\bibinfo{person}{Kexin Pei}, \bibinfo{person}{Zhou Xuan}, \bibinfo{person}{Junfeng Yang}, \bibinfo{person}{Suman Jana}, {and} \bibinfo{person}{Baishakhi Ray}.} \bibinfo{year}{2020}\natexlab{}.
\newblock \showarticletitle{Trex: Learning execution semantics from micro-traces for binary similarity}.
\newblock \bibinfo{journal}{\emph{arXiv preprint arXiv:2012.08680}} (\bibinfo{year}{2020}).
\newblock


\bibitem[Roziere et~al\mbox{.}(2023)]%
        {roziere2023code}
\bibfield{author}{\bibinfo{person}{Baptiste Roziere}, \bibinfo{person}{Jonas Gehring}, \bibinfo{person}{Fabian Gloeckle}, \bibinfo{person}{Sten Sootla}, \bibinfo{person}{Itai Gat}, \bibinfo{person}{Xiaoqing~Ellen Tan}, \bibinfo{person}{Yossi Adi}, \bibinfo{person}{Jingyu Liu}, \bibinfo{person}{Tal Remez}, \bibinfo{person}{J{\'e}r{\'e}my Rapin}, {et~al\mbox{.}}} \bibinfo{year}{2023}\natexlab{}.
\newblock \showarticletitle{Code llama: Open foundation models for code}.
\newblock \bibinfo{journal}{\emph{arXiv preprint arXiv:2308.12950}} (\bibinfo{year}{2023}).
\newblock


\bibitem[Shang et~al\mbox{.}(2024)]%
        {shang2024far}
\bibfield{author}{\bibinfo{person}{Xiuwei Shang}, \bibinfo{person}{Shaoyin Cheng}, \bibinfo{person}{Guoqiang Chen}, \bibinfo{person}{Yanming Zhang}, \bibinfo{person}{Li Hu}, \bibinfo{person}{Xiao Yu}, \bibinfo{person}{Gangyang Li}, \bibinfo{person}{Weiming Zhang}, {and} \bibinfo{person}{Nenghai Yu}.} \bibinfo{year}{2024}\natexlab{}.
\newblock \showarticletitle{How Far Have We Gone in Binary Code Understanding Using Large Language Models}. In \bibinfo{booktitle}{\emph{2024 IEEE International Conference on Software Maintenance and Evolution (ICSME)}}. IEEE, \bibinfo{pages}{1--12}.
\newblock


\bibitem[Team(2024)]%
        {qwen2.5}
\bibfield{author}{\bibinfo{person}{Qwen Team}.} \bibinfo{year}{2024}\natexlab{}.
\newblock \bibinfo{title}{Qwen2.5: A Party of Foundation Models}.
\newblock
\urldef\tempurl%
\url{https://qwenlm.github.io/blog/qwen2.5/}
\showURL{%
\tempurl}


\bibitem[Team(2025)]%
        {qwen3technicalreport}
\bibfield{author}{\bibinfo{person}{Qwen Team}.} \bibinfo{year}{2025}\natexlab{}.
\newblock \bibinfo{title}{Qwen3 Technical Report}.
\newblock
\showeprint[arxiv]{2505.09388}~[cs.CL]
\urldef\tempurl%
\url{https://arxiv.org/abs/2505.09388}
\showURL{%
\tempurl}


\bibitem[Torrey and Shavlik(2010)]%
        {torrey2010transfer}
\bibfield{author}{\bibinfo{person}{Lisa Torrey} {and} \bibinfo{person}{Jude Shavlik}.} \bibinfo{year}{2010}\natexlab{}.
\newblock \showarticletitle{Transfer learning}.
\newblock In \bibinfo{booktitle}{\emph{Handbook of research on machine learning applications and trends: algorithms, methods, and techniques}}. \bibinfo{publisher}{IGI global}, \bibinfo{pages}{242--264}.
\newblock


\bibitem[Touvron et~al\mbox{.}(2023)]%
        {touvron2023llama}
\bibfield{author}{\bibinfo{person}{Hugo Touvron}, \bibinfo{person}{Louis Martin}, \bibinfo{person}{Kevin Stone}, \bibinfo{person}{Peter Albert}, \bibinfo{person}{Amjad Almahairi}, \bibinfo{person}{Yasmine Babaei}, \bibinfo{person}{Nikolay Bashlykov}, \bibinfo{person}{Soumya Batra}, \bibinfo{person}{Prajjwal Bhargava}, \bibinfo{person}{Shruti Bhosale}, {et~al\mbox{.}}} \bibinfo{year}{2023}\natexlab{}.
\newblock \showarticletitle{Llama 2: Open foundation and fine-tuned chat models}.
\newblock \bibinfo{journal}{\emph{arXiv preprint arXiv:2307.09288}} (\bibinfo{year}{2023}).
\newblock


\bibitem[Vaswani et~al\mbox{.}(2017)]%
        {vaswani2017attention}
\bibfield{author}{\bibinfo{person}{Ashish Vaswani}, \bibinfo{person}{Noam Shazeer}, \bibinfo{person}{Niki Parmar}, \bibinfo{person}{Jakob Uszkoreit}, \bibinfo{person}{Llion Jones}, \bibinfo{person}{Aidan~N Gomez}, \bibinfo{person}{{\L}ukasz Kaiser}, {and} \bibinfo{person}{Illia Polosukhin}.} \bibinfo{year}{2017}\natexlab{}.
\newblock \showarticletitle{Attention is all you need}.
\newblock \bibinfo{journal}{\emph{Advances in neural information processing systems}}  \bibinfo{volume}{30} (\bibinfo{year}{2017}).
\newblock


\bibitem[Xie et~al\mbox{.}(2024)]%
        {xie2024resym}
\bibfield{author}{\bibinfo{person}{Danning Xie}, \bibinfo{person}{Zhuo Zhang}, \bibinfo{person}{Nan Jiang}, \bibinfo{person}{Xiangzhe Xu}, \bibinfo{person}{Lin Tan}, {and} \bibinfo{person}{Xiangyu Zhang}.} \bibinfo{year}{2024}\natexlab{}.
\newblock \showarticletitle{Resym: Harnessing llms to recover variable and data structure symbols from stripped binaries}. In \bibinfo{booktitle}{\emph{Proceedings of the 2024 on ACM SIGSAC Conference on Computer and Communications Security}}. \bibinfo{pages}{4554--4568}.
\newblock


\bibitem[Xu et~al\mbox{.}(2023a)]%
        {xu2023improving}
\bibfield{author}{\bibinfo{person}{Xiangzhe Xu}, \bibinfo{person}{Shiwei Feng}, \bibinfo{person}{Yapeng Ye}, \bibinfo{person}{Guangyu Shen}, \bibinfo{person}{Zian Su}, \bibinfo{person}{Siyuan Cheng}, \bibinfo{person}{Guanhong Tao}, \bibinfo{person}{Qingkai Shi}, \bibinfo{person}{Zhuo Zhang}, {and} \bibinfo{person}{Xiangyu Zhang}.} \bibinfo{year}{2023}\natexlab{a}.
\newblock \showarticletitle{Improving Binary Code Similarity Transformer Models by Semantics-Driven Instruction Deemphasis}.
\newblock  (\bibinfo{year}{2023}).
\newblock


\bibitem[Xu et~al\mbox{.}(2023b)]%
        {xu2023lmpa}
\bibfield{author}{\bibinfo{person}{Xiangzhe Xu}, \bibinfo{person}{Zhuo Zhang}, \bibinfo{person}{Shiwei Feng}, \bibinfo{person}{Yapeng Ye}, \bibinfo{person}{Zian Su}, \bibinfo{person}{Nan Jiang}, \bibinfo{person}{Siyuan Cheng}, \bibinfo{person}{Lin Tan}, {and} \bibinfo{person}{Xiangyu Zhang}.} \bibinfo{year}{2023}\natexlab{b}.
\newblock \showarticletitle{LmPa: Improving Decompilation by Synergy of Large Language Model and Program Analysis}.
\newblock \bibinfo{journal}{\emph{arXiv preprint arXiv:2306.02546}} (\bibinfo{year}{2023}).
\newblock


\bibitem[Yang et~al\mbox{.}(2021)]%
        {yang2021codee}
\bibfield{author}{\bibinfo{person}{Jia Yang}, \bibinfo{person}{Cai Fu}, \bibinfo{person}{Xiao-Yang Liu}, \bibinfo{person}{Heng Yin}, {and} \bibinfo{person}{Pan Zhou}.} \bibinfo{year}{2021}\natexlab{}.
\newblock \showarticletitle{Codee: A tensor embedding scheme for binary code search}.
\newblock \bibinfo{journal}{\emph{IEEE Transactions on Software Engineering}} \bibinfo{volume}{48}, \bibinfo{number}{7} (\bibinfo{year}{2021}), \bibinfo{pages}{2224--2244}.
\newblock


\bibitem[Yang et~al\mbox{.}(2023)]%
        {yang2023finchain}
\bibfield{author}{\bibinfo{person}{Xinze Yang}, \bibinfo{person}{Chunkai Zhang}, \bibinfo{person}{Yizhi Sun}, \bibinfo{person}{Kairui Pang}, \bibinfo{person}{Luru Jing}, \bibinfo{person}{Shiyun Wa}, {and} \bibinfo{person}{Chunli Lv}.} \bibinfo{year}{2023}\natexlab{}.
\newblock \showarticletitle{FinChain-BERT: A High-Accuracy Automatic Fraud Detection Model Based on NLP Methods for Financial Scenarios}.
\newblock \bibinfo{journal}{\emph{Information}} \bibinfo{volume}{14}, \bibinfo{number}{9} (\bibinfo{year}{2023}), \bibinfo{pages}{499}.
\newblock


\bibitem[Zhang et~al\mbox{.}(2021)]%
        {zhang2021osprey}
\bibfield{author}{\bibinfo{person}{Zhuo Zhang}, \bibinfo{person}{Yapeng Ye}, \bibinfo{person}{Wei You}, \bibinfo{person}{Guanhong Tao}, \bibinfo{person}{Wen-chuan Lee}, \bibinfo{person}{Yonghwi Kwon}, \bibinfo{person}{Yousra Aafer}, {and} \bibinfo{person}{Xiangyu Zhang}.} \bibinfo{year}{2021}\natexlab{}.
\newblock \showarticletitle{Osprey: Recovery of variable and data structure via probabilistic analysis for stripped binary}. In \bibinfo{booktitle}{\emph{2021 IEEE Symposium on Security and Privacy (SP)}}. IEEE, \bibinfo{pages}{813--832}.
\newblock


\end{thebibliography}
